 \documentclass[11pt]{elsarticle}

\usepackage{lineno,hyperref}

\journal{Journal of \LaTeX\ Templates}

\usepackage{amsmath,amssymb,amsfonts,mathrsfs}
\usepackage{color}
\usepackage{braket}
\usepackage{url}

\newtheorem{theorem}{Theorem}[section]

\newtheorem{corollary}{Corollary}[section]

\newtheorem{definition}{Definition}[section]
\newtheorem{example}{Example}[section]
\newtheorem{exercise}{Exercise}[section]
\newtheorem{lemma}{Lemma}[section]

\newtheorem{problem}{Problem}[section]
\newtheorem{proposition}{Proposition}[section]
\newtheorem{remark}{Remark}[section]
\newcommand{\bthm}{\begin{theorem}}
\newcommand{\ethm}{\end{theorem}}
\newcommand{\blem}{\begin{lemma}}
\newcommand{\elem}{\end{lemma}}
\newcommand{\bex}{\begin{example}}
\newcommand{\eex}{\end{example}}
\newcommand{\beg}{\begin{exercise}}
\newcommand{\eeg}{\end{exercise}}
\newcommand{\bprop}{\begin{proposition}}
\newcommand{\eprop}{\end{proposition}}
\newcommand{\bplm}{\begin{problem}}
\newcommand{\eplm}{\end{problem}}
\newcommand{\bmrk}{\begin{remark}}
\newcommand{\emrk}{\end{remark}}
\newcommand{\bdfn}{\begin{definition}}
\newcommand{\edfn}{\end{definition}}
\newcommand{\bcor}{\begin{corollary}}
\newcommand{\ecor}{\end{corollary}}

\newcommand{\beq}{\begin{equation}}
\newcommand{\eeq}{\end{equation}}
\newcommand{\beqm}{\begin{equation*}}
\newcommand{\eeqm}{\end{equation*}}
\newcommand{\beqn}{\begin{eqnarray}}
\newcommand{\eeqn}{\end{eqnarray}}
\newcommand{\beqnm}{\begin{eqnarray*}}
\newcommand{\eeqnm}{\end{eqnarray*}}
\newcommand{\bea}{\begin{aligned}}
\newcommand{\eea}{\end{aligned}}
\newcommand{\beam}{\begin{aligned*}}
\newcommand{\eeam}{\end{aligned*}}
\newcommand{\bs}{\begin{subequations}}
\newcommand{\es}{\end{subequations}}

\newcommand{\bei}{\begin{itemize}}
\newcommand{\eei}{\end{itemize}}
\newcommand{\bed}{\begin{description}}
\newcommand{\eed}{\end{description}}
\newcommand{\bee}{\begin{enumerate}}
\newcommand{\eee}{\end{enumerate}}
\newcommand{\bey}{\begin{array}}
\newcommand{\eey}{\end{array}}
\newcommand{\bec}{\begin{center}}
\newcommand{\eec}{\end{center}}
\newcommand{\la}{\label}

\newcommand{\mbb}{\mathbb}
\newcommand{\mbf}{\mathbf}
\newcommand{\mrm}{\mathrm}

\def\l{\langle}
\def\r{\rangle}
\def\ff{\frac}

\def\bf[#1]{\mathbf{#1}}
\def\mm[#1]{{\rm #1}}

\newcommand\Tstrut{\rule{0pt}{2.6ex}}         

\begin{document}

\begin{frontmatter}

\title{Linear quantum systems: a tutorial}

\author[mymainaddress,mysecondaryaddress]{Guofeng Zhang}
\ead{guofeng.zhang@polyu.edu.hk}
\address[mymainaddress]{Department of Applied Mathematics, The Hong Kong Polytechnic University, Hung Hom, Kowloon, Hong Kong SAR, China}
\address[mysecondaryaddress]{Shenzhen Research Institute, The Hong Kong Polytechnic University, Shenzhen, China}

\author[mythirdaddress]{Zhiyuan Dong}
\ead{dongzhiyuan@hit.edu.cn}
\address[mythirdaddress]{School of Science, Harbin Institute of Technology, Shenzhen, China}

\begin{abstract}
The purpose of this tutorial is to give a brief introduction to linear quantum control systems. The mathematical model of linear quantum control systems is presented first, then some fundamental control-theoretic notions such as stability, controllability and observability are given, which are closely related to several important concepts in quantum information science such as decoherence-free subsystems, quantum non-demolition variables, and back-action evasion measurements. After that, quantum Gaussian states are introduced, in particular, an information-theoretic uncertainty relation is presented which often gives a better bound for mixed Gaussian states than the well-known Heisenberg uncertainty relation. The quantum Kalman filter is presented for quantum linear systems, which is the quantum analogy of the Kalman filter for classical (namely, non-quantum-mechanical) linear systems. The quantum Kalman canonical decomposition for quantum linear systems is recorded, and its application is illustrated by means of a recent experiment. As single- and multi-photon states are useful resources in quantum information technology, the response of quantum linear systems to these types of input is presented. Finally, coherent feedback control of quantum linear systems is briefly introduced, and a recent experiment is used to demonstrate the effectiveness of quantum linear systems and networks theory.
\end{abstract}


\end{frontmatter}


\tableofcontents

   \section{Introduction}

The dynamics of quantum systems are governed by quantum-mechanical laws. The temporal evolution of a quantum system can be described by its state evolution in the Schr\"odinger picture. Alternatively,  it can also be described by the evolution of  system variables for example position and momentum of a quantum  harmonic oscillator, this is the so-called Heisenberg picture. System variables are operators in a Hilbert space, instead of ordinary functions. Therefore, the operations of these system variables may not commute.  Specifically, let $\mbf{A},\mbf{B}$ be two system variables (operators), $\mbf{AB}\neq \mbf{BA}$ may occur. Non-commutativity renders quantum systems fundamentally different from classical systems where system variables are functions of time and two system variables always commute.

A \textit{linear} quantum system is a quantum system whose temporal evolution
in the Heisenberg picture can be described by a set of \textit{linear}
differential equations of system variables.  Many physical systems can be well
modeled as linear quantum systems,  for instance, quantum optical
systems \cite{WM94,GZ00,WM08,Mabuchi08,WM10,ZJ12,P14,CKS17,NY17,PJU+20,BNC+21}, circuit
quantum electro-dynamical (circuit QED) systems \cite{MJP+11,BLS+11,KAK13,BGGW21},
cavity QED systems \cite{DJ99,SDZ+11,ASD+12}, quantum opto-mechanical
systems \cite{TC10,MHP+11,HM12,DFK+12,MCP+12,NY13,NY14,AKM14,ODP+16,TBCGKP2020,KPS+21,PVB+21,LOW+21},
atomic
ensembles \cite{SvHM04,NJP09,NY13,ANP+17,TBCGKP2020}, and
quantum memories \cite{XDL07,HRG+09,HCH+13,YJ14,NG15}.

Quantum linear systems have been studied extensively, and many results have
been  recorded in the well-known books \cite[][chapter 7]{WM08}, \cite[]{NY17}, \cite[][chapter 6]{WM10},  and a recent survey paper \cite{P18}.
The aim of this tutorial is to give a concise introduction to quantum linear
systems with emphasis on recent development.

This tutorial is organized as follows. Quantum linear systems are introduced in Section
\ref{sec:system}. Some important structural properties, such as stability,
controllability and observability, are summarized in Section  \ref{sec:concepts}. It
is shown that these concepts, widely used in systems and control theory, are
closely related to important properties of quantum linear systems such as
decoherence-free subsystems, quantum non-demolition variables, quantum
mechanics-free subsystems and quantum back-action evasion measurement.  In Section
\ref{sec:Gaussian state}, quantum Gaussian states are introduced. The Wigner
function is given, and an example is used to demonstrate the Heisenberg
uncertainty relation. Skew information and an information-theoretic uncertainty
relation is presented. In Section  \ref{sec:kalman filter}, the quantum Kalman filter
is introduced. A general introduction to quantum filters is first presented in Subsection
\ref{subsec:quantum filter}, after that the quantum Kalman filter for quantum
linear systems as well as a derivation procedure is given in Subsection
\ref{subsec:quantum kalman filter}. The purpose of providing a derivation
procedure is to illustrate some commonly used techniques in the study of
quantum linear systems such as Eqs.~\eqref{dec19-6} and \eqref{eq:XYZ}.   An example is
given in Subsection  \ref{subsec:quantum kalman filter example} which illustrates the
quantum Kalman filter and also demonstrates measurement back-action effect. In Section
\ref{sec:kalman}, several interesting structural properties of quantum linear
systems are summarized, then the quantum Kalman canonical form is presented. An
example, taken from a recent experiment \cite[Fig. 1(A)]{LOW+21}, is analyzed
in Subsection  \ref{subsec: canonical example}. In  Subsection \ref{sec:single-photon-response},
continuous-mode single-photon states are introduced, and the response of
quantum linear systems to this type of quantum states is given.  In Section
\ref{sub:feedback}, a general form of quantum coherent feedback linear
networks is presented in Subsection  \ref{subsec:coherent feedback},  and a recent
experiment is analyzed based on the proposed theory in Subsection  \ref{subsec:coherent
feedback example}. Some concluding remarks are given in Section  \ref{sec:conclusion}.

\medskip

\textit{Notation}.
\begin{itemize}
\item $\imath =\sqrt{-1}$ is the imaginary unit. $I_{k}$ is the identity matrix and $0_{k}$ the zero matrix in $\mathbb{C}^{k \times k}$. $\delta_{ij}$ denotes the Kronecker delta;
i.e.,~$I_k=[\delta_{ij}]$. $\delta(t)$ is the Dirac delta function.

\item $x^{\ast}$ denotes the complex conjugate of a complex number $x$ or
the adjoint of an operator $x$. Clearly. $(xy)^\ast = y^\ast x^\ast$.  Given two operators $\bf[x]$ and $\bf[y]$, their commutator is defined to be $[\bf[x],\bf[y]] \triangleq \bf[x]\bf[y]-\bf[y]\bf[x]$.

\item For a matrix $X=[x_{ij}]$ with  number or operator entries,  $X^{\top}=[x_{ji}]$ is the matrix transpose. Denote $X^{\#}=[x_{ij}^{\ast}]$, and $X^{\dagger}=(X^{\#})^{\top}$. For a vector $x$, we define $\breve{x}\triangleq \bigl[
\begin{smallmatrix}
x \\
x^{\#}
\end{smallmatrix}
\bigr]$.

\item Given two \textit{column} vectors of operators $\bf[X]$ and $\bf[Y]$ of the same length,   their commutator
is defined as
\beq
[\bf[X],\bf[Y]^\top] \triangleq ([\bf[X]_j,\bf[Y]_k] ) =\bf[X]\bf[Y]^\top- (\bf[Y]\bf[X]^\top)^\top.
\eeq
If $\bf[X]$ is a \textit{row} vector of operators of length $m$ and $\bf[Y]$ is a \textit{column} vector of operators of length $n$, their commutator is defined as
\begin{equation}
\label{dec19-6}
[\mbf{X},\mbf{Y}] \triangleq \left(\begin{array}{@{}ccc@{}}                               [\mbf{x}_1,\mbf{y}_1] & \cdots & [\mbf{x}_m,\mbf{y}_1] \\
                               {\vdots} & \ddots & \vdots \\
                               {[\mbf{x}_1,\mbf{y}_n]} & \cdots & [\mbf{x}_m,\mbf{y}_n]
\end{array}
                           \right)_{n\times m}=(\mbf{X}^\top\mbf{Y}^\top)^\top-\mbf{Y}\mbf{X}.
\end{equation}

\item Let $J_{k} \triangleq \mathrm{diag}(I_k,-I_k)$. For a matrix $X\in
\mathbb{C}^{2k\times 2r}$, define its $\flat$-adjoint by $X^{\flat }
\triangleq J_{r}X^{\dagger}J_{k}$. The $\flat$-adjoint operation enjoys the following  properties:
\beq
(x_1 A + x_2 B)^{\flat}=x_1^{*}
A^{\flat} + x_2^{*} B^{\flat}, \ \ (AB)^{\flat}=B^{\flat} A^{\flat}, \ \
(A^{\flat})^{\flat}=A,
\eeq
where $x_1,x_2\in \mathbb{C}$.

\item Given two matrices $U$, $V\in \mathbb{C}^{k\times r}$, define their
 \emph{doubled-up} \cite{GJN10} as  $\Delta
(U,V) \triangleq
\bigl[
\begin{smallmatrix}
U & V \\
V^{\#} & U^{\#}
\end{smallmatrix}
\bigr]$.  The set
of doubled-up matrices is closed under addition, multiplication and $\flat$ adjoint operation.

\item A matrix $T \in \mathbb{C}^{2k\times 2k}$ is called \emph{Bogoliubov}
if it is doubled-up and satisfies $TT^{\flat}=T^{\flat}T=I_{2k}$. The set of Bogoliubov matrices
forms a complex non-compact Lie group known as the Bogoliubov group.

\item Let $\mathbb{J}_{k} \triangleq \bigl[
\begin{smallmatrix}
0_{k} & I_k \\
-I_k & 0_{k}
\end{smallmatrix}
\bigr]$. For a matrix $X\in \mathbb{C}^{2k\times 2r}$, define its $\sharp$-
\emph{adjoint} $X^{\sharp}$ by $X^{\sharp} \triangleq -\mathbb{J}_{r}X^{\dagger}
\mathbb{J}_{k}$. The $\sharp$-\emph{adjoint} satisfies properties similar to
the usual adjoint, namely
\beq
(x_1 A + x_2 B)^{\sharp}=x_1^{*} A^{\sharp} + x_2^{*}
B^{\sharp}, \ \ (AB)^{\sharp}=B^{\sharp} A^{\sharp},  \ \ (A^{\sharp})^{
\sharp}=A.
\eeq

\item A matrix $\mathbb{S} \in \mathbb{C}^{2k\times 2k}$ is called \emph{symplectic},
if $\mathbb{S}\mathbb{S}^{\sharp}=\mathbb{S}^{\sharp}\mathbb{S}=I_{2k}$. Symplectic
matrices forms a complex non-compact group known as the symplectic group.
The subgroup of real symplectic matrices is one-to-one homomorphic to the
Bogoliubov group.
\end{itemize}

   \section{Quantum linear systems}
\label{sec:system}

\begin{figure}
\includegraphics{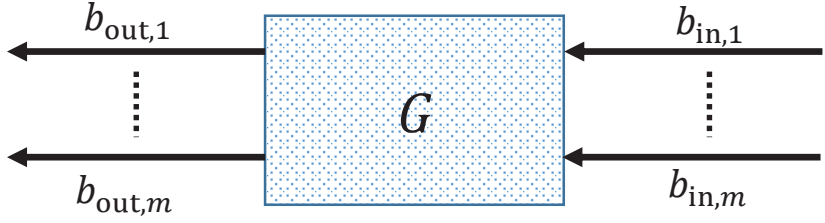}
\centering
\caption{A quantum linear system.\label{fig:sys}}
\end{figure}

Mathematically, a linear quantum system, as shown in Fig. \ref{fig:sys},  describes the dynamics of a collection of $n$ quantum harmonic oscillators which are driven by $m$ bosonic fields, for example light fields. The $j$th quantum harmonic oscillator is represented by its annihilation operator $\mbf{a}_j$ and creation operator $\mbf{a}_j^\ast$ (the Hilbert space adjoint of $\mbf{a}_j$). If the $j$th harmonic oscillator is  in the number state $\ket{k}$ for $k\in \mathbb{Z}^+$, then $\mbf{a}_j\ket{k} = \sqrt{k}\ket{k-1}$ and  $\mbf{a}_j^\ast\ket{k} = \sqrt{k+1}\ket{k+1}$. In particular,  $\ket{0}$ is the vacuum state and  $\mbf{a}_j\ket{0}=0$. From these it is easy to see that the commutator $[\mbf{a}_j, \mbf{a}_j^\ast]=1$.  In general, the operators $\mbf{a}_{j},\mbf{a}_{k}^{\ast }$ satisfy the \emph{canonical
commutation relations}
\beq\label{eq:CCR_a}
\begin{split}
& [\mbf{a}_{j}(t),~\mbf{a}_{k}(t)]= [\mbf{a}_{j}^{\ast}(t),~\mbf{a}_{k}^{\ast}(t)]=0,\;[\mbf{a}_{j}(t),~\mbf{a}_{k}^{\ast}(t)]=\delta _{jk}, \ \ \  \forall
j,k=1,\ldots,n, \forall t \in \mathbb{R}.
\end{split}
\eeq
 Let $\mbf{a}=[ \mbf{a}_{1}~\cdots ~\mbf{a}_{n}]^{\top}$ and $\breve{\mbf{a}}=[
\mbf{a}^{\top}~(\mbf{a}^{\#})^{\top}]^{\top}$ as introduced in the \textit{Notation} part. Then Eq.~\eqref{eq:CCR_a} can be rewritten as
\beq\label{eq:CCR_a_2}
[\breve{\mbf{a}}(t), \breve{\mbf{a}}^\dagger(t)] = J_n.
\eeq
 The Hamiltonian of  the quantum linear system is at most quadratic in terms of $\mbf{a}$ and $\mbf{a}^\#$, thus it is of the form
\beq
 \mbf{H}=\frac{1}{2}\breve{\mbf{a}}
^{\dagger}\Omega \breve{\mbf{a}}+\breve{\mbf{a}}^\dagger K \breve{v} +\breve{v}^\dagger K^\dagger  \breve{\mbf{a}},
\eeq
where  $\Omega
=\Delta (\Omega _{-},\Omega _{+})$ is a
Hermitian matrix with $\Omega _{-},\Omega _{+}\in \mathbb{C}^{n\times n}$, $K =\bigl[
\begin{smallmatrix}
K_1 & K_2\\
K_3 & K_4
\end{smallmatrix}
\bigr]\in \mathbb{C}^{2n\times 2l}$, and $\breve{v}\in \mathbb{C}^{2l\time 1}$  is  a vector
of  classical signals used to model the laser or other classical signals that
drive the system (\cite[Eq. (20)]{EB05}, \cite[Chapter 3]{GZ00}, \cite[Section 5.1.1]{NY17}, \cite[Eq. (1)]{PVB+21}, \cite[Eq. (C6)]{WSU+18}, \cite[Eq.
(9)]{WD05}).  For example, the quantum system can be a particle moving in a
potential well whose potential can be controlled. In the literature of
measurement-based quantum feedback control \cite{WM10},  $\breve{\mbf{a}}^\dagger K \breve{v} +\breve{v}^\dagger K^\dagger  \breve{\mbf{a}}$ are often
called the control Hamiltonian. For example, a quantum system can be measured
and the measurement data can be processed by a classical controller that
generates the classical control signal $v$ which is sent back to the
quantum system to modulate its dynamics.    The system is coupled to the input
fields via the operator $\mbf{L}=[C_{-}\ C_{+}] \breve{\mbf{a}}$, with $C_{-},C_{+}\in
\mathbb{C}^{m\times n}$. The input boson field
$k$, $k=1,\ldots, m$, is
described in terms of an annihilation operator $\mbf{b}_{\mathrm{in},k}(t)$ and a
creation operator $\mbf{b}_{\mathrm{in},k}^{\ast }(t)$, which is the adjoint operator of $
\mbf{b}_{\mathrm{in},k}(t)$. If there are no photons in an input channel, this input channel is in the vacuum state denoted $\ket{\Phi_0}$. Annihilation and creation operators of these free traveling fields satisfy the following commutation relations
\beq\label{eq:CCR_b}
\begin{aligned}
&\mbf{b}_{\mathrm{in}, j}(t)\ket{\Phi_0}=0,\  [\mbf{b}_{\mathrm{in}, j}(t),\mbf{b}_{\mathrm{in}, k}(r)]=[\mbf{b}_{\mathrm{in}, j}^\ast(t),\mbf{b}_{\mathrm{in}, k}^\ast(r)]=0,
\\
&[\mbf{b}_{\mathrm{in}, j}(t),\mbf{b}_{\mathrm{in}, k}^\ast(r)]=\delta_{jk}\delta(t-r),  \ \  \forall j,k=1,\ldots,m, \ t,r\in\mbb{R}.
\end{aligned}
\eeq
Eq.~\eqref{eq:CCR_b} can be re-written in the vector form
\beq \label{eq:CCR_b2}
\mbf{b}_{\mathrm{in}}(t)\ket{\Phi_0}=0, \ \;[\breve{\mbf{b}}_{\rm in}(t), \ \breve{\mbf{b}}^\dagger_{\rm in}(r)] = \delta(t-r)J_m, \ t,r\in\mbb{R}.
\eeq
Before interacting with the system, the input fields may pass through some static devices, for example beamsplitters and phase shifters. This is modeled by a unitary matrix $S\in\mathbb{C}^{m\times m}$. $S$ is often referred to as \textit{scattering matrix} in the quantum optics literature.

The integrated input annihilation, creation, and gauge processes (counting processes) are given by
\beq\label{eq:gauge}
\mbf{B}_{\rm in}(t) = \int_{t_0}^t \mbf{b}_{\rm in}(r)dr,  \ \   \mbf{B}^\#(t)=\int_{t_0}^t \mbf{b}_{\rm in}^\#(r)dr, \ \ \mbf{\Lambda}_{\rm in}(t) =\int_{t_0}^t \mbf{b}_{\rm in}^\# (r)\mbf{b}_{\rm in}^\top(r) dr.
\eeq
respectively, where $t_0$ is the initial time when the system and the input fields start interaction. In this article, the input fields are assumed to be \textit{canonical} fields which satisfy the It\^o table (\ref{table:Ito}).

\begin{table}[htp]
\caption{Quantum It\^{o} table}
\begin{center}
\begin{tabular}{c|cccc}
\hline
$\times$ & $d\mbf{B}_{{\rm in},l}(t)$ & $d\mbf{B}_{{\rm in},m}^\ast(t)$ & $d\mbf{\Lambda}_{{\rm in},lm}(t)$ & $dt$ \\
\hline
$d\mbf{B}_{{\rm in},j}(t)$ &  0 & $\delta_{jm}dt$  & $\delta_{jl}d\mbf{B}_m(t)$  & 0 \\
$d\mbf{B}_{{\rm in},j}^\ast(t)$  & 0 & 0 & 0 &0 \\
$d\mbf{\Lambda}_{{\rm in},jk}(t)$    & 0  & $\delta_{km}d\mbf{B}_{{\rm in},j}^\ast(t)$ & $\delta_{kl}d\mbf{\Lambda}_{{\rm in},jm}(t)$ &0\\
$dt$ & 0 & 0 & 0 &0 \\
\hline
\end{tabular}
\end{center}
\label{table:Ito}
\end{table}%

In the Heisenberg picture,  in terms of the triple of parameters $(S, \mbf{L}, \mbf{H})$ introduced above, the dynamics of the open  quantum   system in  \ref{fig:sys} is governed by a \textit{unitary} operator $\mbf{U}(t)$ that is the solution to the following It\^o quantum  stochastic differential equation (QSDE), first developed by \cite{HP84},
\beq\la{U}
\bea
d\mbf{U}(t)&=\; \bigm\{-(\mbf{L}^{\dagger}\mbf{L}/2+\imath\mbf{H}) dt+d\mbf{B}_{\rm in}^{\dagger}(t)\mbf{L}
\\
& \hskip 7ex -\mbf{L}^{\dagger}S d\mbf{B}_{\rm in}(t)+\mathrm{Tr}[(S-I)d\mbf{\Lambda}_{\rm in} ^{\top}(t)]\bigm\} \mbf{U}(t), \ \ t\geq t_0
\eea
\eeq
with the initial condition $U(t_0)=I$ (the identity operator).  Let $\mbf{X}$ be a system operator. In the Heisenberg picture, we have
\beq \la{eq:X_t}
 \mbf{X}(t)=\mbf{U}^\ast(t)(\mbf{X}\otimes I_{\mathrm{field}})\mbf{U}(t), \ \ t\geq t_0.
\eeq
According to Eq.~\eqref{U} and by quantum It\^o calculus, we have
\beq\la{eq:X}
\bea
d\mbf{X}(t)
=
&\; \mathcal{L}_{\mbf{L},\mbf{H}}(\mbf{X}(t))dt+d\mbf{B}_{\rm in}^{\dagger}(t)S^{\dagger}[\mbf{X}(t),\mbf{L}(t)]
\\
&\;+[\mbf{L}^{\dagger}(t), \mbf{X}(t)] S d\mbf{B}_{\rm in}(t)+\mathrm{Tr}[(S^\dagger \mbf{X}(t) S-\mbf{X}(t))d\mbf{\Lambda}_{\rm in} ^{\top}(t)],
\eea
\eeq
where the superoperator
\begin{equation}
 \label{eq:L_LH}
\mathcal{L}_{\mbf{L},\mbf{H}}(X(t)) \triangleq -\imath[\mbf{X}(t),\mbf{H}(t)]+\frac{1}{2}\mbf{L}^{\dagger}(t)[\mbf{X}(t),\mbf{L}(t)]+\frac{1}{2}[\mbf{L}^{\dagger}(t), \mbf{X}(t)]\mbf{L}(t).
\end{equation}
(Notice that even initially $X$ is an operator on the system's state space, $X(t)$ is an operator on the state space of the joint system-field system).

On the other hand, after interaction, the integrated output annihilation operators and gauge processes
\beq \la{eq:output}
\bea
\mbf{B}_{\mathrm{out}}(t) =&\; \mbf{U}^\ast(t)(I_{\mathrm{system}}\otimes \mbf{B}_{\rm in}(t))\mbf{U}(t),
\\
 \mbf{\Lambda}_{\mathrm{out}}(t)=&\; \mbf{U}^\ast(t)(I_{\mathrm{system}}\otimes \mbf{\Lambda}_{\rm in}(t))\mbf{U}(t)
\eea
\eeq
are generated, and their dynamical evolution  is given by
\beq\label{eq_B_out and Lambda_out}
\bea
d\mbf{B}_{\mathrm{out}}(t)=&\; \mbf{L}(t)dt+Sd\mbf{B}_{\rm in }(t),
\\
d \mbf{\Lambda}_{\mathrm{out}}(t)=&\; \mbf{L}^\#(t)\mbf{L}^\top(t)dt+S^\#d\mbf{B}_{\rm in}^\#(t)L^\top(t)
\\
&\; +\mbf{L}^\#(t)d\mbf{B}_{\rm in}^\top(t)S^\top(t)+S^\#d \mbf{\Lambda}_{\rm in}(t)S^\top.
\eea
\eeq

 \bmrk \label{rem:signal+noise}
{\rm
According to Eq.~\eqref{eq:CCR_b2}, the quantum  It\^o table is  satisfied by
the Bosonic fields that are in the vacuum state  $\ket{\Phi_0}$. In quantum
optics, the input $\mbf{B}_{\rm in}$  may be of the from $d\mbf{B}_{\rm in}(t) = w(t)dt+d\tilde{\mbf{B}}_{\rm in}(t)$, where
$w(t)$ can be an operator of another quantum linear system, and
$\tilde{\mbf{B}}_{\rm in}(t)$ is in the vacuum state \cite{GZ00,JNP08}. For example,
$\mbf{B}_{\rm in}$ may be the output of a quantum linear system $K$.  In
this case, in analogy to Eq.~\eqref{eq_B_out and Lambda_out},  $\mbf{B}_{\rm in}(t)$
will be of the form $\mbf{L}_K(t)dt+S_Kd\mbf{B}_{{\rm in},K }(t)$ with $\mbf{L}_K, S_K, \mbf{B}_{{\rm in},K }(t)$ being the coupling operator,
scattering matrix and input filed for the quantum system $K$. Thus,
$w(t) = \mbf{L}_K(t)$ and  $d\tilde{\mbf{B}}_{\rm in}(t) = S_Kd\mbf{B}_{{\rm in},K }(t)$.  More detail can be found in Section \ref{sub:feedback}. The input field $d\mbf{B}_{\rm in}(t)$ of the above form also
satisfies the quantum It\^o table. Moreover, as will be shown in Subsection
\ref{subsec:photon_state}, fields in  continuous-mode single-photon states
also satisfy the quantum It\^o table. However, there do exist fields  that do
not satisfy the quantum It\^o table. For example, the coherent state
$\ket{\alpha}$ with $\alpha\neq 0$ to be introduced in Section  \ref{sec:Gaussian state}
does not satisfy the quantum  It\^o table. In such cases, another quantum
system, often referred to as a modulating filter may be designed which is
driven by a vacuum field and generates the desired input fields. Consequently,
by cascading the modulating filter and the original quantum system, the quantum
It\^o QSDE \eqref{U} still holds. The problem of generating  various
nonclassical quantum input field states with modulating filters has been
discussed in \cite{GJN14,GZ15b}, and \cite{LZW21}.
}
\emrk

By means of Eqs.~\eqref{eq:X} and \eqref{eq_B_out and Lambda_out}, it can be shown that  the evolution of the quantum linear system is governed  by a system of QSDEs
\beq\la{eq:sys}
\bea
\dot{\breve{\mbf{a}}}(t)
=&\;
\mathcal{A}\breve{\mbf{a}}(t)+ \mathcal{E} \breve{v}(t) +
\mathcal{B}\breve{\mbf{b}}_{\rm in}(t),
\\
\breve{\mbf{b}}_{\mathrm{out}}(t)
=&\;
\mathcal{C}\breve{\mbf{a}}(t)+ \mathcal{D}\breve{\mbf{b}}_{\rm in}(t),  \ \ t\geq t_0,
\eea
\eeq
where the constant system matrices are given by
\beq \la{eq:sys_ABCD}
\bea
&\; \mathcal{D}=\Delta (S,0), ~\mathcal{C}=\Delta(C_{-},\ C_{+}), ~\mathcal{B}=-
\mathcal{C}^{\flat }\mathcal{D},
\\
&\; \mathcal{A}=-\imath J_{n}\Omega-\frac{1}{2}\mathcal{C}^{\flat }\mathcal{C}, \ \  \mathcal{E} = -\imath \left(J_n K+\mathbb{J}_n K^\# \left[
\bey{@{}cc@{}}
0 & I_n\\
I_n &0
\eey
\right]\right).
\eea
\eeq

The form of the system Eq.~\eqref{eq:sys} is often referred to as  the quantum Langevin equation. It should be understood in the It\^o form
\beq
\bea
d\breve{\mbf{a}}(t)
=&\;
\mathcal{A}\breve{\mbf{a}}(t)dt+ \mathcal{E} \breve{v}(t)dt +
\mathcal{B}d\breve{\mbf{B}}_{\rm in}(t),
\\
d\breve{\mbf{B}}_{\mathrm{out}}(t)
=&\;
\mathcal{C}\breve{\mbf{a}}(t)dt+ \mathcal{D}d\breve{\mbf{B}}_{\rm in}(t),  \ \ t\geq t_0,
\eea
\eeq

 \bmrk{\rm
Notice that $v(t)$ in system \eqref{eq:sys} is a \textit{classical}
signal, for example photocurrent. On the other hand, in the literature of
quantum coherent feedback control, two quantum systems may directly couple to
each other via some interaction Hamiltonian, where the involved signals are all
\textit{quantum} operators; see, e.g, \cite[Figs.
8(b) and 16]{SY21}, \cite[Eqs. (3.1) and
(4.1)]{WM94}, \cite[Section II.B]{ZJ11}, and \cite[Fig. 3.1]{ZLH+12}.
}
\emrk

The constant  system matrices in Eq.~\eqref{eq:sys_ABCD} are parametrized by the physical parameters $\Omega_{-},\Omega_{+},C_{-},C_{+}, S$ and they satisfy the following \textit{physical realizability} conditions
\beq \label{eq:PR_a_b}
\bea
{\mathcal{A}}+{\mathcal{A}}^{\flat }+ {\mathcal{B}}{\mathcal{B}}^{\flat } =&\; 0,
\\
{\mathcal{B}} =&\; -{\mathcal{C}}^{\flat }{\mathcal{D}}.
\eea
\eeq

 \bmrk{\rm
Broadly speaking, if its system parameters satisfy the physical realizability conditions~\eqref{eq:PR_a_b}, the system \eqref{eq:sys} can be physically realized  by a genuine quantum-mechanical system. For example, the fundamental commutation relations  are preserved during temporal evolution
\beq \label{eq:phys_realiz}
\bea
\lbrack\breve{\mbf{a}}(t),~\breve{\mbf{a}}^{\dagger}(t)] =&\;
[ \breve{\mbf{a}}(t_0),~\breve{\mbf{a}}^{\dagger}(t_0)],~~~t\geq t_0,
\\
\lbrack \breve{\mbf{a}}(t),\breve{\mbf{b}}_{\mathrm{out}
}^{\dagger}(r)] =&\; 0,~t_0\leq r<t.
\eea
\eeq
The problem of physical realization of quantum linear systems was first
addressed in \cite{JNP08} and \cite{NJD09}. A comprehensive study of physical
realization of quantum linear systems is nicely summarized in \cite[Chapter
3]{NY17}, see also \cite{MP11}, \cite{SP12}, \cite{GZ15} and \cite[Sections 2.4 and 3]{P18}. Further
development can be found in \cite{WNZJ13,WNZJ18}.  The problem  of
physical realization of quantum nonlinear systems has been studied
in \cite{DEMP+16} and \cite{EWP18}. This physical realization theory for
quantum systems can be regarded as an generalization of the network analysis
and synthesis theory for classical systems \cite{AV73}.
}
\emrk

As in the classical linear systems theory, the \emph{impulse response function}  from $\breve{\mathbf{b}}_{\rm in}(t)$ to  $\breve{\mathbf{b}}_{\rm out}(t)$ is defined as
\begin{equation}\label{eq:gg}
g_{G}(t)\triangleq\left\{
\begin{array}{ll}
\delta(t){\cal D}-{\cal C}e^{{\cal A}t}\cal{C}^{\flat}{\cal D}, & t\geq 0,   \\
0, & t<0.
\end{array}
\right.
\end{equation}
Define matrix functions
\beq\label{eq:io}
\bea
g_{G^{-}}(t) \triangleq&\; \left\{\begin{array}{@{}ll@{}}\delta(t)S-[\begin{array}{@{}ll@{}}C_{-} & C_{+}
\end{array}
]e^{{\mathcal{A}}t}\left[\begin{array}{@{}c@{}}C_{-}^{\dagger} \\
-C_{+}^{\dagger}
\end{array}
\right]S, & t\geq 0,   \\
0, & t<0,\end{array}\right.
\\
\ g_{G^{+}}(t) \triangleq& \; \left\{\begin{array}{@{}ll@{}}-[\begin{array}{@{}ll@{}}C_{-} & C_{+}
\end{array}
]e^{{\mathcal{A}}t}\left[\begin{array}{@{}c@{}}-C_{+}^{T} \\
C_{-}^{T}
\end{array}
\right] , & t\geq 0,  \\
0 , & t<0.\end{array}\right.
\eea
\eeq
It is easy to show that the impulse response function $g_{G}(t)$ defined in Eq.~\eqref{eq:gg} has a nice  structure of the form
\begin{equation}
\label{eq:impulse}
g_{G}(t)=\Delta\left( g_{G^{-}}(t),g_{G^{+}}(t)\right).
\end{equation}

Define the bilateral Laplace transform  \cite[Chapter 10]{WRL61}
\beq \label{eq:a[s]}
\bf[a][s] \triangleq \int_{-\infty}^\infty e^{-st} \bf[a](t)dt
\eeq
Conjugating both sides of Eq.~\eqref{eq:a[s]} yields
\beq
\bf[a][s]^\# = \int_{-\infty}^\infty e^{-s^\ast t} \bf[a]^\#(t)dt
\eeq
Denote
\beq \label{eq:notation_s}
\bf[a]^\#[s] \equiv \bf[a][s^\ast]^\#, \ \ \ \breve{\mathbf{a}}[s] \equiv \left[
\bey{@{}c@{}}
\bf[a][s]\\
\bf[a]^\#[s]
\eey
\right].
\eeq
Then
\beq
\breve{\mathbf{a}}[s] = \int_{-\infty}^\infty e^{-st} \breve{\mathbf{a}}(t)dt.
\eeq
Using similar definitions and notations for other operators or functions, the transfer function from $f\breve{\mathbf{b}}_{\rm in}[s]$ to  $\breve{\mathbf{b}}_{\rm out}[s]$ is
\beq\label{eq:tf_G}
\Xi_G[s] = \mathcal{C} (sI-\mathcal{A})^{-1} \mathcal{B} + \mathcal{D}.
\eeq

\bmrk{\rm
The notation in Eq.~\eqref{eq:notation_s} is consistent with that
in \cite{GJN10} and \cite[Section 2.3.3]{NY17}, but  is slightly different
from that in \cite{SY21}. For example, $b_2^\ast(s^\ast)$ in \cite[Eq. (3)]{SY21}
is $\bf[b]_{\mathrm{in},2}^\ast[s]$ in our notation. The same is true for the other operators.
Also, in this article we use $[s]$ to indicate the frequency domain  and
$(t)$ to indicate the time domain, as have been adopted
in \cite{Z21,ZJ13} and \cite{ZGPG18}.}
\emrk

Due to the structure of the impulse response function \eqref{eq:impulse}, the transfer function $\Xi_G[s]$ defined in Eq.~\eqref{eq:tf_G} enjoys the following nice properties
\begin{equation}
\label{eq:G_flat}
 \Xi_G[-s^\ast]^\flat  \Xi_G[s]=  I_{2m},~~ \forall s \in \mathbb{C},
\end{equation}
and
\begin{equation}
\label{eq:unitary}
\Xi_G[\imath \omega]^\flat  \Xi_G[\imath\omega] = \Xi_G[\imath\omega] \Xi_G[\imath\omega]^\flat = I_{2m},~~ \forall \omega \in \mathbb{R} .
\end{equation}
(A derivation of Eq.~\eqref{eq:G_flat} can be seen in \cite[Section
VI.H]{GJN10}, and Eq.~\eqref{eq:unitary} follows Eq.~\eqref{eq:G_flat} by
setting $s=\imath \omega$.)

If $C_+=0$, $\Omega_+=0$, $K_2 =0$ and $K_3=0$,  the resulting
quantum linear system is said to be
\textit{passive} \cite{ZJ11,GZ15,NY17,ZGPG18}. Specifically, the It\^o QSDEs
for a passive linear
quantum system are
\begin{equation}
\label{eq:passive_sys_a}
\begin{split}
d\mbf{a}(t)
=&\;  A \mbf{a}(t)dt+Ev(t)dt + Bd\mbf{B}_{\rm in}(t),
 \\
d\mbf{B}_{\mathrm{out}}(t)
=&\;  C\mbf{a}(t)dt+ Dd\mbf{B}_{\rm in}(t),
\ \ t\geq t_0,
\end{split}
\end{equation}
where
\[  \label{eq:passive_sys_ABCD}
A=-\imath\Omega _{-}-\frac{1}{2}C_{-}^{\dagger }C_{-}, \ \ E = -\imath (K_1+K_4^\#), \ \  B=-C_{-}^{\dagger }S,\ C=C_{-}, \  D = S.
\]
In the passive case, the physical realizability conditions~\eqref{eq:PR_a_b} reduce to
\begin{equation}
  \label{eq:passive_PR}
A+A^{\dagger }+BB^{\dagger }=0, ~B=-C^{\dagger }S.
\end{equation}
Moreover, in the passive case,  $\Xi_{G^+}[s] \equiv 0 $ and
\begin{equation}
 \Xi_{G^-}[s]  = S -C_-(sI+\imath\Omega _{-}+\frac{1}{2}C_{-}^{\dagger }C_{-})^{-1}C_-^\dagger S.
\end{equation}
In other words,  the dynamics of a quantum linear passive system are completely characterized by its  annihilation operators.  Finally, it can be easily verified that for a quantum linear passive system, the following holds
\beq
 \Xi_{G^-}[\imath\omega]^\dagger  \Xi_{G^-}[\imath\omega] \equiv I_m, ~~~ \forall \omega\in \mathbb{R}.
\eeq
As a result, a quantum linear passive system does not change the amplitude of the input signal, but modifies its phase.

Besides the annihilation--creation operator representation \eqref{eq:sys}, a quantum linear system can also be described in the (real) quadrature operator representation.  For a positive integer $k$,  define the unitary matrix
\beq \label{eq:unitary_V}
V_{k}\triangleq \frac{1}{\sqrt{2}}\left[\begin{array}{@{}cc@{}}I_{k} & I_{k} \\
-\imath I_{k} & \imath I_{k}
\end{array}
\right].
\eeq
The following unitary transformations
\beq  \label{complex_to_real_trans}
\begin{aligned}
\left[\begin{array}{@{}c@{}}\mbf{q} \\
\mbf{p}
\end{array}
\right] \equiv \; & \mbf{x} \triangleq V_{n}\breve{\mbf{a}},  \ \
      u \triangleq V_l \breve{v},
 \\
\left[\begin{array}{@{}c@{}}\mbf{q}_{\mathrm{in}} \\
\mbf{p}_{\mathrm{in}}
\end{array}
\right] \equiv\; & \mbf{u} \triangleq V_{m}\breve{\mbf{b}},\
\left[\begin{array}{@{}c@{}}\mbf{q}_{\mathrm{out}} \\
\mbf{p}_{\mathrm{out}}
\end{array}
\right] \equiv \mbf{y} \triangleq V_{m}\breve{\mbf{b}}_{
\mathrm{out}},
\end{aligned}
\eeq
generate real quadrature operators of the system, the classical signal and the fields. The counterparts of the  commutation relations \eqref{eq:CCR_a_2} and \eqref{eq:CCR_b} are
\beq\label{eq:CCR_x}
[\mbf{x}, \mbf{x}^\top] =\imath \mathbb{J}_n,
\eeq
and
\beq \label{eq:CCR_b3}
[\mbf{u}(t), \ \mbf{u}^\top(r)] = \imath\delta(t-r)\mathbb{J}_m, \ t,r\in\mbb{R}
\eeq
respectively.

\bmrk
In some works, for example \cite{JNP08,NJP09}, and \cite{NY17}, the
transformations $\mbf{q}= \mbf{a}+\mbf{a}^\#$, and  $\mbf{p}=-\imath \mbf{a}+\imath\mbf{a}^\#$ are adopted. In this case,
Eq.~\eqref{eq:CCR_x} becomes
\beq\label{eq:CCR_x_2}
[\mbf{x}, \mbf{x}^\top] =2\imath \mathbb{J}_n.
\eeq
\emrk

In terms of unitary transformations in Eq. \eqref{complex_to_real_trans}, the coupling operator $\mbf{L}$ and the Hamiltonian $\mbf{H}$ are transformed to
\beq\la{apr31}
\bea
\mbf{L} =&\;  \Lambda\mbf{x},
\\
\mbf{H}=&\; \frac{1}{2}\mbf{x}
^\top\mathbb{H} \mbf{x}+\mbf{x}^\top \mathbb{K} u +u^\top \mathbb{K}^\dagger  \mbf{x},
\eea
\eeq
where
\beq \label{dec19-2}
\Lambda   = [C_{-}\ C_{+}]V_n^\dagger, \ \
\mathbb{H}  = V_n \Omega V_n^\dagger,
 \ \
\mathbb{K} =V_n K  V_l^\dagger.
\eeq

The
QSDEs that describe the dynamics of the linear quantum system in  \ref{fig:sys} in the real quadrature operator representation are the following:
\beq\la{eq:real_sys}
\bea
\dot{\mbf{x}} =&\; \mathbb{A} \mbf{x} +\mathbb{E}u+ \mathbb{B} \mbf{u},
 \\
\mbf{y} =&\; \mathbb{C} \mbf{x} + \mathbb{D} \mbf{u},
\eea
\eeq
where
\beq \label{eq:real_sys_ABCD}
\bea
\mathbb{D} =&\; V_{m} \mathcal{D} V_{m}^{\dagger} =\left[
\bey{@{}cc@{}}
\mathrm{Re}(S) & -\mathrm{Im}(S) \\
\mathrm{Im}(S)  & \mathrm{Re}(S)
\eey
\right],
\\
\mathbb{C} =&\; V_{m} \mathcal{C}
V_{n}^{\dagger}=
\left[
\bey{@{}ll@{}}
\mathrm{Re}(C_-+C_+) & \mathrm{Im}(-C_-+C_+)\\
\mathrm{Im}(C_-+C_+) & \mathrm{Re}(C_--C_+)
\eey
\right],
 \\
\mathbb{B} =&\; V_{n} \mathcal{B} V_{m}^{\dagger} = -\mathbb{C}^{\sharp}\mathbb{D},
\\
   \mathbb{A} =&\; V_{n} \mathcal{A} V_{n}^{\dagger} = \mathbb{J}_n \mathbb{H} -\frac{1}{2}\mathbb{C}^{\sharp}\mathbb{C},
\\
  \mathbb{E} =&\; \mathbb{J}_{n} (\mathbb{K}+\mathbb{K}^\#).
\eea
\eeq
It is easy to verify that $\mathbb{D} \mathbb{D}^\sharp = I_{2m}$.
           Define It\^o increments
\beq\label{eq:Q_P}
\bea
& d\mbf{Q}_{\rm in}(t) = \int_t^{t+dt} \mbf{q}_{\rm in}(r)dr, \ \ d\mbf{P}_{\rm in}(t) = \int_t^{t+dt} \mbf{p}_{\rm in}(r)dr,
\\
& d\mbf{Q}_{\rm out}(t) = \int_t^{t+dt} \mbf{q}_{\rm out}(r)dr, \ \ d\mbf{P}_{\rm out}(t) = \int_t^{t+dt} \mbf{p}_{\rm out}(r)dr.
\eea
\eeq
And denote
\beq
\mathcal{U}(t) = \left[
\bey{@{}c@{}}
\mbf{Q}_{\rm in}(t) \\
\mbf{P}_{\rm in}(t)
\eey
\right], \ \ \ \mathcal{Y}(t) = \left[
\bey{@{}c@{}}
\mbf{Q}_{\rm out}(t) \\
\mbf{P}_{\rm out}(t)
\eey
\right].
\eeq
Then system \eqref{eq:real_sys} can be re-written as
\beq\la{eq:real_sys_2}
\bea
d\mbf{x}(t) =&\; \mathbb{A} \mbf{x}(t)dt +\mathbb{E}u dt+ \mathbb{B}d\mathcal{U}(t),
 \\
d\mathcal{Y}(t)=&\; \mathbb{C}
\mbf{x}(t)dt + \mathbb{D}d\mathcal{U}(t)
\eea
\eeq

   \section{Hurwitz stability, controllability and observability}
\label{sec:concepts}
Hurwitz stability, controllability and observability are fundamental concepts
of classical linear systems \cite{Kalman63,KS72,Kimura96,ZDG96}.
Interestingly, these concepts can naturally be generalized to linear quantum
systems. In the following discussions of this section, we assume the classical
signal $u=0$ in Eq.~\eqref{eq:real_sys}.

If we take expectation on both sides of Eq. \eqref{eq:real_sys} with respect to the initial joint system-field state
 we get a \textit{classical} linear system
\beq \la{eq:real_sys_mean}
\bea
\frac{d\l \bf[x](t) \r}{dt} &=& \mathbb{A} \l\mbf{x}(t)\r + \mathbb{B} \l\mbf{u}(t)\r,
 \\
\l\mbf{y}(t)\r &=& \mathbb{C} \l\mbf{x}(t) \r+ \mathbb{D} \l\mbf{u}(t)\r.
\eea
\eeq
Thus we can define controllability, observability, and Hurwitz stability for the quantum linear system \eqref{eq:real_sys}  using those for the classical linear system \eqref{eq:real_sys_mean}.

\begin{definition}
\label{def:stab_ctrb_obsv}
The quantum linear system \eqref{eq:real_sys}  is said to be {Hurwitz stable} (resp. {controllable}, {observable}) if the corresponding classical linear system \eqref{eq:real_sys_mean} is {Hurwitz stable} (resp. {controllable}, {observable}).
\end{definition}

Decoherence-free subsystems for linear quantum systems have recently been
studied in
e.g., \cite{WC12a,DFK+12,NY13,NY14,GZ15,PDP17,HLL+21,ZDL21} and \cite{ZGPG18}
and references therein. It turns out that decoherence-free subsystems are
uncontrollable/unobservable subspaces in the linear quantum systems setting.

\begin{definition}[{\cite[Definition 2.1]{ZGPG18}}]
\label{def:DFS} The linear span of the system variables related to the uncontrollable/unobservable subspace of a linear quantum
system is called its {decoherence-free subsystem (DFS)}.
\end{definition}

In quantum information science, decoherence-free subspaces are widely used for protecing ueful quantum informaiton; see for example \cite{LCW98} and \cite{TV08}. The relation between decoherence-free subsystems and decoherence-free subspaces are discussed in \cite{DZWW21}.

In principle, an observable can be measured.  However, the measurement may
perturb the future evolution of this observable; this is the so-called quantum
measurement back-action. Interestingly, sometimes one can engineer a quantum
system so that measurement will not affect the evolution of the desired
observable. Observables having this property are referred to as quantum
non-demolition (QND) variables; see
e.g., \cite{TDC+78,BVT80,HMW95,TC12,NY14,LOW+21} and \cite{ZGPG18}.

\begin{definition}
\label{def:QND}
 An observable $\mbf{F}$ is called a continuous-time  QND variable if
\begin{equation}
  \label{eq:QND}
[\mbf{F}(t_1),\mbf{F}(t_2)]=0
\end{equation}
for all time instants $t_1,t_2 \in \mathbb{R}^+$.
\end{definition}

A natural extension of the notion of a QND variable is the following concept
\cite{TC12}.

\begin{definition}[{\cite{TC12}; \cite[Definition 2.3]{ZGPG18}}]
\label{QMFS} The span of a set of observables $\mbf{F}_{i}$, $i=1,\ldots,r$, is
called a \emph{quantum mechanics-free subsystem} (QMFS) if
\begin{equation}
  \label{eq:QMFS}
[\mbf{F}_{i}(t_1),\mbf{F}_{j}(t_2)]=0
\end{equation}
for all time instants $t_1,t_2 \in \mathbb{R}^+$, and $i,j=1,\ldots,r$.
\end{definition}

QND variables and QMFS subsystems  are $\boldsymbol{p}_{h}$ in the Kalman canonical form of a quantum linear system; see Theorem  \ref{thm:kalman} below.

Examples of physical realization of QMFS subsystems can be found
in \cite{LOW+21} and \cite[Example 5.2]{ZGPG18}.

The transfer function of the quantum linear system  \eqref{eq:real_sys} from $\mbf{u}$ to $\mbf{y}$ is
\begin{equation}
  \label{tf_uy}
\Xi_{\mbf{u}\to{\mbf{y}}} [s] = \mathbb{D}-\mathbb{C}(sI-\mathbb{A})^{-1}\mathbb{B}.
\end{equation}
The transfer function relates the overall input $\mbf{u}$ to the
overall output $\mbf{y}$. However, in many applications, we are
interested in a particular subvector $\mbf{u}^\prime$ of the input
vector $\mbf{u}$ and a particular subvector $\mbf{y}^\prime$
of the output vector $\mbf{y}$. This motivates us to introduce the
following concept.

\begin{definition}[{\cite[Definition 2.4]{ZGPG18}}]
\label{def:BAE} For the linear quantum system  \eqref{eq:real_sys}, let $\Xi_{\mbf{u}^\prime \to{\mbf{y}}^\prime}[s]$ be the transfer function from a subvector $\mbf{u}^\prime$
of the input vector $\mbf{u}$ to a subvector $\mbf{y}^\prime$
of the output vector $\mbf{y}$. We say that system  \eqref{eq:real_sys} realizes the {back-action
evasion (BAE) measurement} of the output $\mbf{y}^\prime$ with
respect to the input $\mbf{u}^\prime$ if $\Xi_{\mbf{u}^\prime
\to{\mbf{y}}^\prime}[s]=0$ for all $s$.
\end{definition}

More discussions on BAE measurements can be found in,
e.g., \cite{HMW95,TC10,WC13,NY14,ODP+16,MTV+17,VP22},  and the references therein.

We shall see that all of these notions can be nicely revealed by
the Kalman decomposition of a linear quantum system, see Section  \ref{sec:kalman}.

   \section{Quantum Gaussian states}
 \label{sec:Gaussian state}

In this section, quantum Gaussian states are briefly introduced. More
discussions can be found in,
e.g., \cite{GZ00,PIS+03,EB05,WD05,KRP10,WM10,MFvL11,KY12,ZJ13,NY17,VPM19,MWPY21,MWP21,BQD21}
and references therein.

  \subsection{An introduction}
 \label{sec:Gaussian state: an introduction}

Define the \textit{displacement operator}
\beq \la{eq:displacement_2}
\mbf{D}(\alpha) \triangleq \exp(\breve{\mbf{a}}^\dagger J_n \breve{\alpha}), \ \ \forall \alpha\in\mathbb{C}^{n}.
\eeq
Define the counterpart of $\breve{\alpha}$ in the real domain,
\beq
\beta = V_n \breve{\alpha} \in \mathbb{R}^{2n},
\eeq
where $V_n$ is the unitary matrix defined in Eq.~\eqref{eq:unitary_V}. Then the displacement operator defined in Eq.~\eqref{eq:displacement_2} can be re-written as
\beq \la{eq:displacement_1}
\mbf{D}(\alpha) = \exp(\imath \mbf{x}^\top \mathbb{J}_n \beta).
\eeq
Given a density matrix $\rho$ of a quantum linear system, define its \textit{quantum characteristic function} to be
\beq \label{eq:chara}
\chi_\rho \triangleq \mathrm{Tr}[\rho \mbf{D}(\alpha)].
\eeq
For Gaussian states, we have, \cite[Eq. (3.1)]{KRP10},
\beq \label{eq:gaussian_chara}
 \chi_\rho
 =
 \exp\left(-\imath\mu^\top \mathbb{J}_n \beta -\frac{1}{2} \beta^\top \mathbb{V} \beta \right),
\eeq
where
\beq \label{eq:mu_V}
\bea
\mu =&\;  \mathrm{Tr}[\rho\mbf{x}]  \in \mathbb{R}^{2n},
\\
\mathbb{V} =&\; \frac{1}{2}\mathrm{Tr}\{\rho  [(\mbf{x}_t-\mu)(\mbf{x}_t-\mu)^\top + ((\mbf{x}_t-\mu)(\mbf{x}_t-\mu)^\top]^\top]\}  \in \mathbb{R}^{2n\times 2n}.
\eea
\eeq
are the mean and covariance, respectively. Define the complex domain counterpart of $\mu$ and $\mathbb{V}$ to be
\beq\label{eq:Pi}
\bea
\breve{\gamma} \triangleq&\;  V_n^\dagger \mu,
\\
 \Pi  \triangleq&\;  V_n^\dagger \mathbb{V} V_n.
 \eea
\eeq
Then the quantum characteristic function  in Eq.~\eqref{eq:gaussian_chara} can be re-written as
\beq \label{eq:gaussian_chara2}
  \chi_\rho  =
 \exp\left(-\breve{\gamma}^\dagger J_n \breve{\alpha} - \frac{1}{2}\breve{\alpha}^\dagger \Pi \breve{\alpha}  \right),
\eeq
  Moreover, for $\rho$ to be a Gaussian state of a quantum
linear system, it is required that, \cite[][Theorem 3.1]{KRP10},
\begin{eqnarray}\label{eq:V_J_n}
\mathbb{V} &\geq& \pm\frac{\imath}{2}\mathbb{J}_n;
\end{eqnarray}
or equivalently,
\begin{eqnarray}\label{eq:apr20_temp1}
\Pi &\geq& \pm\frac{1}{2}J_n.
\end{eqnarray}

In terms of the characteristic function $ \chi_\rho$ defined in Eq.~\eqref{eq:chara}, we can define the \textit{Wigner function}   via the multi-dimensional Fourier transform
\beq \label{eq:wigner function}
W_\rho(\mathrm{w}) = \frac{1}{\sqrt{(2\pi)^{2n}}}\int_{\mathbb{R}^{2n}} \exp(-\imath \mathrm{w}^\top \mathbb{J}\beta)\mathrm{Tr}[\rho  \exp(\imath \mbf{x}^\top \mathbb{J}_n \beta)] d\beta, \ \ \forall \mathrm{w}\in \mathbb{R}^{2n}.
\eeq
 In particular, if $\rho$ is a Gaussian state with the characteristic function in Eq. \eqref{eq:gaussian_chara}, the Wigner function is of the form
\beq \label{eq:wigner function_Gauss}
W_\rho(\mathrm{w}) = \frac{1}{\sqrt{(2\pi)^{2n}\mathrm{det}(\mathbb{V})}}\exp\left(-\frac{1}{2}(\mathrm{w}-\mu)^\top \mathbb{V}^{-1} (\mathrm{w}-\mu)\right),
\eeq
with the mean $\mu$ and covariance matrix $\mathbb{V}$ given in Eq.~\eqref{eq:mu_V}. In other words, a Gaussian state is uniquely determined by its first and second moments.

   \bex
When $n=1$ and the system state is the quantum vacuum state $\ket{0}$, then $\rho =\ket{0}\bra{0}$, and
\beq \label{eq:moments}
\mu=\mathrm{Tr}(\rho \mbf{x})=\left[
\bey{@{}c@{}}
0 \\
0
\eey
\right], \ \  \mathrm{Tr}(\rho \mbf{q}^2) =  \mathrm{Tr}(\rho \mbf{p}^2)  = \frac{1}{2}, \ \
\mathrm{Tr}(\rho \mbf{p}\mbf{q}) = -\frac{\imath}{2}.
\eeq
The covariance matrix $\mathbb{V}$ in Eq.~\eqref{eq:mu_V} becomes
\begin{eqnarray}
\mathbb{V}
&=& \frac{1}{2}\left[
\bey{@{}ll@{}}
1 &0\\
0 &1
\eey
\right].
\label{eq:V2}
\end{eqnarray}
Thus,
\beq
\mathbb{V}  \pm \frac{\imath}{2}\mathbb{J}_n = \frac{1}{2}
\left[
\bey{@{}ll@{}}
1 &\pm \imath\\
\mp \imath &1
\eey
\right]\geq0,
\eeq
which verifies Eq.~\eqref{eq:V_J_n}. Moreover, the Wigner function \eqref{eq:wigner function_Gauss} is now
\beq
W_\rho(w) = \frac{1}{\pi}\exp\left(-\frac{1}{2}w^\top w\right).
\eeq
Finally, from Eq.~\eqref{eq:moments} we have
\beq \label{eq:jan3_1}
\sqrt{\mathrm{Tr}(\rho \mbf{q}^2)} \sqrt{\mathrm{Tr}(\rho \mbf{p}^2)}  =\frac{1}{2}.
\eeq
For any state $\rho$ and observables $\mbf{X}$ and $\mbf{Y}$,  the Heisenberg uncertainty relation is
\beq\label{eq:Heisenberg}
\sqrt{\mathrm{Tr}(\rho \mbf{X}^2)-(\mathrm{Tr}(\rho\mbf{X}))^2} \sqrt{\mathrm{Tr}(\rho \mbf{Y}^2)-(\mathrm{Tr}(\rho\mbf{Y}))^2}  \geq \frac{1}{2}\vert \mathrm{Tr}\left(\rho [\mbf{X}, \mbf{Y}]\right)\vert .
\eeq
According to Eq.~\eqref{eq:jan3_1}, the vacuum state $\ket{0}$  saturates the Heisenberg uncertainty relation \eqref{eq:Heisenberg} when $\mbf{X}=\mbf{q}$ and $\mbf{Y}=\mbf{p}$. In the literature, states saturating the Heisenberg uncertainty relation \eqref{eq:Heisenberg} are often called \textbf{minimum uncertainty states}. For example,  a special type of Gaussian states, coherent states, defined as
\beq
\ket{\alpha} \triangleq  e^{-\frac{\vert \alpha\vert ^2}{2}}\sum_{k=0}^\infty \frac{\alpha^k}{\sqrt{k!}}\ket{k} , \ \  \alpha\in \mathbb{C},
\eeq
are minimum uncertainty states as Eq.~\eqref{eq:Heisenberg} is saturated when $\mbf{X}=\mbf{q}$ and $\mbf{Y}=\mbf{p}$.
\eex

 \bex
In this example, we show that  the vacuum state $\ket{0}$ is a Gaussian state. For simplicity, we look at the single-oscillator case ($n=1$).  In this case, the displacement operator $\mbf{D}(\alpha) $ defined in Eq.~\eqref{eq:displacement_2} becomes
\beq
\mbf{D}(\alpha) = \exp(\alpha \mbf{a}^\ast - \alpha^\ast \mbf{a}) .
\eeq
 If two operators $\mbf{A}$ and $\mbf{B}$ satisfy  $[\mbf{A},[\mbf{A},\mbf{B}]] = [\mbf{B},[\mbf{A},\mbf{B}]]=0$, then the Baker-Campbell-Hausdorff formula is
\beq
e^{\mbf{A}+\mbf{B}} = e^{\mbf{A}} e^{\mbf{B}} e^{-\frac{1}{2}[\mbf{A},\mbf{B}]} =   e^{\mbf{B}} e^{\mbf{A}} e^{\frac{1}{2}[\mbf{A},\mbf{B}]}.
\eeq
Therefore,
\beq
\mbf{D}(\alpha) =e^{\alpha \mbf{a}^\ast} e^{-\alpha^\ast \mbf{a}}  e^{-\frac{\vert \alpha\vert ^2}{2}}.
\eeq
As
\beq
e^{\alpha \mbf{a}^\ast} e^{-\alpha^\ast \mbf{a}}  \ket{0}= e^{\alpha \mbf{a}^\ast}  \ket{0}= \sum_{k=0}^\infty \frac{\alpha^k}{\sqrt{k!}}\ket{k},
\eeq
we have
\beq
\mbf{D}(\alpha) \ket{0}= e^{-\frac{\vert \alpha\vert ^2}{2}}\sum_{k=0}^\infty \frac{\alpha^k}{\sqrt{k!}}\ket{k}  = \ket{\alpha}.
\eeq
Consequently, the characteristic function for the vacuum state $\ket{0}$ is
\beq
\chi = \braket{0\vert \mbf{D}(\alpha)\vert 0} =  e^{-\frac{\vert \alpha\vert ^2}{2}},
\eeq
which is of the form Eq. \eqref{eq:gaussian_chara2} with
\beq
\gamma =0, \ \  \Pi = \frac{1}{2}\left[
\bey{@{}ll@{}}
1 &0\\
0 &1
\eey
\right].
\eeq
Clearly, $\Pi$ derived above satisfies the inequality \eqref{eq:apr20_temp1}, hence the vacuum state $\ket{0}$ is a Gaussian state.
\eex

In \cite{Gough05}, \cite[Section 2.7]{NY17} and \cite{ZJ13}, another form of
characteristic functions is defined for  a  Gaussian system state $\rho $,
which is
\begin{equation}
\label{eq:system-gaussian_multi}
\mathrm{Tr}[\rho\exp(i\breve{z}^\dagger\breve{\mbf{a}})] = \exp(\imath \breve z^\dagger \breve \beta-\frac{1}{2}\breve{z}^\dagger \Sigma \breve{z}
),  ~~ \forall z \in \mathbb{C}^n ,
\end{equation}
where $\breve{\gamma} = \mathrm{Tr}[\rho \breve{\mbf{a}}]$, and
 $\Sigma =  \mathrm{Tr}[\rho ( \breve{\mbf{a}} - \breve{\gamma}) (\breve{\mbf{a}}  -  \breve{\gamma}   )^\dagger]$  is a non-negative Hermitian matrix.
 In general, $\Sigma$ has the form
\begin{equation}
 \label{eq:Sigma}
\Sigma = \left[ \begin{array}{@{}cc@{}}             I_n +N^T  & M \\
             M^\dagger & N
\end{array}
  \right] .
\end{equation}
  In particular, the \emph{ground} or\ \emph{vacuum} state $\ket{0}$ is specified by $\gamma =0$ and $  \Sigma = \left[ \begin{array}{@{}cc@{}}             I_n & 0 \\
             0 & 0_n
\end{array}
  \right]$. Clearly,
\beq
\frac{1}{2}\Sigma \geq \frac{\imath}{2}J_n.
\eeq
However,  the following is not true:
\beq
\frac{1}{2}\Sigma \geq -\frac{\imath}{2}J_n.
\eeq
Consequently, to be consistent with $\mathbb{V}$ for the real quadrature operator representation, it is better to use the covariance matrix $\Pi$ given in Eq.~\eqref{eq:Pi}, instead of $\Sigma$ in Eq.~\eqref{eq:Sigma}.

  \subsection{Pure Gaussian state generation}
\label{subsec: gaussian state generation}
Gaussian states are very useful resources in quantum signal
processing, \cite{BvL05,NvLG+06,FvL11} and \cite{PIS+03,WPG+12}.
Thus, the problem of Gaussian state generation has been studied intensively in
the quantum control literature. In this subsection, we  present one result for
pure Gaussian state generation by means of environment engineering.

A Gaussian state is a \textbf{pure} state is the determinant  of its associated
covariance $\mathbb{V}$ satisfies $\det(\mathbb{V}) = 1/2^{2n}$, where $n$ is the
number of the oscillators.  The covariance matrix $\mathbb{V}$ of a
\textit{pure} Gaussian state can be decomposed as (\cite[Eq. (16)]{KY12}; \cite[Eq.
(2.18)]{MFvL11})
\beq\label{eq:V_graph}
\mathbb{V} = \frac{1}{2}
\mathbb{S}\, \mathbb{S}^\top,
\eeq
where
\beq
\mathbb{S} = \left[
\bey{@{}cc@{}}
Y^{-1/2}  & 0 \\
XY^{-1/2} & Y^{1/2}
\eey
\right]
\eeq
with $X=X^\top\in\mathbb{R}^n $ and $Y=Y^\top\in \mathbb{R}^n$ being positive definite. As $\mathbb{S} \mathbb{J}_n \mathbb{S}^\top = \mathbb{J}_n$, $\mathbb{S}$ is symplectic (See the \textit{Notation} part).
Let $Z=X+\imath Y$.

 \bthm [{\cite{KY12}}]
The pure Gaussian state associated with the covariance matrix $\mathbb{V}$ in Eq.~\eqref{eq:V_graph} can be generated by the linear quantum system \eqref{eq:real_sys} if and only if
\beq
\mathbb{H} =
\left[
\bey{@{}cc@{}}
XRX + YRY-\Gamma Y^{-1}X-XY^{-1}\Gamma^\top  & -XR + \Gamma Y^{-1}\\
-RX + Y^{-1}\Gamma^\top & R
\eey
\right].
\eeq
and
\beq
\Lambda = P^\top [-Z \ \ I],
\eeq
where $R=R^\top$, $\Gamma = -\Gamma^\top$, and the matrix pair $(P,Q)$ is controllable with  $Q=-\imath RY+Y^{-1}\Gamma$.
\ethm

The problem of Gaussian state generation by means of environment engineering
has been studied   insensitively by the quantum control community. Interest
reader may refer to \cite[Section 6.1]{NY17} and further
development \cite{Ma17,MWP+18,MWJ+19,MWPY21,MWP21,BQD21}.

   \subsection{Skew information and information-theoretic uncertainty  relation }
\label{subsec:information}

In addition to Heisenberg's uncertainty relation \eqref{eq:Heisenberg}, an
information-theoretic uncertainty relation was proposed in \cite{Luo05} based
on skew information. In what follows, we use the notation in \cite{Luo05}.
Given a density operator $\rho$ and an observable $\mbf{X}$, the
Wigner--Yanase skew information \cite{WY63} is defined as
\beq\label{eq:skew}
I(\rho, \mbf{X}) \triangleq -\frac{1}{2} \mathrm{Tr}([\sqrt{\rho},\mbf{X}]^2).
\eeq

Skew information was originally defined for Hamiltonians of closed (namely,
isolated) quantum systems \cite{WY63}, and was later generalized to arbitrary
observables of open quantum systems. Roughly speaking,  $I(\rho, \mbf{X})$ measures
the quantum uncertainty of  the observable $\mbf{X}$ with respect to the
density operator $\rho$.

The variance of $\mbf{X}$ with respect to the density operator $\rho$ is
\beq
V(\rho, \mbf{X}) = \mathrm{Tr}(\rho \mbf{X}^2) - (\mathrm{Tr}(\rho \mbf{X}) )^2.
\eeq
When the state is pure, it is easy to see that $I(\rho, \mbf{X}) = V(\rho, \mbf{X}) $. However,
$I(\rho, \mbf{X}) \leq V(\rho, \mbf{X}) $ when the state is a mixed one. To quantify quantum uncertainty,
the following quantity has been defined in \cite{Luo05}:
\beq
U(\rho, \mbf{X}) \triangleq \sqrt{V^2(\rho, \mbf{X})-(V(\rho, \mbf{X})-I(\rho,\mbf{X}))^2}.
\eeq
The information-theoretic uncertainty  relation   is \cite[Eq. (2)]{Luo05}
\beq\label{eq:Heisenberg_luo}
U(\rho, \mbf{X}) U(\rho, \mbf{Y}) \geq \frac{1}{4}\vert \mathrm{Tr}(\rho [\mbf{X},\mbf{Y}])\vert ^2 .
\eeq

Interestingly, it is proved in \cite{FLZ20} that when $n=1$ for the
single-mode oscillator case,  all Gaussian states, pure or mixed, are minimum
uncertainty states, i.e.,~states that saturate \eqref{eq:Heisenberg_luo}. On
the other hand, minimum uncertainty states are Gaussian states.  However, in
general mixed Gaussian states do not saturate the Heisenberg's uncertainty
relation \eqref{eq:Heisenberg}. In this sense, the  information-theoretic
uncertainty  relation  \eqref{eq:Heisenberg_luo} better characterizes Gaussian
states. More studies on quantum skew-information and information-theoretic
uncertainty  relation can be found in,
e.g., \cite{GI+01,Luo+04,Luo+05,Chen+05,Hansen+08,LFO+12,KCF+14,LS+17,SML+17,LS+18,LZ+19,FLZ20}
and references therein.

   \section{Quantum Kalman filter}
\label{sec:kalman filter}

Simply speaking, a quantum filter describes the temporal evolution of an open quantum system under repeated measurement. A general form of quantum filters is first presented in Subsection  \ref{subsec:quantum filter}, after that the quantum filter for linear quantum systems is given in Subsection  \ref{subsec:quantum kalman filter}, which is of the form of the Kalman filter for classical linear systems. Finally,  an example is given in  Subsection \ref{subsec:quantum kalman filter example} for demonstration. In this section, it is assumed that $S=I_m$ and the initial time $t_0=0$.

  \subsection{Quantum filter}
\label{subsec:quantum filter}

In this subsection, we present a quantum filter for open quantum systems.

By Eq.~\eqref{complex_to_real_trans}, we have
\beq
\bea
{[\mathbf{q}_{\rm in} (t_1), \mathbf{q}^\top_{\rm in} (t_2) ]}
=&\;
\frac{1}{2} [\mbf{b}_{\mm[in]}(t_1) +\mbf{b}^\#_{\mm[in]}(t_1), \mbf{b}^\dagger_{\mm[in]}(t_2) +\mbf{b}^\top_{\mm[in]}(t_2)  ]
\\
=&\; \frac{1}{2} [\mathbf{b}_{\rm in} (t_1), \mbf{b}^\dagger_{\mm[in]}(t_2)] +  \frac{1}{2} [\mbf{b}^\#_{\mm[in]}(t_1),\mbf{b}^\top_{\mm[in]}(t_2)   ]
\\
=&\; \frac{1}{2} \delta(t_1-t_2) I_m -\frac{1}{2} \delta(t_1-t_2) I_m
\\
 =&\; 0.
\eea
\eeq
According to Eq.~\eqref{eq:output},
\beq
\mbf{q}_{\mm[out]} (t)  = U(t)^\ast (I_{\mathrm{system}}\otimes\mbf{q}_{\mm[in]} (t))  U(t).
\eeq
Moreover, the unitary operator $U(t)$ has the following property
\cite[Section 5.2]{BvHJ07}
\beq \label{eq:U_t1_t2}
U(t_2)^\ast (I_{\mathrm{system}}\otimes\mbf{q}_{\mm[in]} (t_1))  U(t_2) = U(t_1)^\ast (I_{\mathrm{system}}\otimes\mbf{q}_{\mm[in]} (t_1))  U(t_1), \ \ \ t_1\leq t_2.
\eeq
Consequently, $\mathbf{q}_{\rm out}(t)$ enjoys the self-non-demolition property:
\beq \label{eq:q_t1_t2}
[\mathbf{q}_{\rm out} (t_1), \mathbf{q}^\top_{\rm out} (t_2) ] = 0,~~0\leq t_1\leq t_2.
\eeq
Moreover, from Eqs.~\eqref{eq:phys_realiz} and \eqref{eq:U_t1_t2} the following  non-demolition property can be derived:
\beq\la{eq: non-demolition}
[\mbf{X}(t),\mathbf{q}_{\rm out}(r)^\top]=0,~~t_0\leq r\leq t.
\eeq

It can be easily verified that the integrated quadrature operator  $\mbf{Q}_{\rm out} (t)$ defined in Eq.~\eqref{eq:Q_P} also enjoys the self-non-demolition property \eqref{eq:q_t1_t2} and non-demolition property \eqref{eq: non-demolition}. Due to the self-non-demolition property \eqref{eq:q_t1_t2},  $\{\mbf{Q}_{\rm out}(r): 0\leq r\leq t\}$ can be regarded as a \textit{classical} stochastic process. (Strictly speaking, measuring $\mbf{Q}_{\rm out} (t)$ gives rise to a classical stochastic process.) Moreover, due to the non-demolition property \eqref{eq: non-demolition}, $\mbf{X}(t)$ lives in the $\sigma$-field generated by this classical stochastic process. Hence, it is meaningful to define the expectation of $\mbf{X}(t)$ conditioned on this  $\sigma$-field. We denote this conditional expectation by $E[\mbf{X}(t)\vert \{\mbf{Q}_{\rm out}(r): 0\leq r\leq t\}]$.  Then, one can define the conditional density operator $\rho_{\rm c}(t)$ by means of
\beq \la{eq:rho_c}
\mathrm{Tr}(\rho_{\rm c}(t)\mbf{X}) = E[\mbf{X}(t)\vert \{\mbf{Q}_{\rm out}(r): 0\leq r\leq t\}].
\eeq
Clearly, $\rho_{\rm c}(0)$ is the initial joint system density matrix denoted $\rho_{\rm S}(0)$.
The dynamics of the conditioned density operator is given by  the {stochastic
master equation (SME) (also called  quantum
trajectories \cite{Carmichael93},  \cite{Carmichael09})
\beq\label{eq:conditioned_rho}
\bea
d\rho_{\rm c}(t) &=\; \mathcal{L}^\star_{\mbf{L},\mbf{H}}(\rho_{\rm c}(t))dt
\\
&+ \{\mbf{L}^\top \rho_{\rm c}(t) + \rho_{\rm c}(t) \mbf{L}^\dag - \mrm{Tr}[\rho_{\rm c}(t) (\mbf{L}^\top + \mbf{L}^\dag)]\rho_{\rm c}(t) \} d\nu(t),
\eea
\eeq
where
\beq \label{eq:innovation_nv}
d\nu(t) \triangleq d\mbf{Q}_{\rm out}(t)-\pi_t(\mbf{L}+\mbf{L}^\#)dt
\eeq
is an innovation process, and  the superoperator
\beq
 \mathcal{L}^\star_{\mbf{L},\mbf{H}}(\rho_{\rm c}(t)) = -\imath[\mbf{H},\rho_{\rm c}(t)] + \mbf{L}^\top \rho_{\rm c}(t) \mbf{L}^\# - \frac{1}{2}  \mbf{L}^\dagger \mbf{L} \rho_{\rm c}(t) -  \frac{1}{2}  \rho_{\rm c}(t) \mbf{L}^\dagger  \mbf{L}.
\eeq
For a given system operator $\mbf{X}$, define  the conditioned mean vector
\beq
\pi_t(\mbf{X}) \triangleq \mathrm{Tr}(\rho_{\rm c}(t)\mbf{X}).
\eeq
 It turns out that $\pi_t(\mbf{X})$ is the solution to the  \textit{Belavkin
quantum filtering equation}, which is a \textit{classical} stochastic
differential equation\cite{B80}, \cite{belavkin1989nondemolition}
\beq\label{dec19_1}
\bea
 d\pi_t(\mbf{X})
&= \;  \pi_t(\mathcal{L}_{\mbf{L},\mbf{H}}(\mbf{X}))dt
\\
&+ [\pi_t(\mbf{X}\mbf{L}^\top+\mbf{L}^\dag \mbf{X}) -\pi_t(\mbf{L}^\top+\mbf{L}^\dag)\pi_t(\mbf{X})]d\nu(t),
\eea
\eeq

with the initial condition  $\pi_0(\mbf{X}) = \mathrm{Tr}(\rho_{\rm S}(0)\mbf{X})$, where $\mathcal{L}_{\mbf{L},\mbf{H}}(\mbf{X})$ is the superoperator defined in Eq.~\eqref{eq:L_LH}.

 \bmrk{\rm
In classical  control systems theory, measurement noise is always supposed to be decoupled from the system dynamics. This is no longer true in the quantum regime. As shown by Eq.~\eqref{eq:conditioned_rho}, measurement affects the dynamics of the system which is being monitored. This is often called measurement back-action. Moreover,   Eq.~\eqref{eq:real_sys} show linear dynamics of the system. However, its conditioned dynamics \eqref{eq:conditioned_rho} is \textit{nonlinear}.   This is essentially different from classical linear dynamics.
}
\emrk

 \bmrk{\rm
The measurement used in this section is homodyne measurement. There are other
types of quantum measurements used in quantum filtering and feedback control,
for instance, heterodyne measurement, photodetection and general positive
operator valued measurements (POVMs).  The experimental realization of a
real-time POVM measurement-based feedback control of the 2012 Nobel prize
winning photon-box is described in \cite{ASD+12} and \cite{SDZ+11}.  A
comprehensive study of quantum measurement and feedback control is presented
in \cite{BvHJ07} and \cite{WM10}.
}
\emrk

Finally, denote
\beq
\rho(t) \triangleq E[\rho_{\rm c}(t)].
\eeq
 Then the \textit{unconditioned} system dynamics are given by the Lindblad master equation
\beq\label{eq:unconditioned_rho}
d\rho(t) =  \mathcal{L}^\star_{\mbf{L},\mbf{H}}(\rho(t)) dt.
\eeq

More discussions of quantum filters can be found
in \cite{B80,DJ99,DHJ+00,vHSM05,vH07,BvHJ07,BvH08,VbHJ09,RR15b,LDP+21,GZP19,GDP+19,GZP20,CGL+21,YYD+21},
and \cite{WM10}, among others.

  \subsection{Quantum Kalman filter}
\label{subsec:quantum kalman filter}

The above formulations hold for general open quantum systems, in this subsection we present their specific forms for linear quantum systems. In this case, the quantum filter consists of Eqs.~\eqref{eq:filter_mean} and \eqref{eq:filter_var} to be given below, which is in the same form of a classical Kalman filter.

Define the conditional covariance matrix
\beq\la{eq:condi_cov_V}
V_t \triangleq {\rm Tr}\left[\rho_c(t)\ff{(\mbf{x}-\pi_t(\bf[x]))(\mbf{x}-\pi_t(\bf[x]))^\top
+((\mbf{x}-\pi_t(\bf[x]))(\mbf{x}-\pi_t(\bf[x]))^\top)^\top}{2}\right].
\eeq

 \bthm\la{thm:kalman filter}
The quantum Kalman filter for the quantum linear system \eqref{eq:real_sys} is of the form
\beq \la{eq:filter_mean}
d\pi_t(\bf[x])=\mathbb{A}\pi_t(\bf[x])dt + \mathbb{E}udt + (V_t\mathbb{C}_1^\top+M)d\nu(t),
\eeq
with the initial condition $\pi_0(\bf[x]) ={\rm Tr}\left[\rho_{{\rm S}}(0)\bf[x]\right]$, where the matrices
\begin{equation}
\mathbb{C}_1 = [I_m \ \ 0_m]\mathbb{C}, \ \ M=  \ff{1}{\sqrt{2}}\mathbb{B} \left[
\bey{@{}c@{}}
I_m\\
0_m
\eey
\right],
\end{equation}
and the conditional covariance matrix $V_t$ solves the following differential Riccati equation
\beq\la{eq:filter_var}
\dot{V}_t = \mathbb{A}V_t+V_t\mathbb{A}^\top+\frac{1}{2}\mathbb{B}\mathbb{B}^\top-(V_t\mathbb{C}_1^\top+M)(V_t\mathbb{C}_1^\top+M)^\top.
\eeq
\ethm

The quantum Kalman filter appears very close to the Kalman filter for classical
linear systems \cite[Chapter 3]{KS72,AM79}.

In what follows, a proof of Theorem  \ref{thm:kalman filter} is given,

\newproof{pot1}{Proof of  Theorem \ref{thm:kalman filter}}
\begin{pot1}
Look at Eq.~\eqref{eq:filter_mean} first.

\textit{Step 0.}  Let $\mbf{X}$, $\mbf{Y}$ and $\mbf{Z}$ be vectors of operators of dimension $l$, $m$, and $n$, respectively. Let $M\in\mathbb{C}^{m\times n}$. If the commutators $[a,b]\in \mathbb{C}$ where $a$ and $b$ are arbitrary elements of the vectors $\mbf{X}$, $\mbf{Y}$ and $\mbf{Z}$. Then
\beq\label{eq:XYZ}
[\mbf{X}, \mbf{Y}^\top M \mbf{Z}] = [\mbf{X},\mbf{Y}^\top]M\mbf{Z} +  [\mbf{X},\mbf{Z}^\top]M^\top\mbf{Y}.
\eeq

\textit{Step 1.} Substituting the vector $\mbf{x}(t)$ in Eq.~\eqref{eq:real_sys} into Eq.~\eqref{eq:L_LH} and using Eqs.~\eqref{dec19-6} and \eqref{eq:XYZ} we get
\begin{equation}
\label{dec19-10}
\begin{aligned}
&\pi_t(-\imath[\mbf{x}(t),\mbf{H}(t)]+\frac{1}{2}\mbf{L}^{\dagger}(t)[\mbf{x}(t),\mbf{L}(t)]+\frac{1}{2}[\mbf{L}^{\dagger}(t), \mbf{x}(t)]\mbf{L}(t)) \\
=&\; \mathbb{J}_n\mathbb{H}\pi_t(\mbf{x})+\mathbb{J}_n(\mathbb{K}+\mathbb{K}^\#)u
+\frac{1}{2\imath}\mathbb{J}_n(\Lambda^\dagger\Lambda-\Lambda^\top\Lambda^\#)\pi_t(\mbf{x}) \\
=&\; \left(\mathbb{J}_n\mathbb{H}-\frac{1}{2}\mathbb{C}^\sharp\mathbb{C}\right)\pi_t(\mbf{x})
+\mathbb{J}_n(\mathbb{K}+\mathbb{K}^\#)u\\
=&\; \mathbb{A}\pi_t(\mbf{x}) + \mathbb{E}u.
\end{aligned}
\end{equation}

\textit{Step 2}.
 By Eqs.~\eqref{eq:CCR_x_2} and \eqref{eq:condi_cov_V}, we have
\begin{equation}
\label{dec29-1}
\begin{aligned}
V_t= \pi_t(\mbf{x}\mbf{x}^\top)-\frac{\imath}{2}\mathbb{J}_n-\pi_t(\mbf{x})\pi_t(\mbf{x})^\top.
\end{aligned}
\end{equation}

\textit{Step 3}.  Noticing that
\begin{equation}
\label{dec19-11}
\begin{aligned}
&\mbf{L}^\dagger\mbf{q}=\mbf{q}\mbf{x}^\top\Lambda^\dagger-\imath\left[\begin{array}{@{}cc@{}}                                                                    \mbf{0} & I
\end{array}
                                                                \right]\Lambda^\dagger, \\
&\mbf{L}^\dagger\mbf{p}=\mbf{p}\mbf{x}^\top\Lambda^\dagger+\imath\left[\begin{array}{@{}cc@{}}                                                                    I & \mbf{0}
\end{array}
                                                                \right]\Lambda^\dagger,
\end{aligned}
\end{equation}
we have
\begin{equation}
\label{dec19-12}
\mbf{L}^\dagger\mbf{x}=\mbf{x}\mbf{L}^\dagger-\imath\mathbb{J}_n\Lambda^\dagger.
\end{equation}

\textit{Step 4}.  By Eqs.~\eqref{dec29-1} and \eqref{dec19-12}, we get
\begin{equation}
\label{dec19-14}
\begin{aligned}
&\pi_t(\mbf{x}\mbf{L}^\top+\mbf{L}^\dagger\mbf{x})-\pi_t(\mbf{L}^\top+\mbf{L}^\dagger)\pi_t(\mbf{x}) \\
=&\; \pi_t(\mbf{x}\mbf{x}^\top)\Lambda^\top+\pi_t(\mbf{x}\mbf{L}^\dagger-\imath\mathbb{J}_n\Lambda^\dagger)
-\pi_t(\mbf{L}^\top+\mbf{L}^\dagger)\pi_t(\mbf{x}) \\
=&\; \pi_t(\mbf{x}\mbf{x}^\top)(\Lambda^\top+\Lambda^\dagger)-\imath\mathbb{J}_n\Lambda^\dagger
-\pi_t(\mbf{L}^\top+\mbf{L}^\dagger)\pi_t(\mbf{x}) \\
=&\; (\langle\mbf{x}\mbf{x}^\top\rangle-\pi_t(\mbf{x})\pi_t(\mbf{x})^\top)(\Lambda^\top+\Lambda^\dagger)-\imath\mathbb{J}_n\Lambda^\dagger \\
=&\; (V_t+\frac{\imath}{2}\mathbb{J}_n)(\Lambda^\top+\Lambda^\dagger)-\imath\mathbb{J}_n\Lambda^\dagger \\
=&\; V_t(\Lambda^\top+\Lambda^\dagger)+\frac{\imath}{2}\mathbb{J}_n(\Lambda^\top-\Lambda^\dagger),
\end{aligned}
\end{equation}
where
\begin{equation}
\label{dec19-15}
\begin{aligned}
\pi_t(\mbf{L}^\top+\mbf{L}^\dagger)\pi_t(\mbf{x}) =\pi_t(\mbf{x})\pi_t(\mbf{L}^\top+\mbf{L}^\dagger) =\pi_t(\mbf{x})\pi_t(\mbf{x}^\top)(\Lambda^\top+\Lambda^\dagger)
\end{aligned}
\end{equation}
has been used in the derivation.

Combining \eqref{dec19-10}, \eqref{dec19-14} with \eqref{eq:innovation_nv}, we have
\begin{equation}
\label{dec19-17}
\begin{aligned}
d\pi_t(\mbf{x})=&(\mathbb{A}\pi_t(\mbf{x})+\mathbb{E}u)dt
+\left(V_t\mathbb{C}_1^\top+\frac{\imath}{2}\mathbb{J}_n(\Lambda^\top-\Lambda^\dagger)\right)d\nu_t \\
=&(\mathbb{A}\pi_t(\mbf{x})+\mathbb{E} u)dt+(V_t\mathbb{C}_1^\top+M)d\nu_t,
\end{aligned}
\end{equation}
which is Eq.~\eqref{eq:filter_mean}.

Next, we derive Eq.~\eqref{eq:filter_var}.

By Eqs. \eqref{eq:rho_c} and \eqref{eq:condi_cov_V}, we have
\begin{align}
V_t =&\; E\bigg[\ff{(\mbf{x}_t-\pi_t(\bf[x]))(\mbf{x}_t-\pi_t(\bf[x]))^\top
+((\mbf{x}_t-\pi_t(\bf[x]))(\mbf{x}_t-\pi_t(\bf[x]))^\top)^\top}{2}
\nonumber
\\
&\hspace{1cm} \bigg|\{\mbf{Q}_{\rm out}(r): 0\leq r\leq t\}\bigg].
\end{align}
As a result, by the property of conditional expectation,
\beqn
&&E[V(t)]
\label{eq:apr22_EV}
\\
 &=& E\left[\ff{(\mbf{x}_t-\pi_t(\bf[x]))(\mbf{x}_t-\pi_t(\bf[x]))^\top
+((\mbf{x}_t-\pi_t(\bf[x]))(\mbf{x}_t-\pi_t(\bf[x]))^\top)^\top}{2}\right]
\nonumber\\
&=&
\mrm{Tr}\left[\rho_{S\otimes F}(0)\ff{(\mbf{x}_t-\pi_t(\bf[x]))(\mbf{x}_t-\pi_t(\bf[x]))^\top
+((\mbf{x}_t-\pi_t(\bf[x]))(\mbf{x}_t-\pi_t(\bf[x]))^\top)^\top}{2}\right],
\nonumber
\eeqn
where $\rho_{S\otimes F}(0)$ is the initial joint system-field density matrix. By the
following property of Gaussian random variables: $V(t) = E[V(t)]$ almost surely
(a.s.), see e.g., \cite[Chapter 10]{LS77}, we have
\beq\la{eq:condi_cov_V2}
V_t ={\rm Tr}\left[\rho_{{\rm S}\otimes {\rm F}}(0)\ff{(\mbf{x}_t-\pi_t(\bf[x]))(\mbf{x}_t-\pi_t(\bf[x]))^\top
+((\mbf{x}_t-\pi_t(\bf[x]))(\mbf{x}_t-\pi_t(\bf[x]))^\top)^\top}{2}\right].
\eeq

For convenience, in the following, we denote $\l \mbf{X} \r = \mathrm{Tr}[\rho_{S\otimes F}(0)\mbf{X}]$ for an operator $\mbf{X}$.

\textit{Step 0.}  Differentiating $\langle(\mbf{x}-\pi_t(\mbf{x}) )(\mbf{x}-\pi_t(\mbf{x}) )^\top\rangle$ with respect to $t$, yields
\begin{equation}
\label{dec19-18}
\begin{aligned}
&d\left\langle(\mbf{x}_t-\pi_t(\mbf{x}) )(\mbf{x}_t-\pi_t(\mbf{x}) )^\top\right\rangle \\
=&\langle d(\mbf{x}_t-\pi_t(\mbf{x}) )(\mbf{x}_t-\pi_t(\mbf{x}) )^\top\rangle + \langle(\mbf{x}_t-\pi_t(\mbf{x}) )d(\mbf{x}_t-\pi_t(\mbf{x}) )^\top\rangle \\
&+ \ \  \langle d(\mbf{x}_t-\pi_t(\mbf{x}) )d(\mbf{x}_t-\pi_t(\mbf{x}) )^\top\rangle.
\end{aligned}
\end{equation}
According to Eqs. \eqref{eq:real_sys} and \eqref{dec19-17}, we have
\begin{equation}
\label{dec24-2}
\begin{aligned}
d(\mbf{x}_t-\pi_t(\mbf{x}) )=\mathbb{A}(\mbf{x}_t-\pi_t(\mbf{x}) )dt+\mathbb{B}d\mbf{u}-(V_t\mathbb{C}_1^\top+M)d\nu_t.
\end{aligned}
\end{equation}
Moreover, it can be easily checked that
\begin{equation}
\label{dec24-3}
\begin{aligned}
\mathrm{Tr}[\rho_{\rm S\otimes F}(0)\mathbb{B}d\mbf{u} d\nu_t^\top] &= -\imath\mathbb{J}_n\Lambda^\dagger dt, \\
\mathrm{Tr}[\rho_{\rm S\otimes F}(0)d\nu_t d\mbf{u}^\top\mathbb{B}^\top] &= \imath\Lambda\mathbb{J}_n^\top dt, \\
\mathrm{Tr}[\rho_{\rm S\otimes F}(0)\mathbb{B}d\mbf{u} d\mbf{u}^\top\mathbb{B}^\top] &= \mathbb{J}_n \Lambda^\dagger \Lambda \mathbb{J}_n^\top dt.
\end{aligned}
\end{equation}
By Eqs.~\eqref{dec24-3}, \eqref{dec19-18} can be calculated as
\begin{equation}\label{dec24-4}\begin{aligned}
&\frac{d}{dt}\left\langle(\mbf{x}_t-\pi_t(\mbf{x}) )(\mbf{x}_t-\pi_t(\mbf{x}) )^\top\right\rangle \\
=&\mathbb{A}\left\langle(\mbf{x}_t-\pi_t(\mbf{x}) )(\mbf{x}_t-\pi_t(\mbf{x}) )^\top\right\rangle
-(V_t\mathbb{C}_1^\top+M)\mathbb{C}_1\left\langle(\mbf{x}_t-\pi_t(\mbf{x}) )(\mbf{x}_t-\pi_t(\mbf{x}) )^\top\right\rangle \\
&+\left\langle(\mbf{x}_t-\pi_t(\mbf{x}) )(\mbf{x}_t-\pi_t(\mbf{x}) )^\top\right\rangle\mathbb{A}^\top
-\left\langle(\mbf{x}_t-\pi_t(\mbf{x}) )(\mbf{x}_t-\pi_t(\mbf{x}) )^\top\right\rangle\mathbb{C}_1^\top(V_t\mathbb{C}_1^\top+M)^\top \\
&+\mathbb{J}_n\Lambda^\dagger\Lambda\mathbb{J}_n^\top+(V_t\mathbb{C}_1^\top+M)(V_t\mathbb{C}_1^\top+M)^\top \\
&+\imath\mathbb{J}_n\Lambda^\dagger(V_t\mathbb{C}_1^\top+M)^\top-\imath(V_t\mathbb{C}_1^\top+M)\Lambda\mathbb{J}_n^\top.
\end{aligned}\end{equation}

{\it Step 1.}  Transposing both sides of \eqref{dec24-4}, yields
\begin{equation}\label{dec24-5}\begin{aligned}
&\frac{d}{dt}\left\langle(\mbf{x}_t-\pi_t(\mbf{x}) )(\mbf{x}_t-\pi_t(\mbf{x}) )^\top\right\rangle^\top \\
=&\left\langle(\mbf{x}_t-\pi_t(\mbf{x}) )(\mbf{x}_t-\pi_t(\mbf{x}) )^\top\right\rangle^\top\mathbb{A}^\top
-\left\langle(\mbf{x}_t-\pi_t(\mbf{x}) )(\mbf{x}_t-\pi_t(\mbf{x}) )^\top\right\rangle^\top\mathbb{C}_1^\top(V_t\mathbb{C}_1^\top+M)^\top \\
&+\mathbb{A}\left\langle(\mbf{x}_t-\pi_t(\mbf{x}) )(\mbf{x}_t-\pi_t(\mbf{x}) )^\top\right\rangle^\top
-(V_t\mathbb{C}_1^\top+M)\mathbb{C}_1\left\langle(\mbf{x}_t-\pi_t(\mbf{x}) )(\mbf{x}_t-\pi_t(\mbf{x}) )^\top\right\rangle^\top \\
&+\mathbb{J}_n\Lambda^\top\Lambda^\#\mathbb{J}_n^\top+(V_t\mathbb{C}_1^\top+M)(V_t\mathbb{C}_1^\top+M)^\top \\
&+\imath(V_t\mathbb{C}_1^\top+M)\Lambda^\#\mathbb{J}_n^\top-\imath\mathbb{J}_n\Lambda^\top(V_t\mathbb{C}_1^\top+M)^\top.
\end{aligned}\end{equation}

{\it Step 2.}  Combine \eqref{dec24-4} with \eqref{dec24-5}, and recall the form of  conditional covariance matrix $V_t$ in Eq. \eqref{eq:condi_cov_V2}, we have
\begin{equation}\label{dec24-6}\begin{aligned}
\frac{d}{dt}V_t=&\; \mathbb{A}V_t+V_t\mathbb{A}^\top
+\frac{1}{2}(\mathbb{J}_n\Lambda^\dagger\Lambda\mathbb{J}_n^\top+\mathbb{J}_n\Lambda^\top\Lambda^\#\mathbb{J}_n^\top) \\
&\; +(V_t\mathbb{C}_1^\top+M)(V_t\mathbb{C}_1^\top+M)^\top
-(V_t\mathbb{C}_1^\top+M)\mathbb{C}_1V_t-V_t\mathbb{C}_1^\top(V_t\mathbb{C}_1^\top+M)^\top \\
&\; +\frac{\imath}{2}\mathbb{J}_n(\Lambda^\dagger-\Lambda^\top)(V_t\mathbb{C}_1^\top+M)^\top
+\frac{\imath}{2}(V_t\mathbb{C}_1^\top+M)(\Lambda^\#-\Lambda)\mathbb{J}_n^\top \\
=&\; \mathbb{A}V_t+V_t\mathbb{A}^\top+\frac{1}{2}\mathbb{B}\mathbb{B}^\top-(V_t\mathbb{C}_1^\top+M)(V_t\mathbb{C}_1^\top+M)^\top,
\end{aligned}\end{equation}
which is Eq. \eqref{eq:filter_var}.
\end{pot1}

  \subsection{An example}
\label{subsec:quantum kalman filter example}

In this subsection, we use a simple example to illustrate the quantum Kalman filter.

Consider a single-mode oscillator with system parameters $S=1$, $\mbf{L} =\sqrt{\kappa}\mbf{a}$, and $\mbf{H} = \omega \mbf{a}^\ast \mbf{a}$. Then  by Eq.~\eqref{eq:real_sys_2}
\beq\la{eq:real_sys_2_example}
\bea
d\mbf{x}(t) =&\; \mathbb{A} \mbf{x}(t)dt  + \mathbb{B}d\mathcal{U}(t),
 \\
d\mathcal{Y}(t)=&\; \mathbb{C}
\mbf{x}(t)dt + \mathbb{D}d\mathcal{U}(t)
\eea
\eeq
where
\beq
\mathbb{A} = \left[
\bey{@{}cc@{}}
-\frac{\kappa}{2}  &\omega\\
-\omega & -\frac{\kappa}{2}
\eey
\right],  \ \ \mathbb{B} = -\sqrt{\kappa}I_2, \ \ \mathbb{C} =\sqrt{\kappa}I_2, \ \ \mathbb{D} = I_2.
\eeq
Let the system be initialized in the vacuum state $\ket{0}$. Assume that $\mbf{Q}_{\rm out}(t)$ is continuously measured. Thus, a sequence of measurement data $\{\mbf{Q}_{\rm out}(r), 0\leq r\leq t\}$ is available at the present time $t$.  By the quantum Kalman filter presented in the previous subsection, we have
\begin{subequations}
\begin{align}
d \pi_t(\mbf{q}) =&\;
-\ff{\sqrt{\kappa}}{2} \pi_t(\mbf{q}) dt + \omega \pi_t(\mbf{p}) dt - \sqrt{k} \left(\frac{\sqrt{2}}{2}- V_1\right) d\nu(t)
\nonumber
\\
=&\; \left(\ff{\sqrt{2}\kappa-\sqrt{\kappa}}{2}-\kappa V_1\right) \pi_t(\mbf{q}) dt + \omega \pi_t(\mbf{p}) dt  -\sqrt{k} \left(\frac{\sqrt{2}}{2}- V_1\right) d\mbf{Q}_{\rm out}(t) ,
\label{eq:ex_q_c}
\\
d \pi_t(\mbf{p}) =&\; -\omega \pi_t(\mbf{q}) dt -\ff{\sqrt{\kappa}}{2} \pi_t(\mbf{p}) dt +\sqrt{\kappa}V_2d\nu(t)
\nonumber
\\
=&\; - (\omega+ \sqrt{\kappa}V_2)\pi_t(\mbf{q}) dt -\frac{\kappa}{2}  \pi_t(\mbf{p}) dt +\sqrt{\kappa}V_2 d\mbf{Q}_{\rm out}(t) ,
\label{eq:ex_p_c}
\end{align}
\end{subequations}where  the innovation process is
\beq
 d\nu(t) = d\mbf{Q}_{\rm out}(t) - \sqrt{\kappa}\pi_t(\mbf{q})dt,
\eeq
       and the entries of the covariance matrix $V$ evolve according to
\begin{subequations}
\begin{eqnarray}
\dot{V}_1 &=& (\sqrt{2}-1)\kappa V_1 + 2\omega V_2 - \kappa V_1^2
\label{eq:Ex_V1}
\\
\dot{V}_2 &=& -\kappa(1-\frac{\sqrt{2}}{2})V_2 - \omega(V_1-V_3) -  \kappa V_1 V_2
\label{eq:Ex_V2}
\\
\dot{V}_3   &=&\ff{\kappa}{2}-2\omega V_2 -\kappa V_3 -\kappa V_2^2
\label{eq:Ex_V3}
\end{eqnarray}
\end{subequations}
with the initial condition  $V_1(0) = V_3(0)=1$ and $V_2(0)=0$.

When the detuning $\omega=0$, by Eq.~\eqref{eq:Ex_V2}, $V_2(t)\equiv V_2(0)=0$ provided that the solution is unique.  Then by Eq.~\eqref{eq:ex_p_c},  we have $ \pi_t(\mbf{p}) \equiv   \pi_0(\mbf{p})$, which is not disturbed directly by the continuous measurement of $\mbf{Q}_{\rm out}(t)$. On the other hand, when the detuning  $\omega\neq0$,  $V_2\neq0$.  From Eq.~\eqref{eq:ex_p_c} we can see that the measurement of $\mbf{Q}_{\rm out}(t)$ affects $ \pi_t(\mbf{p}) $, which in the sequel affects $ \pi_t(\mbf{q}) $. This clearly demonstrates the \textit{quantum back-action effect}.

   \section{Quantum Kalman canonical decomposition}
\label{sec:kalman}

In this section, we discuss the quantum Kalman canonical decomposition of
quantum linear systems in Subsection  \ref{subsec:kalman canonical form}. In Subsection
\ref{subsec: canonical example}, an example taken from a recent
experiment \cite{LOW+21} is used to illustrate the  procedures and main
results. Kalman canonical decomposition of classical linear systems can be
found in, e.g., \cite[Section 2.2]{Kimura96} and \cite[Section 3.3]{ZDG96}.

  \subsection{Quantum Kalman canonical form}
\label{subsec:kalman canonical form}

The following result reveals the structure of quantum linear systems;
see \cite{GZ15,ZGPG18} and \cite{ZPL20} for more details.

\begin{proposition}
 \label{prop:3nov}
The quantum linear system \eqref{eq:real_sys}  has the following properties.
\begin{description}
\item[(i)] Its controllability and observability are equivalent to each other;
 see \cite[Proposition 1]{GZ15}, and see \cite{GY15} for the passive
  case.

\item[(ii)] If it is Hurwitz stable, then it is both controllable and
 observable; see \cite[Theorem 3.1]{ZPL20}, and see \cite{GY15} for
  the passive case.

\item[(iii)] In the passive case  Eq. \eqref{eq:passive_sys_a}, its Hurwitz
 stability, controllability and observability are all equivalent;
  see \cite[Lemma 2]{GZ15}.

\item[(iv)] All its poles corresponding to an uncontrollable and unobservable
 subsystem are on the imaginary axis; see \cite[Theorem 3.2]{ZGPG18}.

\end{description}
\end{proposition}

Interestingly,  the equivalence between  stabilizability
and detectability of quantum linear systems is pointed out in the physics
literature \cite{WD05}.

The following result presents the Kalman canonical form of quantum linear systems.

\begin{theorem}[{\cite[Theorem 4.4]{ZGPG18}}]\label{thm:kalman}
 Assume $u=0$ in Eq. \eqref{eq:real_sys}. Also suppose the scattering matrix $S=I_m$.  There exists a real orthogonal and blockwise symplectic matrix $\mathbb{T}$ that facilitates the following
coordinate transformation
\begin{equation}
\left[\begin{array}{@{}c@{}}\boldsymbol{q}_{h} \\
\boldsymbol{p}_{h} \\ \hline
\boldsymbol{x}_{co} \\ \hline
\boldsymbol{x}_{\bar{c}\bar{o}}
\end{array}
\right] = \tilde{\mbf{x}}= \mathbb{T}^{\top} \boldsymbol{x},
  \label{aug31_3}
\end{equation}
and transform the linear quantum system \eqref{eq:real_sys} into the form
\begin{eqnarray}
\left[
\begin{array}{c}
\boldsymbol{\dot{q}}_{h}(t) \\
\boldsymbol{\dot{p}}_{h}(t) \\ \hline
\boldsymbol{\dot{x}}_{co}(t) \\ \hline
\boldsymbol{\dot{x}}_{\bar{c}\bar{o}}(t)
\end{array}
\right] &=& \bar{\mathbb{A}} \left[
\begin{array}{c}
\boldsymbol{q}_{h}(t) \\
\boldsymbol{p}_{h}(t) \\ \hline
\boldsymbol{x}_{co}(t) \\ \hline
\boldsymbol{x}_{\bar{c}\bar{o}}(t)
\end{array}
\right] + \bar{\mathbb{B}} \boldsymbol{u}(t),  \label{real_Kalman_ss} \\
\boldsymbol{y}(t) &=& \bar{\mathbb{C}} \left[
\begin{array}{c}
\boldsymbol{q}_{h}(t) \\
\boldsymbol{p}_{h}(t) \\ \hline
\boldsymbol{x}_{co}(t) \\ \hline
\boldsymbol{x}_{\bar{c}\bar{o}}(t)
\end{array}
\right] +\boldsymbol{u}(t),  \label{real_Kalman_io}
\end{eqnarray}
where the system matrices are
\begin{eqnarray}
\bar{\mathbb{A}} = \left[
\begin{array}{cc|c|c}
A_{h}^{11} & A_{h}^{12} & A_{12} & A_{13} \\
0 & A_{h}^{22} & 0 & 0 \\ \hline
0 & A_{21} & A_{co} & 0 \\ \hline
0 & A_{31} & 0 & A_{\bar{c}\bar{o}}
\end{array}
\right],   \
\bar{\mathbb{B}} = \left[
\begin{array}{c}
B_{h} \\
0 \\ \hline
B_{co} \\ \hline
0
\end{array}
\right], \
\bar{\mathbb{C}}= \left[
\begin{array}{cc|c|c}
0 & C_{h} & C_{co} & 0
\end{array}
\right].  \label{eq:real_Kalman_sys_ABCD}
\end{eqnarray}
 After a re-arrangement,  the system (\ref{real_Kalman_ss})-(\ref{real_Kalman_io}) becomes
\begin{eqnarray}
\left[
\begin{array}{c}
\boldsymbol{\dot{q}}_{h}(t) \\
\boldsymbol{\dot{x}}_{co}(t) \\
\boldsymbol{\dot{x}}_{\bar{c}\bar{o}}(t) \\
\boldsymbol{\dot{p}}_{h}(t)
\end{array}
\right] &=& \left[
\begin{array}{cccc}
A_{h}^{11} & A_{12} & A_{13} & A_{h}^{12} \\
0 & A_{co} & 0 & A_{21} \\
0 & 0 & A_{\bar{c}\bar{o}} & A_{31} \\
0 & 0 & 0 & A_{h}^{22}
\end{array}
\right] \left[
\begin{array}{c}
\boldsymbol{q}_{h}(t) \\
\boldsymbol{x}_{co}(t) \\
\boldsymbol{x}_{\bar{c}\bar{o}}(t) \\
\boldsymbol{p}_{h}(t)
\end{array}
\right]  + \left[
\begin{array}{c}
B_{h} \\
B_{co} \\
0 \\
0
\end{array}
\right] \boldsymbol{u}(t),  \label{real_Kalman_ss_2} \\
\boldsymbol{y}(t)&=& [0 \ C_{co} \ 0 \ C_{h}] \left[%
\begin{array}{c}
\boldsymbol{q}_{h}(t) \\
\boldsymbol{x}_{co}(t) \\
\boldsymbol{x}_{\bar{c}\bar{o}}(t) \\
\boldsymbol{p}_{h}(t)
\end{array}
\right] +\boldsymbol{u}(t).  \label{real_Kalman_io_2}
\end{eqnarray}
A block diagram for the system (\ref{real_Kalman_ss})-(\ref{real_Kalman_io})
is given in Fig.~\ref{fig:kalman}.
\end{theorem}

A refinement of the quantum Kalman canonical form is given in \cite[Theorem
3.3]{ZPL20},

\begin{figure}[tbph]
\centering
\includegraphics[width=0.618\textwidth]{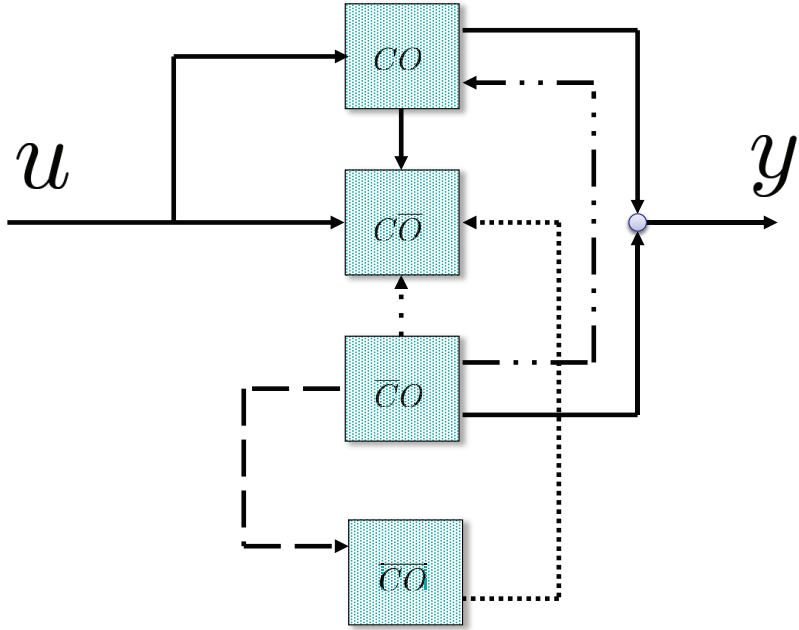}
\caption{Kalman decomposition of a linear quantum system. Taken from Fig. 2 in \cite{ZGPG18}.}
\label{fig:kalman}
\end{figure}

The following result characterize the system parameters corresponding to the Kalman canonical form \eqref{real_Kalman_ss}.

\begin{theorem}\cite[Proposition 2.1]{ZPL20}
Denote
\beq
\tilde{\mathbb{H}} = \mathbb{T}^\top \mathbb{H} \mathbb{T}, \ \ \tilde{\Lambda} = \Lambda \mathbb{T},
\eeq
where the matrices $\mathbb{H}$ and $\Lambda$ are given in Eq. \eqref{dec19-2}.
  The real
symmetric matrix $\tilde{\mathbb{H}}$ corresponding to the Kalman canonical form (\ref{real_Kalman_ss})  must be of the form
\begin{equation}\label{R}
\tilde{\mathbb{H}}=\left[
\begin{array}{cc|c|c}
0 & H_{h}^{12} & 0 & 0 \\
H_{h}^{12^{\top}} & H_{h}^{22} & H_{12} & H_{13}
 \\ \hline
0 & H_{12}^{\top} & H_{co} & 0 \Tstrut
 \\ \hline
0 & H_{13}^{\top} & 0 & H_{\bar{c}\bar{o}} \Tstrut
\end{array}
\right] ,
\end{equation}
and the complex matrix $\tilde{\Lambda} $ corresponding to the Kalman canonical form (\ref{real_Kalman_ss}) must satisfy
\begin{equation}\label{Gamma}
\left[
\begin{array}{c}
\Lambda\\
\Lambda^{\#}
\end{array}
\right]
=
\left[
\begin{array}{cc|c|c}
0 & \Lambda_{h} & \Lambda_{co} & 0
\end{array}
\right] ,
\end{equation}
where
\begin{equation}\label{gamma3a}
\Lambda_{h} =V_{m}^{\dag }C_{h}, \ \
\Lambda_{co} =V_{m}^{\dag }C_{co}.
\end{equation}
\end{theorem}

\begin{remark}{\rm
The matrices $\Lambda_{co}$ and $\Lambda_{h}$ in Eq. \eqref{gamma3a} are respectively of
the form
\begin{subequations}
\begin{equation}\label{Gamma_co0}
\Lambda_{co}
=
\left[
\begin{array}{cc}
\Lambda_{co,q} & \Lambda_{co,p} \\
\Lambda_{co,q}^{\#} & \Lambda_{co,p}^{\#}
\end{array}
\right],
\end{equation}
and
\begin{equation}\label{Gamma_h}
\Lambda_{h} = \left[
\begin{array}{c}
\Lambda_{h,p} \\
\Lambda_{h,p}^{\#}
\end{array}
\right].
\end{equation}
\end{subequations}
}
\end{remark}

 The Kalman canonical form \eqref{real_Kalman_ss} presented above can be used to investigate  BAE measurements of quantum linear systems.

\begin{theorem} \label{thm:BAE}\cite[Theorem 4.1]{ZPL20}
\begin{description}
\item[(i)] The quantum Kalman canonical form (\ref{real_Kalman_ss}) realizes the BAE measurements of $\boldsymbol{q}_{\mathrm{out}}$ with
respect to $\boldsymbol{p}_{\mathrm{in}}$; i.e.,
$
\Xi _{\boldsymbol{p}_{\mathrm{in}}\rightarrow \boldsymbol{q}_{\mathrm{out}}}(s)\equiv 0
$
if and only if
\begin{eqnarray}
&&\left[
\begin{array}{cc}
\mathrm{Re}\left( \Lambda_{co,q}\right) & \mathrm{Re}\left( \Lambda_{co,p}\right)
\end{array}
\right] (sI-\mathbb{J}_{n_{1}}H_{co})^{-1}
 \left[
\begin{array}{c}
\mathrm{Re}(\Lambda_{co,p}^{\top})
\vspace{3pt}
 \\
-\mathrm{Re}(\Lambda_{co,q}^{\top}) %
\end{array}
\right]\equiv0;  \label{temp9}
\end{eqnarray}
\item[(ii)] The quantum Kalman canonical form (\ref{real_Kalman_ss}) realizes the BAE measurements of $\boldsymbol{p}_{\mathrm{out}}$ with
respect to $\boldsymbol{q}_{\mathrm{in}}$; i.e.,
$
\Xi _{\boldsymbol{q}_{\mathrm{in}}\rightarrow \boldsymbol{p}_{\mathrm{out}
}}(s)\equiv0
$
if and only if
\begin{eqnarray}
&&\left[
\begin{array}{cc}
\mathrm{Im}\left( \Lambda_{co,q}\right) & \mathrm{Im}\left( \Lambda_{co,p}\right)
\end{array}
\right] (sI-\mathbb{J}_{n_{1}}H_{co})^{-1}
 \left[
\begin{array}{c}
\mathrm{Im}(\Lambda_{co,p}^{\top})
\vspace{3pt}
\\
-\mathrm{Im}( \Lambda_{co,q}^{\top})%
\end{array}
\right]\equiv0.
  \label{temp9b}
\end{eqnarray}
\end{description}
\end{theorem}

  \subsection{An example}
 \label{subsec: canonical example}

An opto-mechanical system has recently been physically realized
in \cite{LOW+21}. In this subsection, we analyze this system by means of the
quantum Kalman canonical form presented in the previous subsection. An
excellent introduction to quantum opto-mechanical systems can be found
in \cite{AKM14}.

In \cite[Fig. 1(A)]{LOW+21}, an effective positive-mass oscillator and
effective negative-mass oscillator coupled to a cavity are constructed to
generate a quantum mechanics-free subsystem (QMFS). To be specific, the
linearized Hamiltonian is \cite[Eq. (S3)]{LOW+21}
\begin{equation}
\label{dec27-1}
\begin{aligned}
\mbf{H}=&\; \omega (\boldsymbol {a}_1^\ast  \boldsymbol {a}_1- \boldsymbol {a}_2^\ast  \boldsymbol{a}_2)+g_1[(\alpha_{1-}\boldsymbol{a}_1+\alpha_{1+}\boldsymbol{a}_1^\ast )\boldsymbol{a}_3^\ast +(\alpha_{1-}^\ast \boldsymbol{a}_1^\ast +\alpha_{1+}^\ast \boldsymbol{a}_1)\boldsymbol {a}_3] \\
&+ \ \  g_2[(\alpha_{2-}\boldsymbol{a}_2+\alpha_{2+}\boldsymbol{a}_2^\ast )\boldsymbol{a}_3^\ast +(\alpha_{2-}^\ast \boldsymbol{a}_2^\ast +\alpha_{2+}^\ast \boldsymbol{a}_2)\boldsymbol{a}_3],
\end{aligned}
\end{equation}
where the reduced Planck constant is omitted and $\omega$ is the detuning
frequency. $\boldsymbol{a}_j$, $\boldsymbol{a}_j^\ast$, ($j=1,2$) denote the annihilation
and creation operators for the mechanical oscillators, and their corresponding
coupling strengths (to the cavity) are $G_{j\pm}=g_j\alpha_{j\pm}$. The cavity is
characterized by the damping rate $\kappa$ and described by the
annihilation and creation operators $\boldsymbol{a}_3$, $\boldsymbol{a}_3^\ast$. Choose equal
effective couplings $\vert G_{j\pm}\vert  \equiv G$ and let $G_{j\pm}=G e^{-\imath\theta_{j\pm}}$, where $\theta_{j\pm}\in\mathbb{C}$,
$j=1,2$.  Then the interaction Hamiltonian \eqref{dec27-1} can be
re-written as the following quadrature form \cite[Eq. (S4)]{LOW+21}
\begin{equation}
\label{dec27-2}
\begin{aligned}
\mbf{H}_{\mathrm{int}}=&\; \frac{G}{2}\boldsymbol{a}_3[A_-\boldsymbol{q}_- + A_+\boldsymbol{q}_+ + B_-\boldsymbol{p}_- + B_+\boldsymbol{p}_+] \\
&+ \ \  \frac{G}{2}\boldsymbol{a}_3^\dagger[A_-^\ast \boldsymbol{q}_- + A_+^\ast \boldsymbol{q}_+ + B_-^\ast \boldsymbol{p}_- + B_+^\ast \boldsymbol{p}_+],
\end{aligned}
\end{equation}
where
\begin{equation}
\label{dec27-3}
\begin{aligned}
\boldsymbol{q}_j&=\; \frac{\boldsymbol{a}_j^\ast+\boldsymbol{a}_j}{\sqrt{2}}, ~~~~ \boldsymbol{p}_j=\frac{\imath(\boldsymbol{a}_j^\ast-\boldsymbol{a}_j)}{\sqrt{2}}, ~~ j=1,2, \\
\boldsymbol{q}_{\pm}&=\frac{\boldsymbol{q}_1 \pm \boldsymbol{q}_2}{\sqrt{2}}, ~~~~ \boldsymbol{p}_{\pm}=\frac{\boldsymbol{p}_1 \pm \boldsymbol{p}_2}{\sqrt{2}},
\end{aligned}
\end{equation}
and
\begin{equation}
\label{dec27-3_b}
\begin{aligned}
A_-&=e^{-\imath\theta_{1-}}+e^{-\imath\theta_{1+}}-e^{-\imath\theta_{2-}}-e^{-\imath\theta_{2+}}, \\
A_+&=e^{-\imath\theta_{1-}}+e^{-\imath\theta_{1+}}+e^{-\imath\theta_{2-}}+e^{-\imath\theta_{2+}}, \\
B_-&=\imath[-e^{-\imath\theta_{1-}}+e^{-\imath\theta_{1+}}+e^{-\imath\theta_{2-}}-e^{-\imath\theta_{2+}}], \\
B_+&=\imath[-e^{-\imath\theta_{1-}}+e^{-\imath\theta_{1+}}-e^{-\imath\theta_{2-}}+e^{-\imath\theta_{2+}}],
\end{aligned}
\end{equation}
which are \cite[Eq. (S5)]{LOW+21}.

Let $\theta_{1-}=\theta_{2+}=0$ and $\theta_{1+}=\theta_{2-}=\phi$. Eq.~\eqref{dec27-2} becomes, \cite[Eq.
(S6)]{LOW+21}
\begin{equation}
\label{dec27-4}
\mbf{H}_{\mathrm{int}}=2G(e^{-\imath\phi/2} \boldsymbol{a}_3 + e^{\imath\phi/2} \boldsymbol{a}_3^\ast)\left(\boldsymbol{q}_+\cos\frac{\phi}{2}+\boldsymbol{p}_-\sin\frac{\phi}{2}\right).
\end{equation}
Moreover, by omitting the dissipation term and choosing $\phi=0$, Eq.~\eqref{dec27-4} becomes
\begin{equation}
\label{dec27-5}
\mbf{H}_{\mathrm{int}}=2\sqrt{2}G \boldsymbol{q}_c \boldsymbol{q}_+,
\end{equation}
where $\boldsymbol{q}_c=\frac{\boldsymbol{a}_3^\ast + \boldsymbol{a}_3}{\sqrt{2}}$.

\bmrk{\rm
When $\theta_{1-}=\theta_{2+}=\theta_{1+}=\theta_{2-}=0$, all coupling strengths $G_{j\pm}=G$ are equal. This setting is used below.
}
\emrk

Assume that the coupling strengths are equal and by the framework presented in Section  \ref{sec:system}, the system Hamiltonian \eqref{dec27-1} can be written as
\begin{equation}
\label{dec28-1}
\begin{aligned}
\Omega=\Delta(\Omega_-,\Omega_+),
\end{aligned}
\end{equation}
where
\begin{equation}
\label{dec28-2}
\begin{aligned}
\Omega_-=\left[\begin{array}{@{}ccc@{}}                          \omega & 0 & G \\
                          0 & -\omega & G \\
                          G & G & 0
\end{array}
                      \right], ~~~~ \Omega_+=\left[\begin{array}{@{}ccc@{}}                                                              0 & 0 & G \\
                                                              0 & 0 & G \\
                                                              G & G & 0
\end{array}
                                                          \right].
\end{aligned}
\end{equation}
As the optical coupling is $\boldsymbol{L}=\sqrt{\kappa}\boldsymbol{a}_3$ what describes energy dissipation from the cavity, we have
\begin{equation}
\label{dec28-3}
\mathcal{C}=\Delta(C_-,C_+),
\end{equation}
where $C_-=\left[\begin{array}{@{}ccc@{}}               0 & 0 & \sqrt{\kappa}
\end{array}
           \right]$, $C_+=0$. By Eq.~\eqref{eq:sys_ABCD}, the system matrices can be calculated as
\begin{equation}
\label{dec29-2}
\begin{aligned}
\mathcal{B}&=-\mathcal{C}^\flat=-\left[\begin{array}{@{}cccccc@{}}                                    0 & 0 & \sqrt{\kappa} & 0 & 0 & 0 \\
                                    0 & 0 & 0 & 0 & 0 & \sqrt{\kappa}
\end{array}
                                \right]^\top, \\
\mathcal{A}&=-\imath J_3\Omega-\frac{1}{2}\mathcal{C}^\flat\mathcal{C}
=\left[\begin{array}{@{}cccccc@{}}      -\imath\omega & 0 & -\imath G & 0 & 0 & -\imath G \\
      0 & \imath\omega & -\imath G & 0 & 0 & -\imath G \\
      -\imath G & -\imath G & -\frac{\kappa}{2} & -\imath G & -\imath G & 0 \\
      0 & 0 & \imath G & \imath\omega & 0 & \imath G \\
      0 & 0 & \imath G & 0 & -\imath\omega & \imath G \\
      \imath G & \imath G & 0 & \imath G & \imath G & -\frac{\kappa}{2}
\end{array}
  \right].
\end{aligned}
\end{equation}
As a result, the opto-mechanical system composed of two oscillators and a cavity can be described by
\begin{equation}
\label{dec29-3}
\begin{aligned}
\dot{\breve{\boldsymbol{a}}}&=\mathcal{A}\breve{\boldsymbol{a}}+\mathcal{B}\breve{\boldsymbol{b}}_{\rm in}, \\
\breve{\boldsymbol{b}}_{\mathrm{out}}&=\mathcal{C}\breve{\boldsymbol{a}}+\breve{\boldsymbol{b}}_{\rm in},
\end{aligned}
\end{equation}
where $\boldsymbol{a}=\left[\begin{array}{@{}ccc@{}}                          \boldsymbol{a}_1 & \boldsymbol{a}_2 & \boldsymbol{a}_3
\end{array}
                      \right]^\top$. The auxiliary matrix $O_s$ defined in the proof
of \cite[Theorem 3.1]{ZGPG18}, originally in \cite[Eq. 7]{GZ15}, can be
solved by
\begin{equation}
\label{dec29-4}
O_s=\left[\begin{array}{@{}c@{}}        \mathcal{C} \\
        \mathcal{C}(J_3\Omega) \\
        \vdots \\
        \mathcal{C}(J_3\Omega)^{5}
\end{array}
    \right].
\end{equation}

In what follows, $R_{co}$, $R_{\bar{c}\bar{o}}$, $R_{c\bar{o}}$, and $R_{\bar{c}o}$
represent the controllable/observable ($co$),
uncontrollable/unobservable ($\bar{c}\bar{o}$), controllable/unobservable
($c\bar{o}$), and uncontrollable/observable ($\bar{c}o$) subspaces of
system \eqref{dec29-3}, respectively. By \cite[Lemma 4.2]{ZGPG18}, these four
subspaces can be calculated as
\begin{equation}
\label{dec29-5}
\begin{aligned}
&R_{co}=\mathrm{span}\left\{\left[\begin{array}{@{}c@{}}                               0 \\
                               0 \\
                               1 \\
                               0 \\
                               0 \\
                               0
\end{array}
                           \right],\left[\begin{array}{@{}c@{}}                                       0 \\
                                       0 \\
                                       0 \\
                                       0 \\
                                       0 \\
                                       1
\end{array}
                                   \right]\right\}, ~~ R_{\bar{c}\bar{o}}=\emptyset, \\
&R_{c\bar{o}}=\mathrm{span}\left\{\left[\begin{array}{@{}c@{}}                                      0 \\
                                      -1 \\
                                      0 \\
                                      1 \\
                                      0 \\
                                      0
\end{array}
                                  \right],\left[\begin{array}{@{}c@{}}                                              -1 \\
                                              0 \\
                                              0 \\
                                              0 \\
                                              1 \\
                                              0
\end{array}
                                          \right]\right\}, ~~ R_{\bar{c}o}=
\mathrm{span}\left\{\left[\begin{array}{@{}c@{}}                        0 \\
                        1 \\
                        0 \\
                        1 \\
                        0 \\
                        0
\end{array}
                    \right],\left[\begin{array}{@{}c@{}}                                1 \\
                                0 \\
                                0 \\
                                0 \\
                                1 \\
                                0
\end{array}
                            \right]\right\}.
\end{aligned}
\end{equation}
Then by \cite[Lemma 4.3]{ZGPG18} and \cite[Lemma 4.7]{ZGPG18}, the special
orthonormal bases $T_{co}$, $T_{\bar{c}\bar{o}}$, $T_{c\bar{o}}$, and $T_{\bar{c}o}$
can be constructed. From \cite[Eq. (47)]{ZGPG18}, the blockwise Bogoliubov
transformation matrix can be calculated as
\begin{equation}
\label{dec29-6}
T=\left[\begin{array}{@{}cccccc@{}}      \frac{1}{\sqrt{2}} & \frac{1}{\sqrt{2}} & 0 & 0 & 0 & 0 \\
      -\frac{1}{\sqrt{2}} & \frac{1}{\sqrt{2}} & 0 & 0 & 0 & 0 \\
      0 & 0 & 0 & 0 & 1 & 0 \\
      0 & 0 & \frac{1}{\sqrt{2}} & \frac{1}{\sqrt{2}} & 0 & 0 \\
      0 & 0 & -\frac{1}{\sqrt{2}} & \frac{1}{\sqrt{2}} & 0 & 0 \\
      0 & 0 & 0 & 0 & 0 & 1
\end{array}
  \right].
\end{equation}
By \cite[Theorem 4.2]{ZGPG18}, we have the transformed system matrices
\begin{equation}
\label{dec29-7}
\begin{aligned}
\bar{\mathcal{A}}=T^\dagger \mathcal{A} T, ~~ \bar{\mathcal{B}}=T^\dagger \mathcal{B}, ~~ \bar{\mathcal{C}}=\mathcal{C}T.
\end{aligned}
\end{equation}

Recall the dimensions of the four subspaces introduced in \cite[Remark
4.2]{ZGPG18}, we have $n_1=1$, $n_2=0$, $n_3=2$, and
$n_1+n_2+n_3=n=3$ in this example. By \cite[Lemma 4.8]{ZGPG18},
\begin{equation}
\label{dec29-8}
\tilde{V}_n=\mathrm{diag}\left\{\tilde{V}_{n_3},V_{n_1}\right\},
\end{equation}
where $\tilde{V}_{n_3}=\Pi V_{n_3}$, $\Pi=\left[\begin{array}{@{}cccc@{}}        1 & 0 & 0 & 0 \\
        0 & 0 & 0 & -1 \\
        0 & 0 & 1 & 0 \\
        0 & 1 & 0 & 0
\end{array}
    \right]$, and
\begin{equation}
\label{dec29-9}
\begin{aligned}
\bar{A}&=\tilde{V}_n\bar{\mathcal{A}}\tilde{V}_n^\dagger=\left[\begin{array}{@{}cccccc@{}}                                                            0 & -\omega & 0 & 0 & 0 & 0 \\
                                                            \omega & 0 & 0 & 0 & 2\sqrt{2}G & 0 \\
                                                            0 & 0 & 0 & -\omega & 0 & 0 \\
                                                            0 & 0 & \omega & 0 & 0 & 0 \\
                                                            0 & 0 & 0 & 0 & -\frac{\kappa}{2} & 0 \\
                                                            0 & 0 & 0 & -2\sqrt{2}G & 0 & -\frac{\kappa}{2}
\end{array}
                                                        \right], \\
\bar{B}&=\tilde{V}_n\bar{\mathcal{B}}V_1^\dagger=-\left[\begin{array}{@{}cccccc@{}}                                                      0 & 0 & 0 & 0 & \sqrt{\kappa} & 0 \\
                                                      0 & 0 & 0 & 0 & 0 & \sqrt{\kappa}
\end{array}
                                                  \right]^\top, \\
\bar{C}&=V_1\bar{\mathcal{C}}\tilde{V}_n^\dagger=-\bar{B}^\top.
\end{aligned}
\end{equation}
By \cite[Lemma 4.9]{ZGPG18}, the real, orthogonal, and blockwise symplectic
matrix $\mathbb{T} \triangleq V_n T \tilde{V}_n^\dagger$. From \cite[Theorem 4.4]{ZGPG18}, the transformed system
operators
\begin{equation}
\label{dec29-10}
\mathbb{T}^\top V_n \breve{\boldsymbol{a}}=\left[\begin{array}{@{}c@{}}                                              \boldsymbol{q}_- \\
                                              -\boldsymbol{p}_+ \\
                                              \boldsymbol{p}_- \\
                                              \boldsymbol{q}_+ \\ \hline
                                              \boldsymbol{q}_c \\
                                              \boldsymbol{p}_c
\end{array}
                                          \right]\equiv \left[\begin{array}{@{}c@{}}    \boldsymbol{q}_h \\
    \boldsymbol{p}_h \\ \hline
    \boldsymbol{x}_{co}
\end{array}
\right],
\end{equation}
where $\boldsymbol{p}_c=\frac{\imath(\boldsymbol{a}_3^\ast-\boldsymbol{a}_3)}{\sqrt{2}}$. By \cite[Theorem 4.4]{ZGPG18}, the Kalman decomposition
for the opto-mechanical system \eqref{dec29-3} in the real quadrature operator
representation can be expressed as
\begin{equation}
\label{dec29-11}
\begin{aligned}
\dot{\boldsymbol{q}}_h&=\left[\begin{array}{@{}cc@{}}                            0 & -\omega \\
                            \omega & 0
\end{array}
                        \right]\boldsymbol{q}_h + \left[\begin{array}{@{}cc@{}}                                                      0 & 0 \\
                                                      2\sqrt{2}G & 0
\end{array}
                                                  \right]\boldsymbol{x}_{co}, \\
\dot{\boldsymbol{p}}_h&=\left[\begin{array}{@{}cc@{}}                            0 & -\omega \\
                            \omega & 0
\end{array}
                        \right]\boldsymbol{p}_h, \\
\dot{\boldsymbol{x}}_{co}&=-\frac{\kappa}{2}\boldsymbol{x}_{co}-\left[\begin{array}{@{}cc@{}}                                                                    0 & 0 \\
                                                                    0 & 2\sqrt{2}G
\end{array}
                                                                \right]\boldsymbol{p}_h-\sqrt{\kappa}\left[\begin{array}{@{}c@{}}                                                                                                         \boldsymbol{q}_{\mathrm{in}} \\
                                                                                                         \boldsymbol{p}_{\mathrm{in}}
\end{array}
                                                                                                     \right], \\
\left[\begin{array}{@{}c@{}}                   \boldsymbol{q}_{\mathrm{out}} \\
                   \boldsymbol{p}_{\mathrm{out}}
\end{array}
               \right]&=\sqrt{\kappa}\boldsymbol{x}_{co}+\left[\begin{array}{@{}c@{}}                   \boldsymbol{q}_{\mathrm{in}} \\
                   \boldsymbol{p}_{\mathrm{in}}
\end{array}
               \right].
\end{aligned}
\end{equation}
Consequently, by omitting the dissipation term ($\kappa=0$), we have
\begin{equation}
\label{dec29-12}
\begin{aligned}
\dot{\boldsymbol{q}}_-&=\omega\boldsymbol{p}_+, \\
\dot{\boldsymbol{p}}_+&=-\omega\boldsymbol{q}_- - 2\sqrt{2}G \boldsymbol{q}_c, \\
\dot{\boldsymbol{p}}_-&=-\omega\boldsymbol{q}_+, \\
\dot{\boldsymbol{q}}_+&=\omega\boldsymbol{p}_-, \\
\dot{\boldsymbol{q}}_c&=0, \\
\dot{\boldsymbol{p}}_c&=-2\sqrt{2}G\boldsymbol{q}_+,
\end{aligned}
\end{equation}
where $\left\{\boldsymbol{q}_+, ~ \boldsymbol{p}_-\right\}$ forms an isolated QMFS, and Eq.~\eqref{dec29-12} is
consistent with \cite[Eq. (S10)]{LOW+21}.

Moreover, by Eq. \eqref{dec29-11} it can be verified that the opto-mechanical system realizes a BAE measurement of $\boldsymbol{q}_{\mathrm{out}}$ with respect to $\boldsymbol{p}_{\mathrm{in}}$, and a BAE measurement of $\boldsymbol{p}_{\mathrm{out}}$ with respect to $\boldsymbol{q}_{\mathrm{in}}$.

   \section{Response to single-photon states}
\label{sec:single-photon-response}

  \subsection{Continuous-mode single-photon states}
\label{subsec:photon_state}
In this subsection, we introduce continuous-mode single-photon states of a free traveling light field.

We look at the single-channel ($m=1$) case first. Denote $\ket{1_t} = \mbf{b}_{\rm in}^\ast (t)\ket{\Phi_0}$, i.e.,~a photon is generated at the time instant $t$ by the creation operator $ \mbf{b}_{\rm in}^\ast(t)$ from the vacuum field $\ket{\Phi_0}$. By Eq.~\eqref{eq:CCR_b} we get $\braket{1_t\vert 1_\tau} = \delta(t-\tau)$; in other words, the entries in $\{\ket{1_t}: t\in \mathbb{R}\}$ are orthogonal. Actually,  $\{\ket{1_t}: t\in \mathbb{R}\}$ form a complete basis of    continuous-mode single-photon states of a free propagating light field, in the sense that a continuous-mode single-photon state $\ket{1_\xi}$  of the temporal pulse shape $\xi \in L_{2}(\mathbb{R},\mathbb{C})$ can be expressed as
\beq \la{1 photon jan21}
\ket{1_\xi} \equiv \mbf{B}_{\rm in}^\ast(\xi)\ket{\Phi_0} \triangleq \int_{-\infty}^\infty \xi(t) \ket{1_t}d t.
\eeq
(Here, it is assumed that the $L_2$ norm $\| \xi\| \triangleq\sqrt{\int_{-\infty}^\infty \vert \xi(t)\vert ^2}dt=1$.  Then $\braket{1_\xi\vert 1_\xi}=1$.) The physical interpretation of the single-photon state $\ket{1_\xi} $ is that  the probability of detecting the photon in the time bin $[t,t+dt)$ is $\vert \xi(t)\vert ^2 dt$. In the frequency domain, we denote $\ket{1_\omega} \triangleq \mbf{b}_{\rm in}^\ast [i\omega]\ket{0}$.  Hence, in the frequency domain \eqref{1 photon jan21} becomes
\beq \la{1 photon jan28}
\ket{1_\xi} = \int_{-\infty}^\infty \xi[i\omega] \ket{1_\omega}d \omega.
\eeq

\bmrk
Notice
\beqn
&&\l 1_\xi |d\mbf{B}_{\rm in}^\ast(t)  d\mbf{B}_{\rm in} (t) |1_\xi \r
\nonumber
\\
&=&
 \int_{-\infty}^\infty \int_{-\infty}^\infty dr d\tau \xi(r)^\ast \xi(\tau)  \int_{t}^{t+dt}\int_{t}^{t+dt}   dt_1  dt_2  \l\Phi_0| \mbf{b}_{\rm in}(r)  \mbf{b}_{\rm in}^\ast (t_1)  \mbf{b}_{\rm in} (t_2)   \mbf{b}_{\rm in}^\ast  (\tau) \Phi_0\r
 \nonumber
 \\
&=&
 \int_{t}^{t+dt}\int_{t}^{t+dt}   dt_1  dt_2 \int_{-\infty}^\infty \int_{-\infty}^\infty dr d\tau \xi(r)^\ast \xi(\tau)  \delta(t_1-r) \delta(t_2-\tau)
  \nonumber
 \\
&=&
 \int_{t}^{t+dt}    \xi^\ast(t_1) dt_1 \int_{t}^{t+dt}  \xi(t_2)   dt_2.
\eeqn
Thus, for most functions $\xi \in L_{2}(\mathbb{R},\mathbb{C})$,
\beq
\l 1_\xi |d\mbf{B}_{\rm in}^\ast(t)  d\mbf{B}_{\rm in} (t) |1_\xi \r  = ({\cal O}(dt))^2.
\eeq
In It\^{o} stochastic calculus,  $\l 1_\xi |d\mbf{B}_{\rm in}^\ast(t)  d\mbf{B}_{\rm in} (t) |1_\xi \r =0$. This shows that the field with  a continuous-mode single-photon state $\ket{1_\xi}$ is a canonical field.
\emrk

In this tutorial, we investigate single-photon states from a control-theoretic
perspective. For physical implementation of single photon generation, detection
and storing, please refer to  the physics
literature \cite{RL00,LHA+01,YKS+02,MBB+04,HSG+07,WKF07,OFV09,BC09,LR09,SFY10,EBL+11,BRV12,PHC+14,LMS,RR15,QPB+15,BRS+15,NJY16,OOM+16,Peng2016,GKM+17,RWF17,DTK+18,WQD+19,TOS+19}
and references therein. A concise discussion can be found in \cite[Section
3.1]{Z21}.

  \subsection{Response of quantum linear systems to continuous-mode single-photon states}
\label{subsec:response photon_state}

Let the linear quantum system $G$ be initialized in the coherent state $\ket{\alpha}$ (defined in Section   \ref{sec:Gaussian state}) and the input field be initialized in the vacuum state $\ket{\Phi_0}$. Then the initial joint system-field state is $\rho_{0g}\triangleq \ket{\alpha}\bra{\alpha}
\otimes \ket{\Phi_0}\bra{\Phi_0}$ in the form of a density matrix.  Denote
\begin{equation}
\label{eq:rho_inf_g}
\rho_{\infty g} = \lim_{t\rightarrow\infty,t_{0}\rightarrow-\infty}U\left(
t,t_{0}\right)  \rho_{0g}U\left(t,t_{0}\right) ^{\ast}.
\end{equation}
Here, $t_0\to -\infty$ indicates that the interaction starts in the remote past and $t\to\infty$ means that we are interested in the dynamics in the far future. In other words, we look at the steady-state dynamics.
Define
\begin{equation}
\label{eq:rho_field}
\rho_{{\rm field},g}\triangleq  \langle \alpha \vert \rho_{\infty g}\vert  \alpha \rangle.
\end{equation}
In other words, the system is traced off and we focus on the  steady  state of the output field.

Let the $k$th input channel be in a single photon state $\vert 1_{\mu_k} \rangle$, $k=1,\ldots, m$. Thus, the state of the $m$-channel input is given by the tensor product
\begin{equation}\label{eq:multichannel}
\vert \Psi_{\mu} \rangle  = \vert 1_{\mu_1} \rangle \otimes \cdots \otimes  \vert 1_{\mu_m} \rangle .
\end{equation}
Denote $\mu = [\mu_1 \ \ \cdots \ \ \mu_m]^\top$.

\bthm  \label{cor:passive}
Assume that the passive linear quantum system \eqref{eq:passive_sys_a} is Hurwitz stable,  initialized in  the vacuum state $\ket{0}$ and driven by an $m$-photon input state  $\vert \Psi_{\mu} \rangle$ in Eq. \eqref{eq:multichannel}.  Then the steady-state output state is another  $m$-photon $\vert \Psi_{\nu} \rangle$    whose pulse   $\nu = [\nu_1 \ \ \cdots \ \ \nu_m]^\top$ is given by
\beq
\nu[i\omega] =   \Xi_{G^-}[i\omega]  \mu[i\omega].
\eeq
\ethm

If the linear system $G$ is not passive, or is not initialized in the
vacuum state $\ket{0}$, the steady-state output field state $\rho_{\rm out}$ in
general is not a single- or multi-photon state. This new type of states has
been named ``photon-Gaussian'' states in \cite{ZJ13}. Moreover, it has been
proved in \cite{ZJ13} that the class of photon-Gaussian states is invariant
under the steady-state action of a linear quantum system.  In what follows we
present this result.

\begin{definition}\cite[Definition 1]{ZJ13} \label{def:F}
A state $\rho_{\xi, R}$ is said to be a \emph{photon-Gaussian} state if it belongs to the set
\beq\label{class_F}
\bea
\mathcal{F} \triangleq&\; \left\{\rho_{\xi, R} = \prod\limits_{k=1}^{m}\sum_{j=1}^{m}\left(\mbf{B}_{{\rm in}, j}^\ast (\xi_{jk}^-) -\mbf{B}_{{\rm in}, j}(\xi_{jk}^+) \right)\rho_R\left(\prod\limits_{k=1}^{m}\sum_{j=1}^{m}\left(\mbf{B}_{{\rm in}, j}^\ast (\xi_{jk}^-) -\mbf{B}_{{\rm in}, j}(\xi_{jk}^+) \right)\right)^\ast \right.
\\
&  \ \ \ \   \left.  : \mathrm{function~}  \xi=\Delta(\xi^-, \xi^+) \mathrm{~and~ density~matrix~} \rho_R \mathrm{~satisfy ~} \mathrm{Tr}[\rho_{\xi, R}] = 1   \right\}.
\eea
\eeq
\end{definition}

\begin{theorem}\cite[Theorem 5]{ZJ13}   \label{thm:main}
 Let  $\rho_{\xi_\mm[in], R_\mm[in]}  \in \mathcal{F}$ be a photon-Gaussian input state.  Also, assume that $G$ is Hurwitz stable and is initialized in a coherent state $\ket{\alpha}$.   Then the linear quantum system $G$ produces in steady state a photon-Gaussian output state
$\rho_{\xi_{\rm out}, R_{\rm out}} \in \mathcal{F}$, where
\begin{equation}\label{eq:SS}\begin{aligned}
 \xi_{\rm out}[s]  =&\;  \Xi_{G}[s] \xi_\mm[in][s] ,   \\
R_{\rm out}[i\omega] =&\; \Xi_G[ \imath \omega] R_\mm[in][\imath\omega]  \Xi_G[i\omega]^\dag  .
\end{aligned}
\end{equation}
\end{theorem}

  \subsection{Response of quantum linear systems to continuous-mode multi-photon states}
\label{subsec:response multi-photon_state}

Response of quantum linear systems to multi-photon states has been studied
in \cite{Z14,Z17}, as generalization  to  the single-photon case.  In
this subsection, we present one of the main results in these papers.

Let  there  be  $\ell_j$ photons in  the $j$th channel. The state for this channel is
\begin{equation}
\ket{\Psi_j} =
 \frac{1}{\sqrt{N_{\ell_j}}}\int_{\ell_j}\Psi_j(t_1,\ldots,t_{\ell_j})\mbf{b}_{{\rm in},j}^\ast(t_1)\cdots \mbf{b}_{{\rm in},j}^\ast(t_{\ell_j})
dt_{1}\ldots dt_{\ell_j}\ket{\Phi_0},
\label{eq:general_state_-1}
\end{equation}
where $\Psi_j$ is the pulse shape and $\sqrt{N_{\ell_j}}$ is the normalization coefficient. Here,  for any integer $r>1$, we write $\int_r$ for integration in the space $\mathbb{C}^r$.  If $\ell_j=0$, then Eq.~\eqref{eq:general_state_-1} is understood as $\ket{\Psi_j}  = \ket{\Phi_0}$.  Then the state for the $m$-channel input field can be defined as
\begin{equation} \label{eq:general_state_0}
|\Psi\rangle =\prod\limits_{j=1}^m |\Psi_j\rangle  .
\end{equation}

Next, we rewrite this $m$-channel multi-photon state into an alternative form; this will enable us to present the input and output states in a unified form. For $j=1,\ldots,m$, $i=1,\ldots,\ell_j$, and $k_i=1,\ldots,m$, define functions
\beq \label{eq:psi_update}
 \Psi_{j,k_1,\ldots,k_{\ell_j}}(\tau_1,\ldots,\tau_{\ell_j})
\triangleq  \left\{\begin{array}{@{}cc@{}}             \Psi_j(\tau_1,\ldots,\tau_{\ell_j}), & k_1=\cdots=k_{\ell_j} = j, \\
             0, & \mathrm{otherwise}.\end{array}\right.
\eeq
Then we define a class of pure $m$-channel multi-photon states
\begin{align}
& \mathcal{F}_1
\label{eq:F_1}\\
=& \left\{|\Psi\rangle
=
\prod_{j=1}^m \frac{1}{\sqrt{N_{\ell_j}}}\sum_{k_1,\ldots,k_{\ell_j}=1}^m \int_{\ell_j} \Psi_{j,k_1,\ldots,k_{\ell_j}}(\tau_1,\ldots,\tau_{\ell_j}) \right.
\nonumber\\
&~~~~~ \left. \times
\prod_{i=1}^{\ell_j} \mbf{b}_{{\rm in},k_i}^\ast(\tau_i)
d\tau_{1} \ldots d_{\ell_j}\ket{\Phi_0} : \langle\Psi|\Psi\rangle=1\right\} .
\nonumber
\end{align}
Clearly,  $|\Psi\rangle$ in Eq. (\ref{eq:general_state_0}) belongs to $\mathcal{F}_1$.

We need some operations between matrices and tensors. Let $g_G(t)=(g_G^{jk}(t))\in\mathbb{C}^{m\times m}$ be the impulse response function of the  quantum linear system $G$ in Fig. \ref{fig:sys}. For each $j=1,\ldots,m$, let $\mathscr{V}_j(t_1,\ldots,t_{\ell_j})$ be an $\ell_j$-way $m$-dimensional tensor function that encodes the pulse information of the $j$th input field containing $\ell_j$ photons. Denote the entries of $\mathscr{V}_j(t_1,\ldots,t_{\ell_j})$ by $\mathscr{V}_{j,k_1,\ldots,k_{\ell_j}}(t_1,\ldots,t_{\ell_j})$.  Define an  $\ell_j$-way $m$-dimensional tensor $\mathscr{W}_j$ with entries given by the following multiple convolution
\begin{align*}\label{eq:W}
& \mathscr{W}_{j,r_1,\ldots,r_{\ell_j}}(t_1,\ldots,t_{\ell_j})= \sum_{k_1,\ldots,k_{\ell_j}=1}^m  \int_{\ell_j} g_G^{r_1k_1}(t_1-\iota_1)\times\\
 &  \cdots \times g_G^{r_{\ell_j}k_{\ell_j}}(t_{\ell_j}-\iota_{\ell_j})\mathscr{V}_{j,k_1,\ldots,k_{\ell_j}}(\iota_1,\ldots,\iota_{\ell_j})
 d\iota_1\ldots d_{\ell_j} \nonumber
\end{align*}
for all $1\leq r_1,\ldots,r_{\ell_j}\leq m$.
In compact form we write
\begin{equation}
\label{eq:VW_j}
\mathscr{W}_j = \mathscr{V}_j   \times_1 g_G  \times_2 \cdots  \times_{\ell_j} g_G , \ \ \ \forall j=1,\ldots,m,
\end{equation}
cf. \cite{LDW00},\cite[Sec. 2.5]{KB09},\cite{BLS10} and \cite{CML+15}.

\begin{theorem}(\cite[Theorem 12]{Z14})\label{thm:passive_multi_channel}
Suppose that the quantum linear system $G$ is Hurwitz stable and passive. The steady-state output state of  $G$ driven by a state $|\Psi_{\rm in}\rangle\in \mathcal{F}_1$ is another state $|\Psi_{\rm out}\rangle\in \mathcal{F}_1$ with wave packet transfer
\begin{align*}
\Psi_{{\rm out},j} =  \Psi_{{\rm in},j} \times_1 g_{G^-} \times_2\cdots\times_{\ell_j} g_{G^-} , \ \ \ \forall j=1,\ldots,m,
\label{eq:psi_out}
\end{align*}
where the operation between the matrix and tensor is defined in Eq. (\ref{eq:VW_j}).
\end{theorem}

More discussions on continuous-mode multi-photon states can be found  in \cite{Z14,Z17}.  An application to the amplification of optical Schr\"odinger
cat states can be found in \cite{SZW+21}. Simply speakin, the mathematial methods proposed in \cite{Z14,Z17} can be used to study photon-catalyzed quabtunm non-Gaussian states, which are useful resources in quantun information processing \cite{KTS+06,TNT+10,BDS+12} The problem of the response of
quantum nonlinear systems to multi-photon states has been studied
in \cite{GEP+98,DZ02,milburn08,SAL09,RSF10,FKS10,WMS+11,BCB+12,LP14,NKM+15,SCC15, FB15,BCG16,Leong2016,PDZ16,PZJ16,DSC17,ZFHW+17,FTR+18,TFX18,DSC19,NCP+19,MKL+19,DZA19b,ZP20,DZWW21,LDZW22}.
It turns out that the linear systems theory plays a key role in some of these
studies.

A continuous-mode single-photon field studied in Subsection  \ref{subsec:photon_state}
has statistical properties. Thus, it is natural to study the filtering problem
of a quantum system driven by a continuous-mode single-photon states.
Continuous-mode single-photon filters were derived
in \cite{GJN+12} and \cite{GJN+12b} first, and their multi-photon version was
developed in \cite{BC17,DZA18,DZA19a} and \cite{SZX16}. A review of
continuous-mode single or multi-photon states is given in \cite{Z21}.

\section{Feedback control of quantum linear systems}\label{sub:feedback}

Feedback control of quantum systems has been covered in several books, for
example, \cite{Jacobs14,NY17,MR20} and \cite{WM10}. In
particular, the monograph \cite{NY17} is devoted to the feedback control of
quantum linear systems. Depending on whether the underlying quantum system
(plant) is measured and the measurement data is used for the feedback control
of the plant, feedback control methods of quantum systems can be roughly
divided into two categories: measurement feedback control and coherent feedback
control. It is clear that the former makes use of measurement information,
whereas in a coherent feedback network no measurement is involved and thus
coherence of quantum signals is preserved.  To our understanding, measurement
feedback control has been studied intensively and well recorded
in \cite{WM10}, \cite{Jacobs14},  and \cite{MR20}. In contrast, coherent feedback control is
still a bit new to many researchers in the quantum control community, though it
has advantages in many
applications \cite{Gardiner93,WM94,SLloyd00,NWCL00,HJ06,GGY08,GJ09,GJ09b,JG10,petersen11,TNP+12,IYY+12,VP13,NY14,JWW14,Grimsmo15,DZA16,HC16,ANP+17,NY17,CKS17,ZLW+17,XPD17,KJN18,GP18,Mabuchi19,PJU+20,AMD20,SY21,BNC+21,KPS+21,PVB+21,HCS21,Cobarrubia21,WHM+22,DP22,DRM+22}.
Thus, in this section, we describe briefly linear quantum coherent feedback
networks and use a recent experiment as demonstration.

\subsection{Quantum coherent feedback linear networks}

\label{subsec:coherent feedback}
For notational simplicity, in this and the next  subsections, the subscript ``in'' for input fields is omitted.
\begin{figure}[tbph]
\centering
\includegraphics[width=0.70\textwidth]{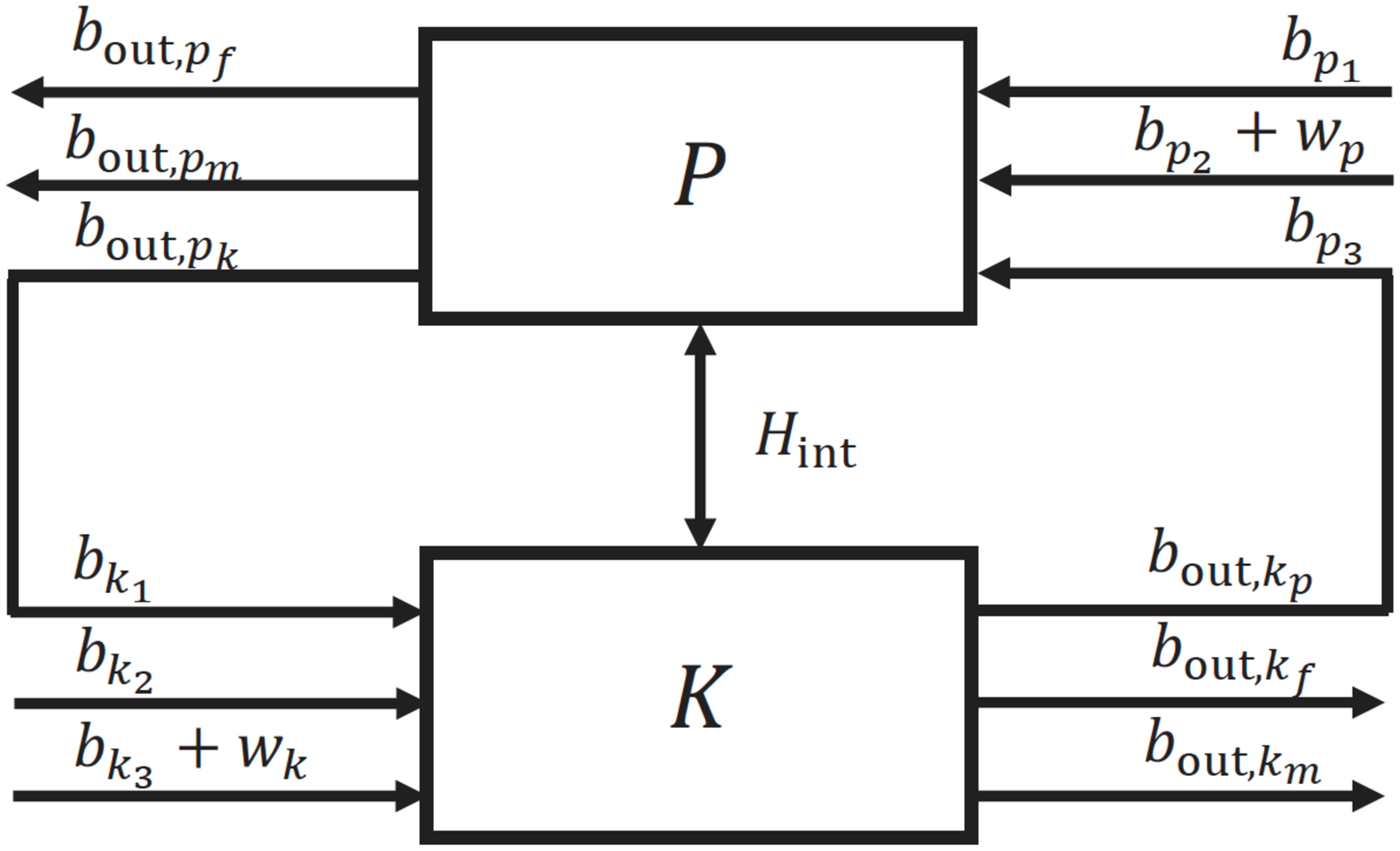}
\caption{Coherent feedback control network.}
\label{sec-PK}
\end{figure}
In  Fig. \ref{sec-PK}, the closed-loop system contains a quantum plant $P$
and a quantum controller $K$ to be designed. We look at the plant
$P$ first, which is a quantum linear system driven by three types of
input channels. To be specific, $\mbf{b}_{p_1}$ describes the free input channels
which may model quantum white noise such as the vacuum or thermal noise. For
example, $\mbf{b}_{p_1}$ can be unmodeled quantum vacuum noise on the quantum
plant due to imperfection of the physical system. As discussed in Remark
\ref{rem:signal+noise},  $\mbf{b}_{p_2}$ models the quantum vacuum fields and
$w_p$ represents quantum or  classical signals. For example, $w_p$ may be some undesired
disturbance on the quantum plant $P$; it may also represent classical
signals from a classical controller such as $u_2$ in \cite[Figure
5.1]{NY17}.   The third input, which is the output $\mbf{b}_{\mathrm{out},k_p}$ of the
controller $K$, is denoted by $\mbf{b}_{p_3}$. Correspondingly, we define
three coupling operators $\mbf{L}_{p_1}$, $\mbf{L}_{p_2}$, $\mbf{L}_{p_3}$ and three
scattering matrices $S_{p_1}$, $S_{p_2}$, $S_{p_3}$ for the
input--output channels. By means of the theory introduced in Section
\ref{sec:system}, the \emph{physical} output channels are $\mbf{b}_{\mathrm{out},p_1}$, $\mbf{b}_{\mathrm{out},p_2}$,
and $\mbf{b}_{\mathrm{out},p_3}$, which in the integral form are given by
\begin{equation}
\label{dec31-1}
\begin{aligned}
d\mbf{B}_{\mathrm{out},p_1}(t)&=\mbf{L}_{p_1}(t)dt +S_{P_1}d\mbf{B}_{p_1}(t), \\
d\mbf{B}_{\mathrm{out},p_2}(t)&=\mbf{L}_{p_2}(t)dt+S_{P_2}(d\mbf{B}_{p_2}(t)+w_p dt), \\
d\mbf{B}_{\mathrm{out},p_3}(t)&=\mbf{L}_{p_3}(t)dt+S_{P_3}d\mbf{B}_{p_3}(t),
\end{aligned}
\end{equation}

These physical channels can be put into three categories  $\mbf{b}_{\mathrm{out},p_f}$, $\mbf{b}_{\mathrm{out},p_m}$, and  $\mbf{b}_{\mathrm{out},p_k}$.  Here,  $\mbf{b}_{\mathrm{out},p_f}$ is a set of  free output channels, $\mbf{b}_{\mathrm{out},p_m}$ represents a collection of output field channels which are to be measured, and $\mbf{b}_{\mathrm{out},p_k}$ is a set of output channels to be sent to the controller $K$. As a result, the dynamics of the plant $P$ can be described  by the following QSDEs
\begin{equation}
\label{dec31-2}
\begin{aligned}
d\breve{\mbf{a}}_p(t)&=\; \mathcal{A}_p \breve{a}_p(t)dt + \mathcal{E}_p \breve{v}_p(t)dt + \mathcal{B}_{p_1}d\breve{\mbf{B}}_{p_1}(t)
\\
&+ \ \  \mathcal{B}_{p_2}(d\breve{\mbf{B}}_{p_2}(t)+\breve{w}_p(t)dt)+\mathcal{B}_{p_3}d\breve{\mbf{B}}_{p_3(t)},
\\
d\breve{\mbf{B}}_{\mathrm{out},p_f}(t)&=\;\mathcal{C}_{p_f}\breve{a}_p(t)dt +\mathcal{D}_{p_{f1}}d\breve{\mbf{B}}_{p_1}(t)
\\
&+ \ \  \mathcal{D}_{p_{f2}}(d\breve{\mbf{B}}_{p_2}(t)+\breve{w}_p(t)dt) +\mathcal{D}_{p_{f3}}d\breve{\mbf{B}}_{p_3}(t),
\\
d\breve{\mbf{B}}_{\mathrm{out},p_m}(t)&=\;\mathcal{C}_{p_m}\breve{a}_p(t)dt +\mathcal{D}_{p_{m1}}d\breve{\mbf{B}}_{p_1}(t)
\\
&+ \ \  \mathcal{D}_{p_{m2}}(d\breve{\mbf{B}}_{p_2}(t)+\breve{w}_p(t)dt) +\mathcal{D}_{p_{m3}}d\breve{\mbf{B}}_{p_3}(t),
\\
d\breve{\mbf{B}}_{\mathrm{out},p_k}(t)&=\;\mathcal{C}_{p_k}\breve{a}_p(t)dt +\mathcal{D}_{p_{k1}}d\breve{\mbf{B}}_{p_1} (t)
\\
&+ \ \  \mathcal{D}_{p_{k2}}(d\breve{\mbf{B}}_{p_2}(t)+\breve{w}_p(t)dt) +\mathcal{D}_{p_{k3}}d\breve{\mbf{B}}_{p_3}(t).
\end{aligned}
\end{equation}

Similarly, the inputs $\mbf{b}_{k_1}$, $\mbf{b}_{k_2}$, $\mbf{b}_{k_3}$ and outputs $\mbf{b}_{\mathrm{out},k_p}$, $\mbf{b}_{\mathrm{out},k_f}$, $\mbf{b}_{\mathrm{out},k_m}$ of the controller $K$ are labeled in  Fig. \ref{sec-PK}, respectively. The QSDEs for the controller $K$ is given by
\begin{equation}
\label{dec31-4}
\begin{aligned}
d\breve{\mbf{a}}_k(t)&=\mathcal{A}_k \breve{\mbf{a}}_k(t)dt + \mathcal{E}_k \breve{v}_k(t)dt + \mathcal{B}_{k_1}d\breve{\mbf{B}}_{k_1}(t)
\\
&+ \ \  \mathcal{B}_{k_2}d\breve{\mbf{B}}_{k_2}(t)+\mathcal{B}_{k_3}(d\breve{\mbf{B}}_{k_3}(t)+\breve{w}_k(t)dt),
\\
d\breve{\mbf{B}}_{\mathrm{out},k_p}(t)&=\mathcal{C}_{k_p}\breve{\mbf{a}}_k(t)dt+\mathcal{D}_{k_{p2}}d\breve{\mbf{B}}_{k_2}(t) +\mathcal{D}_{k_{p3}}(d\breve{\mbf{B}}_{k_3}(t)+\breve{w}_k(t)dt),
\\
d\breve{\mbf{B}}_{\mathrm{out},k_f}(t)&=\mathcal{C}_{k_f}\breve{\mbf{a}}_k(t)dt +\mathcal{D}_{k_{f1}}d\breve{\mbf{B}}_{k_1}(t)
\\
&+ \ \  \mathcal{D}_{k_{f2}}d\breve{\mbf{B}}_{k_2}(t) +\mathcal{D}_{k_{f3}}(d\breve{\mbf{B}}_{k_3}(t)+\breve{w}_k(t)dt),
\\
d\breve{\mbf{B}}_{\mathrm{out},k_m}(t)&=\mathcal{C}_{k_m}\breve{\mbf{a}}_k(t)dt +\mathcal{D}_{k_{m1}}d\breve{\mbf{B}}_{k_1} (t)
\\
&+ \ \  \mathcal{D}_{k_{m2}}d\breve{\mbf{B}}_{k_2}(t) +\mathcal{D}_{k_{m3}}(d\breve{\mbf{B}}_{k_3}(t)+\breve{w}_k(t)dt).
\end{aligned}
\end{equation}

 \bmrk{\rm
As shown in  Fig. \ref{sec-PK},  the input field $\mbf{b}_{k_1}$ of the controller
$K$ is the output field $\mbf{b}_{{\rm out},p_k}$ of the plant $P$, and
the output $\mbf{b}_{{\rm out},k_p}$ of the controller $K$ is  the input
$\mbf{b}_{p_3}$ of the plant $P$. As shown in Eq.~\eqref{dec31-2} the
output field $\mbf{b}_{{\rm out},p_k}$ may correspond to the input field  $\mbf{b}_{p_3}$ which
is the field $\mbf{b}_{{\rm out},k_p}$. To guarantee causality, the field $\mbf{b}_{{\rm out},k_p}$ must
not contain the input field  $\mbf{b}_{k_1}=\mbf{b}_{{\rm out},p_k}$. This is the reason why the evolution
of $\breve{\mbf{B}}_{\mathrm{out},k_p}(t)$ in Eq.~\eqref{dec31-4} depends on the free traveling fields
$\mbf{b}_{k_2}$ and $\mbf{b}_{k_3}$, but not on $\mbf{b}_{k_1}$. There are other
possible configurations which can guarantee causality.  One example is given in Subsection
\ref{subsec:coherent feedback example} to demonstrate one such possible
configuration. In this example,    $\mbf{b}_{{\rm out},p_k}$ corresponds to the input laser
which is $\mbf{b}_{p2}+w_p$.  $\mbf{b}_{{\rm out},p_k}$ is sent to $K$ generating the
corresponding output $\mbf{b}_{{\rm out},k_p}$. However,  $\mbf{b}_{{\rm out},k_p}$ is the input
$\mbf{b}_{p_3}$ of  $P$ whose corresponding output is the output laser
$\mbf{b}_{{\rm out},p_m}$  to be measured.  This fundamental assumption is also used in
quantum circuits \cite[Section 1.3.4]{NC10}.
}
\emrk

The plant $P$ and controller $K$ can also be directly coupled via an interaction Hamiltonian $H_{\mathrm{int}}$ as labeled in  Fig. \ref{sec-PK} with the following form
\begin{equation}
\label{dec31-6}
\mbf{H}_{\mathrm{int}}=\frac{1}{2}\left(\breve{\mbf{a}}_p^\dagger \Xi^\dagger \breve{\mbf{a}}_k + \breve{\mbf{a}}_k^\dagger \Xi \breve{\mbf{a}}_p \right),
\end{equation}
where
$
\Xi=\Delta (\imath K_{-},\imath K_{+})$
for complex matrices $K_{-}$ and $K_{+}$ with suitable dimensions.  It is easy to see that the commutators $[\breve{\mbf{a}}_p, \mbf{H}_{\mathrm{int}}]$ and $[\breve{\mbf{a}}_k, \mbf{H}_{\mathrm{int}}]$ yield
\begin{equation}
 \label{eq:direct-7}
{\mathcal{B}}_{12} = - \Delta( K_-, K_+)^\flat , \ \;{\mathcal{B}}_{21} = - {\mathcal{B}}_{12}^\flat = \Delta( K_-, K_+) .
\end{equation}
On the other hand, indirect coupling refers to the coupling through field
channels  $\breve{\mbf{b}}_{{\rm out},P}$ and  $\breve{\mbf{b}}_{{\rm out},K}$.  More discussions on direct coupling
and indirect coupling can be found in,
e.g., \cite{CKS17} and \cite{ZJ11,ZJ12}.

The controller matrices for direct and indirect couplings are to be found to optimize performance criteria defined in terms of the set of  closed-loop performance variables
\begin{equation}
\breve{z}(t) =[ {\mathcal{C}}_{p} ~~ {\mathcal{C}}_k ]\left[\begin{array}{@{}c@{}}\breve{\mbf{a}}_p(t) \\
\breve{\mbf{a}}_{k}(t)
\end{array}
\right] +  {\mathcal{D}}_z\breve{w}(t) .
\label{syn_performance}
\end{equation}
An example of control  performance variables is $z=\mbf{a}_p$ for a single-mode
cavity. Then $z^\ast z =\mbf{a}_p^\ast \mbf{a}_p = \ff{\mbf{q}^2+\mbf{p}^2-1}{2}$. Minimizing the mean value of $\int z^\ast(t)z(t)dt$ means
cooling the cavity oscillator; see \cite{HM13}. The form of performance
variables for $H^\infty$ control can be found in \cite{JNP08,Mabuchi08}. More
discussions can be found in books \cite[Chapter 6]{WM10} and { \cite{NY17}.

By eliminating the in-loop fields $\mbf{b}_{{\rm out}, p_k}$ and    $\mbf{b}_{{\rm out}, k_p}$,   the overall plant-controller quantum system, including direct and indirect couplings, can be written as
\begin{equation}
\label{dec31-7}
\begin{aligned}
\left[\begin{array}{@{}c@{}}    d\breve{\mbf{a}}_p(t) \\
    d\breve{\mbf{a}}_k(t)
\end{array}
\right]=&\;\left[\begin{array}{@{}cc@{}}            \mathcal{A}_p & \mathcal{B}_{p_3}\mathcal{C}_{k_p}+\mathcal{B}_{12} \\
            \mathcal{B}_{k_1}\mathcal{C}_{p_k}+\mathcal{B}_{21} & \mathcal{A}_k+\mathcal{B}_{k_1}\mathcal{D}_{p_{k3}}\mathcal{C}_{k_p}
\end{array}
        \right]\left[\begin{array}{@{}c@{}}                   \breve{\mbf{a}}_p(t) \\
                   \breve{\mbf{a}}_k(t)
\end{array}
               \right]dt \\
&+ \ \  \left[\begin{array}{@{}cc@{}}                           \mathcal{B}_{p_2} & \mathcal{B}_{p_3}\mathcal{D}_{k_{p3}} \\
                           \mathcal{B}_{k_1}\mathcal{D}_{p_{k2}} & \mathcal{B}_{k_3}+\mathcal{B}_{k_1}\mathcal{D}_{p_{k3}}\mathcal{D}_{k_{p3}}
\end{array}
                       \right]\left[\begin{array}{@{}c@{}}                                  \breve{w}_p(t) \\
                                  \breve{w}_k(t)
\end{array}
                              \right]dt\\
&+ \ \  \left[\begin{array}{@{}cc@{}}                                          \mathcal{E}_p & 0 \\
                                          0 & \mathcal{E}_k
\end{array}
                                      \right]\left[\begin{array}{@{}c@{}}                                                 \breve{v}_p(t) \\
                                                 \breve{v}_k(t)
\end{array}
                                             \right]dt
                                             +{\mathcal{G}}_{\rm cl}\left[\begin{array}{@{}c@{}}                                                                d\breve{\mbf{B}}_{p_1}(t) \\
                                                                d\breve{\mbf{B}}_{p_2}(t) \\
                                                                d\breve{\mbf{B}}_{k_2}(t) \\
                                                                d\breve{\mbf{B}}_{k_3}(t)
\end{array}
                                                            \right],
\end{aligned}
\end{equation}
where
\beq
{\mathcal{G}}_{\rm cl} = \left[\begin{array}{@{}cccc@{}}                                                         \mathcal{B}_{p_1} & \mathcal{B}_{p_2} & \mathcal{B}_{p_3}\mathcal{D}_{k_{p2}} & \mathcal{B}_{p_3}\mathcal{D}_{k_{p3}} \\
                                                         \mathcal{B}_{k_1}\mathcal{D}_{p_{k1}} & \mathcal{B}_{k_1}\mathcal{D}_{p_{k2}} & \mathcal{B}_{k_2}+\mathcal{B}_{k_1}\mathcal{D}_{p_{k3}}\mathcal{D}_{k_{p2}} & \mathcal{B}_{k_3}+\mathcal{B}_{k_1}\mathcal{D}_{p_{k3}}\mathcal{D}_{k_{p3}}
\end{array}
                                                     \right].
\eeq

Because standard matrix algorithms commonly used in $H^\infty$ synthesis and LQG synthesis are for real-valued matrices, in what follows we resort to quadrature representation. Let $\mbf{x}_p$, $\mbf{x}_k$, $\tilde{w}_p$, $\tilde{w}_k$, $u_p$, $u_k$, ${\mathcal{U}}_{p_1}$, ${\mathcal{U}}_{p_2}$, ${\mathcal{U}}_{k_2}$, ${\mathcal{U}}_{k_3}$ be the quadrature counterparts of $\breve{\mbf{a}}_p$, $\breve{\mbf{a}}_k$, $\breve{w}_p$, $\breve{w}_k$, $\breve{v}_p$, $\breve{v}_k$, $\breve{\mbf{B}}_{p_1}$, $\breve{\mbf{B}}_{p_2}$, $\breve{\mbf{b}}_{k_2}$, $\breve{\mbf{B}}_{k_3}$, respectively. Then the closed-loop quantum system in the quadrature representation is given by
\begin{equation}
\label{dec31-9}
\begin{aligned}
\left[\begin{array}{@{}c@{}}    d\mbf{x}_p(t) \\
    d\mbf{x}_k(t)
\end{array}
\right]=&\; \mathbb{A}_{\rm cl}\left[\begin{array}{@{}c@{}}                   \mbf{x}_p(t) \\
                   \mbf{x}_k(t)
\end{array}
               \right] dt+ \mathbb{B}_{\rm cl} \left[\begin{array}{@{}c@{}}                                  \tilde{w}_p(t) \\
                                  \tilde{w}_k(t)
\end{array}
                              \right]dt\\
       &\;                       +\mathbb{E}_{\rm cl}\left[\begin{array}{@{}c@{}}                                                 u_p(t) \\
                                                 u_k(t)
\end{array}
                                             \right] dt +
                                             \mathbb{G}_{\rm cl}\left[\begin{array}{@{}c@{}}                                                                d\mathcal{U}_{p_1}(t) \\
                                                                d\mathcal{U}_{p_2}(t) \\
                                                                d\mathcal{U}_{k_2}(t) \\
                                                                d\mathcal{U}_{k_3}(t)
\end{array}
                                                            \right],
\\
\tilde{z}(t)=&\; \mathbb{C}_{\rm cl}\left[\begin{array}{@{}c@{}}\mbf{x}_p(t) \\
\mbf{x}_{k}(t)
\end{array}
\right]+\mathbb{D}_{\rm cl}\tilde{w}(t) ,
\end{aligned}
\end{equation}
where
\beq
\bea
\mathbb{A}_{\rm cl} =&\; \left[\begin{array}{@{}cc@{}}            \mathbb{A}_p & \mathbb{B}_{p_3}\mathbb{C}_{k_p}+\mathbb{B}_{12} \\
            \mathbb{B}_{k_1}\mathbb{C}_{p_k}+\mathbb{B}_{21} & \mathbb{A}_k+\mathbb{B}_{k_1}\mathbb{D}_{p_{k3}}\mathbb{C}_{k_p}
\end{array}
        \right],
\\
   \mathbb{B}_{\rm cl} =&\; \left[\begin{array}{@{}cc@{}}                           \mathbb{B}_{p_2} & \mathbb{B}_{p_3}\mathbb{D}_{k_{p3}} \\
                           \mathbb{B}_{k_1}\mathbb{D}_{p_{k2}} & \mathbb{B}_{k_3}+\mathbb{B}_{k_1}\mathbb{D}_{p_{k3}}\mathbb{D}_{k_{p3}}
\end{array}
                       \right], \ \
\mathbb{E}_{\rm cl} = \left[\begin{array}{@{}cc@{}}                                          \mathbb{E}_p & 0 \\
                                          0 & \mathbb{E}_k
\end{array}
                                      \right]
\\
\mathbb{G}_{\rm cl}=&\; \left[\begin{array}{@{}cccc@{}}                                                         \mathbb{B}_{p_1} & \mathbb{B}_{p_2} & \mathbb{B}_{p_3}\mathbb{D}_{k_{p2}} & \mathbb{B}_{p_3}\mathbb{D}_{k_{p3}} \\
                                                         \mathbb{B}_{k_1}\mathbb{D}_{p_{k1}} & \mathbb{B}_{k_1}\mathbb{D}_{p_{k2}} & \mathbb{B}_{k_2}+\mathbb{B}_{k_1}\mathbb{D}_{p_{k3}}\mathbb{D}_{k_{p2}} & \mathbb{B}_{k_3}+\mathbb{B}_{k_1}\mathbb{D}_{p_{k3}}\mathbb{D}_{k_{p3}}
\end{array}
                                                     \right],
\\
\mathbb{C}_{\rm cl} =&\; \left[\begin{array}{@{}cc@{}}\mathbb{C}_{p} & \mathbb{C}_{k}
\end{array}
\right] ,~~\mathbb{D}_{\rm cl} = \mathbb{D}_z.
\eea
\eeq

\subsection{An example}

\label{subsec:coherent feedback example}

In this subsection, we use one example to demonstrate the coherent feedback network in  Fig. \ref{sec-PK}.

A hybrid atom-optomechanical system has recently been implemented \cite[Fig.
1A]{TBCGKP2020}, in which a laser beam is used to realize couplings between an
atomic spin ensemble and a micromechanical membrane. Strong coupling between
these two subsystems is successfully realized in a room-temperature
environment. Interesting physical phenomena, such as normal-mode splitting,
coherent energy exchange between the atomic ensemble and the micromechanical
membrane, and two-mode thermal noise squeezing, are observed.

\begin{figure}[tbph]
\centering
\includegraphics[width=0.70\textwidth]{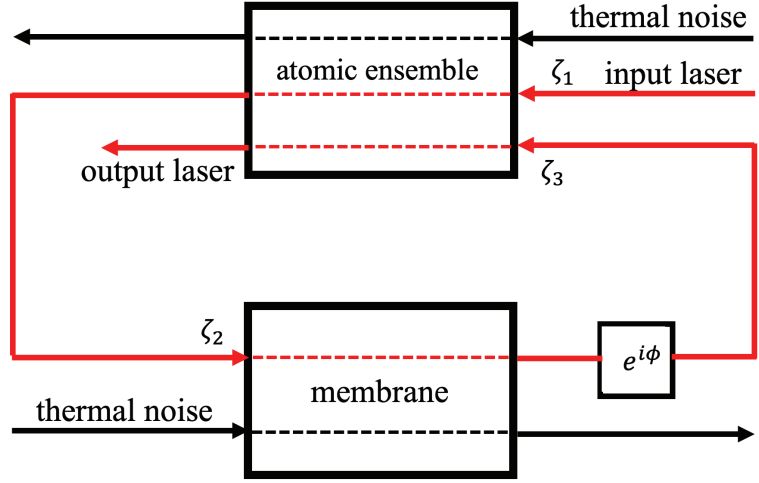}
\caption{The system experimentally realized in \cite{TBCGKP2020}.}
\label{fig:experiment}
\end{figure}

This hybrid system is  depicted in  Fig. \ref{fig:experiment}.  Compared with the coherent feedback network in Fig.   \ref{sec-PK},   the atomic ensemble corresponds to the plant $P$, whereas the membrane corresponds to the controller $K$.  There is no direct coupling Hamiltonian $H_{\mathrm{int}}$ between the atomic ensemble and the membrane.

The atomic ensemble is modeled as a single-mode quantum mechanical oscillator  which is parametrized by
\begin{equation}
\label{apr3-1}
S_p=I_3, ~ \mbf{L}_p=\left[
\bey{@{}c@{}}
\mbf{L}_{p1}\\
\mbf{L}_{p2}\\
\mbf{L}_{p3}
\eey
\right]= \left[\begin{array}{@{}c@{}}                   \sqrt{\gamma_s}\mbf{q}_s \\
                   \sqrt{2\Gamma_s}\mbf{q}_s \\
                   \sqrt{2\Gamma_s}\mbf{q}_s
\end{array}
               \right], ~ \mbf{H}_p=\frac{\Omega_s}{2}(\mbf{q}_s^2+\mbf{p}_s^2),
\end{equation}
where $\mbf{q}_s$ and  $\mbf{p}_s$ are real quadrature operators of the
atomic ensemble. The input laser beam is parametrized by,
\cite{CKS17} and \cite[Appendix C]{WSU+18}
\begin{equation}
\label{apr3-2}
S_l=1, ~ \mbf{L}_l=\sqrt{\kappa_{\mathrm{ext}}}\mbf{a}_l+\alpha I, ~ \mbf{H}_l=0,
\end{equation}
where $\mbf{a}_l$ is the annihilation operator of the laser and $\alpha\in \mathbb{C}$. Thus, $w_p$ in  Fig. \ref{sec-PK} is $\mbf{L}_l$ in Eq.~\eqref{apr3-2}.  By the concatenation product and the series product,\footnote{Given two open quantum systems $G_1\triangleq(\mbf{S}_1,\mbf{L}_1,\mbf{H}_1)$ and $G_2\triangleq(\mbf{S}_2,\mbf{L}_2,\mbf{H}_2)$ where $\mbf{S}_j,\mbf{L}_j,\mbf{H}_j$ are operators on the Hilbert space of the system $G_j$ ($j=1,2$), their concatenation product is defined to be
\beq
G_1\boxplus G_2\triangleq\left(\left[\begin{array}{cc}\boldsymbol{S}_1&0\\0&\boldsymbol{S}_2\end{array}\right],\left[\begin{array}{c}\boldsymbol{L}_1\\ \boldsymbol{L}_2\end{array}\right],\boldsymbol{H}_1+\boldsymbol{H}_2\right),
\eeq
and  their series product is defined to be
\beq\la{eq:Series_product}
G_2\vartriangleleft G_1\triangleq \left(\mbf{S}_2\mbf{S}_1, \mbf{L}_2+\mbf{S}_2 \mbf{L}_1, \mbf{H}_1+\mbf{H}_2+\ff{1}{2\imath}(\mbf{L}_2^\dagger \mbf{S}_2 \mbf{L}_1 - \mbf{L}_1^\dagger \mbf{S}_2^\dagger \mbf{L}_2 )\right).\eeq
In this paper, the term $\ff{1}{2\imath}(\mbf{L}_2^\dagger \mbf{S}_2 \mbf{L}_1 - \mbf{L}_1^\dagger \mbf{S}_2^\dagger \mbf{L}_2 )$ is called the \textit{interaction Hamiltonian}.
See \cite{GJ09b,GJ09}, \cite{ZJ12} and \cite{CKS17} for more detailed discussions
on coherent feedback connections.} the cascaded $P\vartriangleleft \mathrm{laser}$ system  is
\begin{equation}
\label{apr3-3}
\begin{aligned}
&S_c=I_3, ~ \mbf{L}_c =\left[
\bey{@{}c@{}}
\mbf{L}_{c1}\\
\mbf{L}_{c2}\\
\mbf{L}_{c3}
\eey
\right]=\left[\begin{array}{@{}c@{}}                   \sqrt{\gamma_s}\mbf{q}_s \\
                   \sqrt{2\Gamma_s}\mbf{q}_s+\sqrt{\kappa_{\mathrm{ext}}}\mbf{a}_l+\alpha I \\
                   \sqrt{2\Gamma_s}\mbf{q}_s
\end{array}
               \right], \\
& \mbf{H}_c=\frac{\Omega_s}{2}(\mbf{q}_s^2+\mbf{p}_s^2)+\sqrt{\kappa_{\mathrm{ext}}\Gamma_s}\mbf{q}_s\mbf{p}_l.
\end{aligned}
\end{equation}
It can be seen that the last term of $\mbf{H}_c$, namely $\sqrt{\kappa_{\mathrm{ext}}\Gamma_s}\mbf{q}_s\mbf{p}_l$,
describes the interaction Hamiltonian between the atomic ensemble and the
light, which is consistent to the form given in \cite[Eq. (S4)]{TBCGKP2020}.

Let $\mbf{x}_c$ be the real quadrature operators of the cascaded system $P\vartriangleleft \mathrm{laser}$ and $\mbf{u}_c$ be the input quadrature operators. Specifically,
\begin{equation}
\label{apr3-4}
\begin{aligned}
&\mbf{x}_c=\left[\begin{array}{@{}cccc@{}}              \mbf{q}_s & \mbf{q}_l & \mbf{p}_s & \mbf{p}_l
\end{array}
          \right]^\top, \\
&\mbf{u}_c=\left[\begin{array}{@{}cccccc@{}}               \mbf{q}_s^{(\mathrm{th})} & \mbf{q}_{s,\zeta_1} & \mbf{q}_{s,\zeta_3} & \mbf{p}_s^{(\mathrm{th})} & \mbf{p}_{s,\zeta_1} & \mbf{p}_{s,\zeta_3}
\end{array}
           \right]^\top,
\end{aligned}
\end{equation}
where $(\mbf{q}_s^{(\mathrm{th})}, \mbf{p}_s^{(\mathrm{th})})$ is input thermal noise, $(\mbf{q}_{s,\zeta_1},\mbf{p}_{s,\zeta_1})$ (at position $\zeta_1$) and  $(\mbf{q}_{s,\zeta_3}, \mbf{p}_{s,\zeta_3})$ (at position $\zeta_3$) denote the second and third inputs of the atomic spin ensemble, respectively. In the notation used in   \ref{sec-PK}, we have system parameters
\beq\la{eq:apr16_cprresponse_1}
\left[
\bey{@{}c@{}}
 \mbf{q}_s^{(\mathrm{th})}\\
  \mbf{p}_s^{(\mathrm{th})}
\eey
\right] = V_1 \breve{b}_{p1}, \ \
\left[
\bey{@{}c@{}}
 \mbf{q}_{s,\zeta_1}\\
\mbf{p}_{s,\zeta_1}
\eey
\right] = V_1 \breve{b}_{p2}, \ \
\left[
\bey{@{}c@{}}
 \mbf{q}_{s,\zeta_3}\\
\mbf{p}_{s,\zeta_3}
\eey
\right] = V_1 \breve{b}_{p3},
\eeq
where the unitary matrix $V_1$ is defined in Eq.~\eqref{eq:unitary_V}. By the unitary transformations in \eqref{apr31}, we have
\begin{align}
\label{apr3-5}
& \Lambda_c=\left[\begin{array}{@{}cccc@{}}              \sqrt{\gamma_s} & 0 & 0 & 0 \\
              \sqrt{2\Gamma_s} & \frac{\sqrt{2\kappa}}{2} & 0 & \frac{\sqrt{2\kappa}}{2}\imath \\
              \sqrt{2\Gamma_s} & 0 & 0 & 0
\end{array}
          \right],\nonumber\\
          & ~ \mathbb{H}_c=\left[\begin{array}{@{}cccc@{}}                                      \Omega_s & 0 & 0 & \sqrt{\kappa_{\mathrm{ext}}\Gamma_s} \\
                                      0 & 0 & 0 & 0 \\
                                      0 & 0 & \Omega_s & 0 \\
                                      \sqrt{\kappa_{\mathrm{ext}}\Gamma_s} & 0 & 0 & 0
\end{array}
                                  \right], ~\mathbb{K} =0.
\end{align}
Consequently, by Eq. \eqref{eq:real_sys_ABCD} the system matrices can be calculated as
\begin{equation}
\label{apr3-6}
\begin{aligned}
&\mathbb{D}_c = I_6, ~\mathbb{C}_c=\left[\begin{array}{@{}cccc@{}}               \sqrt{2\gamma_s} & 0 & 0 & 0 \\
               2\sqrt{2\Gamma_s} & \sqrt{\kappa_{\mathrm{ext}}} & 0 & 0 \\
               2\sqrt{2\Gamma_s} & 0 & 0 & 0 \\
               0 & 0 & 0 & 0 \\
               0 & 0 & 0 & \sqrt{\kappa_{\mathrm{ext}}} \\
               0 & 0 & 0 & 0
\end{array}
           \right], \\
&\mathbb{B}_c=-\left[\begin{array}{@{}cccccc@{}}                  0 & 0 & 0 & 0 & 0 & 0 \\
                  0 & \sqrt{\kappa_{\mathrm{ext}}} & 0 & 0 & 0 & 0 \\
                  0 & 0 & 0 & \sqrt{2\gamma_s} & 2\sqrt{2\Gamma_s} & 2\sqrt{2\Gamma_s} \\
                  0 & 0 & 0 & 0 & \sqrt{\kappa_{\mathrm{ext}}} & 0
\end{array}
              \right], \\
&\mathbb{A}_c=\left[\begin{array}{@{}cccc@{}}                 0 & 0 & \Omega_s & 0 \\
                 (1-\sqrt{2})\sqrt{\kappa_{\mathrm{ext}}\Gamma_s} & -\frac{\kappa_{\mathrm{ext}}}{2} & 0 & 0 \\
                 -\Omega_s & 0 & 0 & -(1+\sqrt{2})\sqrt{\kappa_{\mathrm{ext}}\Gamma_s} \\
                 0 & 0 & 0 & -\frac{\kappa_{\mathrm{ext}}}{2}
\end{array}
             \right],
\end{aligned}
\end{equation}
which yield the linear QSDEs that describe the dynamics of the atomic spin ensemble in the real quadrature operator representation
\begin{equation}
\label{apr3-7}
\begin{aligned}
\dot{\mbf{q}}_s=&\Omega_s\mbf{p}_s, \\
\dot{\mbf{p}}_s=&-\Omega_s\mbf{q}_s-(1+\sqrt{2})\sqrt{\kappa_{\mathrm{ext}}\Gamma_s}\mbf{p}_l-2\sqrt{2\Gamma_s}\mbf{p}_{s,\zeta_1} \\
&- \ \  2\sqrt{2\Gamma_s}\mbf{p}_{s,\zeta_3}-\sqrt{2\gamma_s}\mbf{p}_s^{(\mathrm{th})}.
\end{aligned}
\end{equation}
According to Eq.~\eqref{eq:real_sys}, the output  of the atomic ensemble is given by
\begin{equation}
\label{apr3-8}
\begin{aligned}
\mathbb{C}_c\mbf{x}_c+\mbf{u}_c=\left[\begin{array}{@{}c@{}}                                \sqrt{2\gamma_s}\mbf{q}_s+\mbf{q}_s^{(\mathrm{th})} \\
                                2\sqrt{2\Gamma_s}\mbf{q}_s+\sqrt{\kappa_{\mathrm{ext}}}\mbf{q}_l+\mbf{q}_{s,\zeta_1} \\
                                2\sqrt{2\Gamma_s}\mbf{q}_s+\mbf{q}_{s,\zeta_3} \\
                                \mbf{p}_s^{(\mathrm{th})} \\
                                \sqrt{\kappa_{\mathrm{ext}}}\mbf{p}_l+\mbf{p}_{s,\zeta_1} \\
                                \mbf{p}_{s,\zeta_3}
\end{array}
                            \right].
\end{aligned}
\end{equation}
As shown in Fig. \ref{fig:experiment}, the second output of the atomic spin ensemble, which is
\beq
\left[\bey{@{}c@{}}2\sqrt{2\Gamma_s}\mbf{q}_s+\sqrt{\kappa_{\mathrm{ext}}}\mbf{q}_l+\mbf{q}_{s,\zeta_1}\\
\sqrt{\kappa_{\mathrm{ext}}}\mbf{p}_l+\mbf{p}_{s,\zeta_1}
  \eey\right] =V_1 \breve{\mbf{b}}_{\mathrm{out},p_k}
\eeq
  with $\mbf{b}_{\mathrm{out},p_k}$ being that in  Fig. \ref{sec-PK},  is sent to the micromechanical membrane. As a result, we have
\begin{equation}
\label{apr3-9}
\begin{aligned}
\left[\begin{array}{@{}c@{}}    \mbf{q}_{m,\zeta_2}(t) \\
    \mbf{p}_{m,\zeta_2}(t)
\end{array}
\right]=\left[\begin{array}{@{}c@{}}            2\sqrt{2\Gamma_s}\mbf{q}_s(t-\tau)+\sqrt{\kappa_{\mathrm{ext}}}\mbf{q}_l(t-\tau)+\mbf{q}_{s,\zeta_1}(t-\tau) \\
            \sqrt{\kappa_{\mathrm{ext}}}\mbf{p}_l(t-\tau)+\mbf{p}_{s,\zeta_1}(t-\tau)
\end{array}
        \right],
\end{aligned}
\end{equation}
where $\tau$ denotes the time delay from the atomic spin ensemble to the micromechanical membrane in the feedback loop.  Here, $
\begin{smallmatrix}
\left[\bey{@{}c@{}}  \mbf{q}_{m,\zeta_2}(t) \\
    \mbf{p}_{m,\zeta_2}(t)
  \eey\right]
\end{smallmatrix}
=V_1 \breve{\mbf{b}}_{k_1}$ with $ \breve{\mbf{b}}_{k_1}$ being that in  Fig. \ref{sec-PK}, is the first input to the micromechanical membrane in Fig.  \ref{fig:experiment}.

On the other hand, the micromechanical membrane in  Fig. \ref{fig:experiment}, which is a single-mode quantum harmonic oscillator too,  can be parametrized by
\begin{equation}
\label{apr3-10}
\begin{aligned}
S_m=I_2, ~   \mbf{L}_m=\left[
\bey{@{}c@{}}
\mbf{L}_{k1}\\
\mbf{L}_{k2}
\eey
\right]=\left[\begin{array}{@{}c@{}}                               -\imath\sqrt{2\Gamma_m}\mbf{q}_m  \\
                               \sqrt{\gamma_m}\mbf{q}_m
\end{array}
                           \right], ~ \mbf{H}_m=\frac{\Omega_m}{2}(\mbf{q}_m^2+\mbf{p}_m^2),
\end{aligned}
\end{equation}
where $\mbf{q}_m$ and  $\mbf{p}_m$ are real quadrature operators of the micromechanical membrane.
     Notice that
\beq
\ff{\mbf{L}_{k1}^\ast \mbf{L}_{c2}- \mbf{L}_{c2}^\ast \mbf{L}_{k1}}{2\imath} = 2\sqrt{\Gamma_s\Gamma_m} \mbf{q}_s\mbf{q}_m +\sqrt{\Gamma_m \kappa_{\mathrm{ext}}}\mbf{q}_m\mbf{q}_l + \sqrt{2\Gamma_m} \mbf{q}_m \ff{\alpha + \alpha^\ast}{2}.
\eeq
The coupling between the second output channel of the atomic ensemble and the
micromechanical membrane generates several interaction Hamiltonian terms, among
which the second term $\sqrt{\Gamma_m \kappa_{\mathrm{ext}}}\mbf{q}_m\mbf{q}_l$  is consistent with the form given
in \cite[Eq. (S14)]{TBCGKP2020}. As the coupling happens at $\zeta_2$ in Fig.
\ref{fig:experiment}, $\mbf{q}_l$ can be written as $\mbf{q}_{l,\zeta_2}$ which
corresponds to $\mbf{X}_{L}(\zeta_m)$ in \cite[Eq. (S14)]{TBCGKP2020}.

Let $\mbf{x}_m$ be the real quadrature operators of the micromechanical membrane $K$ and $\mbf{u}_m$ be the input, i.e.,~\begin{equation}
\label{apr31-4}
\begin{aligned}
&\mbf{x}_m=\left[\begin{array}{@{}cc@{}}              \mbf{q}_m & \mbf{p}_m
\end{array}
          \right]^\top, \\
&\mbf{u}_m=\left[\begin{array}{@{}cccc@{}}               \mbf{q}_{m,\zeta_2} & \mbf{q}_m^{(\mathrm{th})} & \mbf{p}_{m,\zeta_2} & \mbf{p}_m^{(\mathrm{th})}
\end{array}
           \right]^\top,
\end{aligned}
\end{equation}
where $(\mbf{q}_m^{(\mathrm{th})}, \mbf{p}_m^{(\mathrm{th})})$ is input thermal noise, and $(\mbf{q}_{m,\zeta_2}, \mbf{p}_{m,\zeta_2})$ (at position $\zeta_2$) denotes the first input of the micromechanical membrane, as introduced in Eq.~\eqref{apr3-9}. In the notation used in  Fig. \ref{sec-PK}, we have
\beq
\left[
\bey{@{}c@{}}
  \mbf{q}_{m,\zeta_2}\\
   \mbf{p}_{m,\zeta_2}
\eey
\right] = V_1 \breve{b}_{k1}, \ \
\left[
\bey{@{}c@{}}
 \mbf{q}_m^{(\mathrm{th})}\\
 \mbf{p}_m^{(\mathrm{th})}
\eey
\right] = V_1 \breve{b}_{k2},
\eeq
and there is no input channel associated with $b_{k3}$.

Similarly, the system matrices of the micromechanical membrane $K$ can be calculated as
\begin{equation}
\label{apr3-11}
\begin{aligned}
&\mathbb{D}_m = I_4, ~ \mathbb{C}_m=\left[\begin{array}{@{}cc@{}}                 0 & 0 \\
                 \sqrt{2\gamma_m} & 0 \\
                 -2\sqrt{\Gamma_m} & 0 \\
                 0 & 0
\end{array}
             \right], \\
&\mathbb{B}_m=-\left[\begin{array}{@{}cccc@{}}                  0 & 0 & 0 & 0 \\
                  2\sqrt{\Gamma_m} & 0 & 0 & \sqrt{2\gamma_m}
\end{array}
              \right],  \ \
\mathbb{A}_m=\left[\begin{array}{@{}cc@{}}                 0 & \Omega_m \\
                 -\Omega_m & 0
\end{array}
             \right],
\end{aligned}
\end{equation}
which yields the linear QSDEs that describe the dynamics of the micromechanical membrane in the real quadrature operator representation
\begin{equation}
\label{apr3-12}
\begin{aligned}
\dot{\mbf{q}}_m&=\Omega_m\mbf{p}_m, \\
\dot{\mbf{p}}_m&=-\Omega_m\mbf{q}_m-2\sqrt{\Gamma_m}\mbf{q}_{m,\zeta_2}-\sqrt{2\gamma_m}\mbf{p}_m^{(\mathrm{th})}.
\end{aligned}
\end{equation}
Substituting \eqref{apr3-9} into \eqref{apr3-12}, we have
\begin{equation}
\label{apr3-122}
\begin{aligned}
\dot{\mbf{q}}_m(t)=&\Omega_m\mbf{p}_m(t), \\
\dot{\mbf{p}}_m(t)=&-\Omega_m\mbf{q}_m(t)-4\sqrt{2\Gamma_s\Gamma_m}\mbf{q}_s(t-\tau)-2\sqrt{\kappa_{\mathrm{ext}}\Gamma_m}\mbf{q}_l(t-\tau) \\
&- \ \  2\sqrt{\Gamma_m}\mbf{q}_{s,\zeta_2}(t)-\sqrt{2\gamma_m}\mbf{p}_m^{(\mathrm{th})}(t).
\end{aligned}
\end{equation}
Moreover, the  output  of the micromechanical membrane is
\begin{equation}
\label{apr3-13}
\begin{aligned}
\mathbb{C}_m\mbf{x}_m+\mbf{u}_m=\left[\begin{array}{@{}c@{}}                                    \mbf{q}_{m,\zeta_2} \\
                                    \sqrt{2\gamma_m}\mbf{q}_m+\mbf{q}_m^{(\mathrm{th})} \\
                                    -2\sqrt{\Gamma_m}\mbf{q}_m+\mbf{p}_{m,\zeta_2} \\
                                    \mbf{p}_m^{(\mathrm{th})}
\end{array}
                                \right].
\end{aligned}
\end{equation}
The first output of the micromechanical membrane, which is
\beq
\left[\bey{@{}c@{}}\mbf{q}_{m,\zeta_2}\\
-2\sqrt{\Gamma_m}\mbf{q}_m+\mbf{p}_{m,\zeta_2}
  \eey\right]=V_1 \breve{\mbf{b}}_{\mathrm{out},k_p}
\eeq
 with $\mbf{b}_{\mathrm{out},k_p}$ being that in  Fig. \ref{sec-PK},  is sent to the atomic spin ensemble. Noticing the phase shifter $e^{\imath \phi}$ on the way,  we have
\begin{equation}
\label{apr3-14}
\begin{aligned}
b_{p_3}(t)=e^{\imath\phi}b_{\mathrm{out},k_p}(t-\tau).
\end{aligned}
\end{equation}
By Eqs.~\eqref{apr3-13}, \eqref{apr3-14} and noticing Eq.~\eqref{eq:apr16_cprresponse_1}, we get
\begin{equation}
\label{apr3-15}
\begin{aligned}
\mbf{q}_{s,\zeta_3}(t)&=\cos\phi \;\mbf{q}_{m,\zeta_2}(t-\tau)-\sin\phi\left[-2\sqrt{\Gamma_m}\mbf{q}_m(t-\tau)+\mbf{p}_{m,\zeta_2}(t-\tau)\right], \\
\mbf{p}_{s,\zeta_3}(t)&=\cos\phi\left[-2\sqrt{\Gamma_m}\mbf{q}_m(t-\tau)+\mbf{p}_{m,\zeta_2}(t-\tau)\right]+\sin\phi\mbf{q}_{m,\zeta_2}(t-\tau).
\end{aligned}
\end{equation}
Substituting Eq. \eqref{apr3-9} into Eq. \eqref{apr3-15}, yields
\begin{equation}
\label{apr3-16}
\begin{aligned}
\mbf{p}_{s,\zeta_3}(t)=&-2\cos\phi\sqrt{\Gamma_m}\mbf{q}_m(t-\tau)+2\sin\phi\sqrt{2\Gamma_s}\mbf{q}_s(t-2\tau) \\
&+ \ \  \cos\phi\sqrt{\kappa_{\mathrm{ext}}}\mbf{p}_l(t-2\tau)+\sin\phi\sqrt{\kappa_{\mathrm{ext}}}\mbf{q}_l(t-2\tau) \\
&+ \ \  \cos\phi\mbf{p}_{s,\zeta_1}(t-2\tau)+\sin\phi\mbf{q}_{s,\zeta_1}(t-2\tau).
\end{aligned}
\end{equation}
Thus, Eq. \eqref{apr3-7} can be rewritten as
\begin{equation}
\label{apr3-17}
\begin{aligned}
\dot{\mbf{q}}_s(t)=&\Omega_s\mbf{p}_s(t), \\
\dot{\mbf{p}}_s(t)=&-\Omega_s\mbf{q}_s(t)-(1+\sqrt{2})\sqrt{\kappa_{\mathrm{ext}}\Gamma_s}\mbf{p}_l(t)-2\cos\phi\sqrt{2\kappa_{\mathrm{ext}}\Gamma_s}\mbf{p}_l(t-2\tau) \\
&- \ \  2\sin\phi\sqrt{2\kappa_{\mathrm{ext}}\Gamma_s}\mbf{q}_l(t-2\tau)-8\sin\phi\Gamma_s\mbf{q}_s(t-2\tau) \\
&+ \ \  4\cos\phi\sqrt{2\Gamma_s\Gamma_m}\mbf{q}_m(t-\tau) \\
&- \ \  2\sqrt{2\Gamma_s}\left[\mbf{p}_{s,\zeta_1}(t)+\cos\phi\mbf{p}_{s,\zeta_3}(t)+\sin\phi\mbf{q}_{s,\zeta_3}(t)\right]-\sqrt{2\gamma_s}\mbf{p}_s^{(\mathrm{th})}(t).
\end{aligned}
\end{equation}

Despite of the terms containing laser loss $\kappa_{\mathrm{ext}}$, the dynamical
equations of the micromechanical membrane \eqref{apr3-122} and the atomic spin
ensemble \eqref{apr3-17} are consistent with the forms given in \cite[Eqs.
(S54--S58)]{TBCGKP2020}.

\begin{remark}
In the following, we only look into the input--output channel with the laser, i.e.~ignoring the thermal noise inputs.  Notice that $\mbf{L}_{p2}=\sqrt{2\Gamma_s}\mbf{q}_s$ is the coupling between the spin and the input light, and $\mbf{L}_{k1}=-\imath\sqrt{2\Gamma_m}\mbf{q}_m$ is the membrane--light coupling. The coherent feedback loop can be divided into two parts. The first part is from the spin to the membrane, according to Eq.~\eqref{eq:Series_product} whose interaction Hamiltonian is given by
\begin{equation}
\label{apr14-1}
\mbf{H}_{\mathrm{int}}^{(1)}=\frac{1}{2\imath}(\mbf{L}_{k1}^{\ast}\mbf{L}_{p2}-\mbf{L}_{p2}^{\ast}\mbf{L}_{k1})
=2\sqrt{\Gamma_m\Gamma_s}\mbf{q}_m\mbf{q}_s,
\end{equation}
and the cascaded coupling operator $\mbf{L}^{(1)}=\mbf{L}_{k1}+\mbf{L}_{p2}$. The second part is from the membrane to the spin with the phase shifter $e^{\imath \phi}$  on the way, the corresponding interaction Hamiltonian is
\begin{equation}
\label{apr14-2}
\mbf{H}_{\mathrm{int}}^{(2)}=\frac{1}{2\imath}(\mbf{L}_{p3}^{\ast}e^{\imath\phi}\mbf{L}^{(1)}-\mbf{L}^{(1)\ast}e^{-\imath\phi}\mbf{L}_{p3})
=-2\sqrt{\Gamma_m\Gamma_s}\cos\phi\mbf{q}_s\mbf{q}_m+2\sin\phi\Gamma_s\mbf{q}_s^2,
\end{equation}
and the cascaded coupling operator
\begin{equation}
\label{apr14-7}
\mbf{L}^{(2)}=\mbf{L}_{p3}+e^{\imath\phi}\mbf{L}^{(1)}=-\imath e^{\imath\phi}\sqrt{2\Gamma_m}\mbf{q}_m+(1+e^{\imath\phi})\sqrt{2\Gamma_s}\mbf{q}_s,
\end{equation}
which is consistent with the collective jump operator $\mbf{J}$ used
 in \cite[Eq. (1)]{TBCGKP2020}. Combining Eq. \eqref{apr14-1} with Eq.
 \eqref{apr14-2}, the interaction Hamiltonian between the atomic spin
 ensemble and the micromechanical membrane is
\begin{equation}
\label{apr14-3}
\mbf{H}_{sm}=(1-\cos\phi)2\sqrt{\Gamma_m\Gamma_s}\mbf{q}_s\mbf{q}_m+2\sin\phi\Gamma_s\mbf{q}_s^2,
\end{equation}
which is consistent with the effective interaction Hamiltonian $\mbf{H}_{\rm eff}$
 used in \cite[Eq. (1)]{TBCGKP2020}.  In summary, all the essential
 equations in \cite{TBCGKP2020} can be reproduced by means of the
 quantum linear systems and network theory introduced in this tutorial.
\end{remark}

\section{Conclusion}\label{sec:conclusion}

In this tutorial, we have given a concise introduction to linear quantum systems, for example, their mathematical models,  relation between their control-theoretic properties and physical properties,  Gaussian states,  quantum Kalman filter,  Kalman canonical form, and response to continuous-mode single-photon states. Several simple examples are designed to demonstrate some fundamental properties of linear quantum systems.  Pointers to more detailed discussions are given in various places. It is hoped that this tutorial is helpful to researchers in the control community who are interested in quantum control of dynamical systems. Finally, an information-theoretic uncertainty relation has been recorded in this tutorial, which describes uncertainties of mixed quantum Gaussian states better than the Heisenberg uncertainty relation. It is an open question whether this uncertainty relation is useful for mixed quantum Gaussian state engineering.

\medskip
\textbf{Acknowledgement.} We would like to thank Bo Qi, Liying Bao and Shuangshuang Fu for their careful reading and constructive suggestions.

\bibliographystyle{plain}
\bibliography{gzhang}

\end{document}